%% file: thesis.tex
\documentclass[12pt,final]{dukephd}
\usepackage[final]{epsfig}
\usepackage{float}
\usepackage{amsbsy}
\usepackage{afterpage}
\usepackage{amssymb}
\usepackage{amsmath}
\usepackage{indentfirst}

\ssp

\input{symdef}

\begin{document}

  \input{\chaptdir/front}

\addcontentsline{toc}{chapter}{Coulomb Blockade in Quantum Dots}
%
  \input{\chaptdir/chapter1}

%
  \input{\chaptdir/chapter2}

%
  \input{\chaptdir/chapter3}

\addcontentsline{toc}{chapter}{Quantum Computation with Quantum
Dots: Mesoscopic Fluctuations and Decoherence}
%
  \input{\chaptdir/chapter4}

%
  \input{\chaptdir/chapter5}
%
  \input{\chaptdir/chapter6}

%
  \input{\chaptdir/chapter7}

%
  \input{\chaptdir/conclusion}

\addcontentsline{toc}{chapter}{Appendices}
  \appendix
%
  \input{\chaptdir/appendix_ch2}

%
  \input{\chaptdir/appendix_phonons}

%
  \input{\chaptdir/appendix_ch7}

%

  \setlength{\textheight}{8.0in}
  \clearpage
  \ssp
\bibliographystyle{ieeetr}
  \bibliography{all}

  \input{\chaptdir/bio}

\end{document}

%% file: symdef.tex
\gdef\figdir {images}


%% file: chapters/front.tex
\begin{frontmatter}
  \title{Quantum Dots: Coulomb Blockade, \\
         Mesoscopic Fluctuations, and Qubit Decoherence}
  \author{Serguei Vorojtsov}
  \degreeyear{2005}
  \duketitle
  \begin{abstract}
    \addcontentsline{toc}{chapter}{Abstract}
    \input{chapters/abstract}

  \end{abstract}
  \begin{acknowledgements}
    \addcontentsline{toc}{chapter}{Acknowledgments}
    \input{chapters/ack}

  \end{acknowledgements}
  \tableofcontents
  \begin{dedication}
    \vspace*{1in}
  To my grandmother Vorozhtsova, Tatyana Ivanovna (1917-1987)
  \end{dedication}
%
\end{frontmatter}
{\normalsize }

%% file: chapters/abstract.tex
\begin{center}
{\bf Quantum Dots: Coulomb Blockade, \\
Mesoscopic Fluctuations, and Qubit Decoherence}
\end{center}

\ssp
The continuous minituarization of integrated 
circuits is going to affect the underlying physics
of the future computers. 
This new physics first came into play as
the effect of Coulomb blockade 
in electron transport
through small conducting islands. 
Then, as the size of the island $L$ 
continued to shrink further, 
the quantum phase coherence length 
became larger than $L$ 
leading to mesoscopic fluctuations
-- fluctuations of the island's quantum mechanical
properties upon small external perturbations.
Quantum coherence of the mesoscopic systems
is essential for building reliable quantum
computer.
Unfortunately, one can not completely 
isolate the system from the environment and 
its coupling to the environment inevitably
leads to the loss of coherence or decoherence.
All these effects are to be thoroughly investigated 
as the potential of the future applications
is enormous.

In this thesis I find an analytic expression for 
the conductance of a single electron transistor 
in the regime when temperature, level spacing, 
and charging energy of an island are all of the same order. 
I also study the correction to the spacing 
between Coulomb blockade peaks 
due to finite dot-lead tunnel couplings.
I find analytic expressions for both
correction to the spacing
averaged over mesoscopic fluctuations 
and the rms of the correction fluctuations.

In the second part of the thesis
I discuss the feasibility 
of quantum dot based spin- and charge-qubits.
Firstly, I study the effect of mesoscopic fluctuations 
on the magnitude of errors that can occur 
in exchange operations on quantum dot spin-qubits. 
Mid-size double quantum dots, 
with an odd number of electrons 
in the range of a few tens in each dot, 
are investigated through the constant interaction model 
using realistic parameters. 
It is found that 
the number of independent parameters per dot that 
one should tune depends on the configuration and 
ranges from one to four.
Then, I study decoherence of a quantum dot charge qubit 
due to coupling to piezoelectric acoustic phonons 
in the Born-Markov approximation.
After including appropriate form factors, 
I find that phonon decoherence rates are 
one to two orders of magnitude weaker 
than was previously predicted.
My results suggest that mechanisms other than 
phonon decoherence play a more significant role
in current experimental setups.

%% file: chapters/ack.tex
\ssp
First and foremost, I would like to thank my adviser 
Prof.~Harold~U. Baranger for his thoughtful guidance
on every aspect of my research and for the funding 
of my work all these years. 
I am also grateful to Prof.~Konstantin~A. Matveev
for investing so much time in my education and
his guidance on my first research project at Duke.
Prof.~Eduardo~R. Mucciolo has been a wonderful 
collaborator and friend. He gave me the confidence
that I can publish in Physical Review B. Without
him this thesis would have been much thinner.

I am grateful to Profs.~Berndt M\"{u}ller and 
M.~Ronen Plesser for helping me to secure 
TA positions when I needed them most.

I am grateful to the members of my Ph.D. committee:
Profs.~Shailesh Chandrasekharan, Albert~M. Chang, 
Gleb Finkelstein, and Weitao Yang for finding
time to be on my committee and valuable comments 
on the manuscript.

During my years at Duke, I benefited from stimulating
discussions with Alexey Bezryadin, Alexander~M. Finkelstein, 
Martina Hentschel, Alexei Kaminski, Eduardo Novais, 
Stephen~W. Teitsworth, Denis Ullmo, Gonzalo Usaj, 
and Frank~K. Wilhelm.

I would like to thank my friends: Sven Rinke, 
Alex Makarovski, Kostya Sabourov, Anand Priyadarshee, 
Sung Ha Park, Ji-Woo Lee, Martina Hentschel, 
Gonzalo Usaj, Eduardo Novais, Ribhu Kaul, 
and Oleg Tretiakov. 
Thank you, guys, for being there for me 
even when I did not ask. 
It is always fun to be around you!

Finally, a special thank you goes to my parents,
grandparents, and my wife Elena whose patience
helped me to get the job done. 
And, certainly, life makes much more sense
because my daughter Tanya is around.

%% file: chapters/chapter1.tex
\setlength{\textheight}{8.0in}
\clearpage
\chapter{Introduction to Quantum Dot Physics}
\label{ch:ch1}
\thispagestyle{botcenter}
\setlength{\textheight}{8.4in}

\section{Overview}

Quantum dot research~\cite{Kas92,Kas93,Ash96,Kou98} 
has developed into an exciting branch of mesoscopic physics.
Many novel phenomena were observed
in transport measurements through 
quantum dots: Coulomb blockade~\cite{Kou97}, 
even-odd asymmetry in Coulomb blockade 
peak spacings~\cite{Pat98b,Usa02}, 
and Kondo effect~\cite{Gol98a,Gla00}
to name just a few. 
The field has kept researchers busy for 
about twenty years now and still continues 
to surprise us. 

By no means can I provide a detailed introduction 
to the entire field in this thesis (even restricting myself 
to quantum dot physics alone) and I do not think 
it is necessary as there are quite a few nice 
review papers available~\cite{Kou97,Alh00,Del01,Ale02}. 
Instead, I give just enough introductory material 
on 2D lateral and 3D quantum dots 
so that the reader can jump into the chapters 
where the original results of my research are presented.

\section{2D Lateral Quantum Dots}

\begin{figure}
\begin{center}
\includegraphics[width=13.2cm]{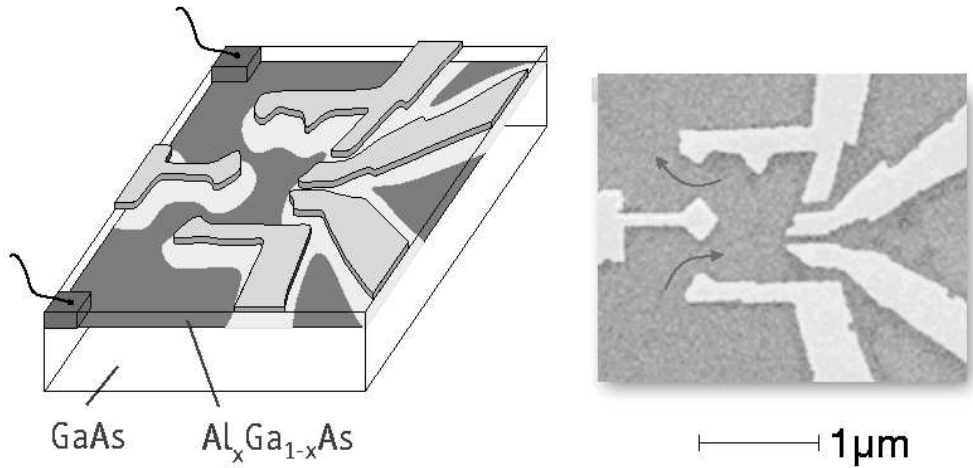}
\caption{2D lateral quantum dot.
Left panel: Schematic picture of the quantum dot setup.
Two-dimensional electron gas is formed 
on the the AlGaAs/GaAs heterostructure interface.
Metal electrodes (or gates) are deposited on top
of the heterostructure. 
Negative voltages are applied to the electrodes
to repel the electron gas underneath -- 
bright regions contain no electrons.
Then, bias voltage is applied between source and drain
leads (see wires coming out of the sample) 
and the current through the quantum dot is measured. 
Right panel: Micrograph of a real quantum dot~\cite{Kou98}. 
Bright regions correspond to the deposited metal 
electrodes. Arrows show the flow of electrons.
Copyright by C.~M.~Marcus, from Ref.~\cite{Kou98}.}
\label{fig:ch1_lateral-QD}
\end{center}
\end{figure}

Recent advances in materials science made possible 
the fabrication of small conducting devices known as 
quantum dots. 
In particular, in 2D lateral quantum dots
from one to several thousand electrons 
are confined to a spatial region whose linear size is 
from about 40$\,$nm to 1$\,\mu$m~\cite{Kas93,Cio00}. 
These quantum dots are typically made by 
(i)\,forming a two-dimensional electron gas 
on the interface of semiconductor heterostructure 
and (ii)\,applying electrostatic potential 
to the metal surface electrodes 
to further confine the electrons to a small region 
(quantum dot) in the interface plane~\cite{Kou98}, 
see Fig.~\ref{fig:ch1_lateral-QD}.

The transport properties of a quantum dot can be measured by coupling it 
to leads and passing current through the dot. The electron's phase is
preserved over distances that are large compared with the size of the  system
(quantum coherence), giving rise to new phenomena not observed in macroscopic
conductors. 

The coupling between a quantum dot and 
its leads can be experimentally controlled. 
In an open dot, the coupling is strong and the movement 
of electrons across dot-lead junctions is classically allowed. 
However, when the point contacts are pinched off, 
effective barriers are formed and
conduction occurs only by tunneling. 
In these almost-isolated or closed quantum dots, 
the charge of the dot is quantized and 
the dot's low-lying energy levels are discrete 
with their widths smaller than the spacing
between them, see Chapter~\ref{ch:ch2}.

The advantage of these artificial systems is that 
their transport properties are readily measured 
and all the parameters -- the strength of the dot-lead 
tunnel couplings, the number of electrons in the dot, 
and the dot's size and shape -- are under experimental control.

To observe quantization of the quantum dot charge, 
two conditions have to be satisfied.
Firstly, the barriers must be high enough 
so that the transmission is small. 
This gives the following condition for the conductance: 
$G \ll e^2/h$, that is, the dot must be almost or 
completely isolated. 
Secondly, the temperature must be low enough 
so that the effects of charge quantization are not washed out. 
The quantum dot's ability to hold charge 
is classically described by its average capacitance $C$. 
Since the energy required to add one electron 
is approximately $e^2/C$, 
we find the following condition: $T \ll e^2/C$. 

The tunneling of an electron onto the dot 
is normally blocked by the classical Coulomb repulsion 
with the electrons already in the dot; 
hence, the conductance is very small. 
This phenomenon is known as the Coulomb blockade. 
However, by changing the voltage of the back-gate $V_{g}$ 
one can compensate for this repulsion and, 
at the appropriate value of $V_{g}$, 
the charge of the dot can fluctuate 
between $n$ and $n+1$ electrons 
leading to a maximum in the conductance. 
Thus, one can observe 
Coulomb-blockade oscillations of the conductance 
as a function of the back-gate voltage, see Fig.~\ref{fig:ch2_gd0many}. 
At sufficiently low temperatures these oscillations
turn into sharp peaks that are spaced almost uniformly in $V_{g}$.
Their separation is approximately equal 
to the charging energy $e^{2}/C$.

\section{Constant Interaction Model and
Single-Particle Hamiltonian}
\label{sec:ch1_CI_model}

Electron-electron interactions in a quantum dot 
in the Coulomb-blockade regime 
are conventionally described by
the constant interaction (CI) model~\cite{Kou97}. 
In this model the Hamiltonian of the system 
is given by a sum of two terms:
(i)\,the electrostatic charging energy, 
which depends only on the total number of electrons 
in the dot $n$ and 
(ii)\,the Hamiltonian of free quasiparticles:
\begin{eqnarray}
H_{CI} = E_{C} {\hat n}^{2} 
+ \sum_{\alpha \sigma}\varepsilon_{\alpha} 
{\hat c}^{\dagger}_{\alpha\sigma}
{\hat c}_{\alpha\sigma},
\end{eqnarray}
where $E_{C} = e^{2}/2C$ and ${\hat c}_{\alpha\sigma}$ 
is the annihilation operator of a quasiparticle (electron) 
on orbital level $\alpha$ with spin $\sigma$.

Single-particle dynamics inside real 
2D lateral quantum dots with more than 40 electrons
has no particular symmetry 
due to irregular boundaries (chaotic quantum dot)
or the presence of impurities (disordered quantum dot).
In both of these cases, free quasiparticles inside the dot
can be described by random matrix theory 
(RMT): The Hamiltonian is chosen ``at random'' 
except for its fundamental space-time 
symmetry~\cite{Wig51,Dys62a,Dys62b,Dys62c,Meh91,Alh00}. 
Random matrix theory is applicable if
the dimensionless conductance of the dot $g$ is large: 
$g = E_{T}/\delta\varepsilon\gg 1$, 
where $E_{T}$ is the Thouless energy and 
$\delta\varepsilon$ is the mean level spacing in the quantum dot. 
For a ballistic quantum dot $E_{T}\sim\hbar v_{F}/L$, 
where $v_{F}$ is the Fermi velocity of the electrons 
and $L$ is the linear size of the dot. 
A large value of $g$ indicates that 
the dot can be treated as a good conductor.

The spacings between conductance peaks 
contain two contributions as well.
The first one, due to the charging energy, 
does not fluctuate much. 
The second contribution is proportional to 
the spacing between discrete energy levels 
in the quantum dot. 
This term does fluctuate and 
obeys the Wigner-Dyson statistics. 
Spin degeneracy of 
each one-particle energy level in the quantum dot 
leads to the even-odd parity effect: 
the second contribution appears only when 
we promote an electron to the next orbital.

\section{Constant Exchange And Interaction Model}
\label{sec:ch1_CEI_model}

More careful treatment shows that 
the interactions between electrons in 
a quantum dot should be correctly described by 
the ``universal'' Hamiltonian~\cite{Kur00,Ale02}. 

In the basis of eigenfunctions $\{\phi_{\alpha}\}$ 
of the free-electron Hamiltonian,
\begin{eqnarray}
\left[ -\frac{1}{2m}{\bf \nabla}^{2} + U({\bf r})\right] 
\phi_{\alpha}({\bf r}) 
= \varepsilon_{\alpha} \phi_{\alpha}({\bf r}), 
\end{eqnarray}
the two-particle interaction takes the form
\begin{eqnarray}
H_{int} = \frac{1}{2} \sum 
H_{\alpha\beta\gamma\delta}\,
{\hat c}_{\alpha\sigma_{1}}^{\dagger}{\hat c}_{\beta\sigma_{2}}^{\dagger} 
{\hat c}_{\gamma\sigma_{2}}{\hat c}_{\delta\sigma_{1}},
\label{eq:ch1_H_int}
\end{eqnarray}
where $U({\bf r})$ is the random potential 
determined by the shape of the quantum dot
and 
$\sigma_{1}$ and $\sigma_{2}$ are 
the spin indices for the fermionic operators.
The generic matrix element of the interaction is
\begin{eqnarray}
H_{\alpha\beta\gamma\delta} 
= \int d{\bf r}_{1} d{\bf r}_{2} V({\bf r}_{1} - {\bf r}_2) 
\phi_{\alpha}({\bf r}_{1})\phi_{\beta}({\bf r}_{2}) 
\phi_{\gamma}^{*}({\bf r}_{2})\phi_{\delta}^{*}({\bf r}_{1}).
\label{eq:ch1_H_abcd}
\end{eqnarray}
The matrix elements of the interaction Hamiltonian 
have a hierarchical structure. Only a few of these elements 
are large and universal, whereas the majority of them 
are proportional to the inverse dimensionless conductance
$1/g$ ($1/g = 2\pi / k_F L$ for a $2D$ quantum dot) 
and, therefore, small~\cite{Kur00,Ale02}.
As a result, the Hamiltonian [Eq.~(\ref{eq:ch1_H_int})]
can be broken in two pieces:
\begin{eqnarray}
H_{int} =  H_{int}^{(0)} + H_{int}^{(1/g)}.
\label{eq:ch1_H2}
\end{eqnarray}
The first term here is universal -- 
it does not depend on the quantum dot geometry 
and does not fluctuate from sample to sample 
(for samples differing only by realization of disorder). 
The second term in Eq.~(\ref{eq:ch1_H2}) does fluctuate 
but it is of order $\delta\varepsilon /g$ and,
hence, small. 
This term only weakly affect 
the low-energy $(E < E_T)$ properties of the system.

The form of the universal term in Eq.~(\ref{eq:ch1_H2})
can be established using the requirement of
compatibility of this term with the RMT~\cite{Kur00,Ale02}. 
Since the random matrix distribution 
is invariant with respect to 
an arbitrary rotation of the basis, 
the operator $ H_{int}^{(0)}$ may include only 
the operators which are invariant under such rotations. 
In the absence of the spin-orbit interaction, 
there are three such operators:
\begin{eqnarray}
{\hat n} = \sum_{\alpha\sigma} 
{\hat c}_{\alpha\sigma}^{\dagger} {\hat c}_{\alpha\sigma}
\end{eqnarray}
-- the total number of electrons,
\begin{eqnarray}
{\hat {\bf S}} = \frac{1}{2} \sum_{\alpha\sigma_{1}\sigma_{2}} 
{\hat c}_{\alpha\sigma_{1}}^{\dagger}
({\bf \sigma})_{\sigma_{1}\sigma_{2}}{\hat c}_{\alpha\sigma_{2}}
\end{eqnarray}
-- the total electron spin of the dot, and
the operator
\begin{eqnarray}
{\hat T} = \sum_{\alpha}{\hat c}_{\alpha\uparrow}{\hat c}_{\alpha\downarrow},
\end{eqnarray}
which corresponds to the interaction in the Cooper channel.

Gauge invariance requires that only 
the product ${\hat T}^{\dagger}{\hat T}$ 
of the operators ${\hat T}$ and ${\hat T}^{\dagger}$ 
may enter the Hamiltonian.
At the same time SU(2) symmetry dictates that 
the Hamiltonian may depend only on ${\hat {\bf S}}^{2}$ 
and not on the separate spin components. 
Taking into account that 
the initial interaction Hamiltonian [Eq.~(\ref{eq:ch1_H_abcd})]
is proportional to ${\hat c}^{4}$ 
we find the ``universal'' Hamiltonian:
\begin{eqnarray}
H_{int}^{(0)} 
= E_{C}{\hat n}^{2} - J_{\rm s}{\hat {\bf S}}^{2} 
+ J_{c}{\hat T}^{\dagger}{\hat T}, 
\label{eq:ch1_H_univ}
\end{eqnarray}
where $E_{C}$ is the redefined value of the charging 
energy~\cite{Kur00,Usa02} 
and $J_{\rm s}>0$ is the exchange interaction constant.
The constants in this Hamiltonian are model-dependent. 
The first two terms represent the dependence of the energy 
on the total number of electrons and the total spin, respectively. 
Because both the total charge and the total spin
commute with the free-electron Hamiltonian, 
these two terms do not have any dynamics for a closed dot. 
The situation changes as one couples the dot to the leads.

The third term vanishes 
in the Gaussian Unitary (GUE) random matrix ensemble.
GUE corresponds to the absence of time-reversal invariance, 
or placing the quantum dot into an external magnetic field. 
One can say that the Cooper channel 
is suppressed by a weak magnetic field 
(it is sufficient to thread a unit quantum flux $\Phi_{0}$ 
through the cross-section of the dot).

Thus, in the absence of superconducting correlations, 
the universal part of the interaction Hamiltonian
consists of two parts:
\begin{eqnarray}
V_{CEI} 
= E_{C}{\hat n}^{2} - J_{\rm s}{\hat {\bf S}}^{2}. 
\label{eq:ch1_H_CEI}
\end{eqnarray}
This is the so-called constant exchange and interaction (CEI) 
model~\cite{Usa01,Usa02}.
Its dominant part depends on the QD charge number $n$ -- 
the corresponding energy scale $E_{C} = e^{2}/2C$ 
is related to the capacitance of the QD, $C$, and
exceeds parametrically the mean level spacing 
$\delta\varepsilon$. 
The second part depends on the total spin ${\bf S}$ -- 
the corresponding energy scale $J_{\rm s}$ 
is less than $\delta\varepsilon$. 

If the level spacings did not fluctuate, 
then the smallness of $J_{\rm s}$ would automatically imply 
that the spin of the QD can only take the values of 
$0$ (if $n$ is even) or $\frac{1}{2}$ (if $n$ is odd).
Fluctuations in the level spacings may lead to 
a violation of this periodicity~\cite{Bro99,Bar00,Kur00}.
However, the Stoner criterion; 
$J_{\rm s}<\delta\varepsilon$, guarantees that 
the total QD spin is not macroscopically large; 
that is, it does not scale with the QD volume. 

In the presence of the back-gate electrode 
capacitively coupled to the QD, 
the CEI model and free quasiparticle Hamiltonian become 
\begin{eqnarray}
H^{(0)} = \sum_{k\sigma}\varepsilon_{k}{\hat n}_{k\sigma} 
+ E_{C}\left( {\hat n}-{\mathcal N} \right)^{2} - J_{\rm s}{\hat {\bf S}}^{2}, 
\end{eqnarray}
where ${\mathcal N}=C_{g}V_{g}/e$ 
is the dimensionless back-gate voltage 
and $C_{g}$ is the dot-backgate capacitance.

\section{3D Quantum Dots}

A 3D quantum dot is just a metal nanograin.
There are numerous methods to synthesize them~\cite{Shi00}.
At the time of writing, metal nanoparticles 
of down to about 1~nm in diameter are commercially
available.

To form a single electron transistor 
one has to trap a nanograin between two leads. 
To create a narrow gap between two conductors, 
the {\it electromigration} technique is often implemented~\cite{Bol04}.
In this technique one takes a conductor 
of a small cross-section and gradually increases 
the voltage and, hence, the current through it
until cracking occurs. Using this method, 
one can create very narrow nanometer size gaps.
To trap a nanoparticle the potential difference 
between two leads is applied. 
Then the nanoparticle gets polarized and 
attracted to the region of high electric field.
This is called the {\it electrostatic trapping} technique.
Thus, the nanoparticle is attached to two leads
via oxide tunnel barriers. 
As an example, in the recent experiment 
by Bolotin and coworkers 
an ultra-small gold nanograin, 5~nm in diameter, 
was incorporated into a gap between two leads. 
Thus, a single-electron transistor was formed, 
and Coulomb blockade oscillations were observed 
for more than ten charge states of the grain. 

Although these metal nanograins are similar 
to semiconductor quantum dots, 
there are a number of important differences~\cite{Del01}. 
(i)\,Metals have much higher densities of states 
than semiconductors; 
hence, they require much smaller sample sizes 
(less than 10~nm) before discrete energy levels 
in the QD become resolvable.
The ratio $E_{C}/\delta\varepsilon$ is 
usually larger for nanograins; 
therefore, mesoscopic fluctuations 
have significantly less impact on their quantum properties. 
(ii)\,For nanograins, the tunnel barriers 
to the leads are formed by insulating oxide layers. 
Therefore, they are insensitive to applied voltages, 
whereas for 2D quantum dots they can be tuned 
by changing voltages on the electrodes.

%% file: chapters/chapter2.tex
\setlength{\textheight}{8.0in}
\clearpage
\chapter{Coulomb Blockade Oscillations of Conductance 
at Finite Energy Level Spacing in a Quantum Dot}
\label{ch:ch2}
\thispagestyle{botcenter}
\setlength{\textheight}{8.4in}

\section{Overview}

In this chapter
we find an analytical expression for the conductance 
of a single electron transistor in the regime when
temperature, level spacing, and charging energy
of a grain are all of the same order. We consider 
the model of equidistant energy levels in a grain 
in the sequential tunneling approximation.In the case 
of spinless electrons our theory describes transport 
through a dot in the quantum Hall regime. In the case 
of spin-$\frac{1}{2}$ electrons we analyze the line shape 
of a peak, shift in the position of the peak's maximum as 
a function of temperature, and the values of the conductance 
in the odd and even valleys.

\section{Introduction}

\begin{figure}
\begin{center}
\includegraphics[width=10.0cm]{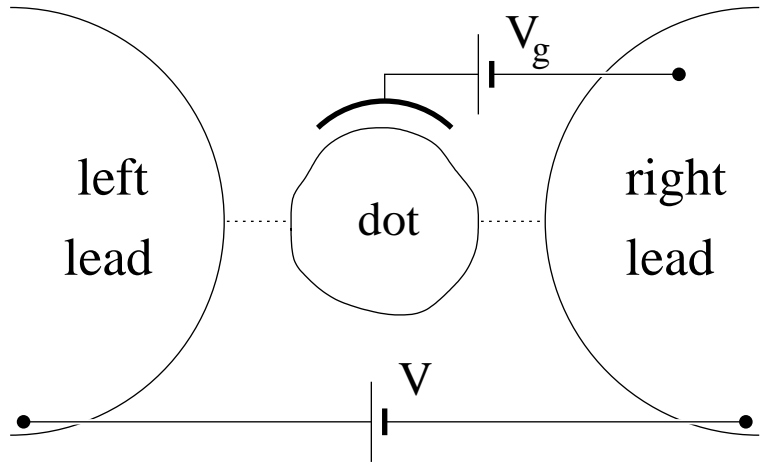}
\caption{Scheme of the Coulomb blockade setup.}
\label{fig:ch2_Fig1}
\end{center}
\end{figure}

Recent progress in mesoscopic fabrication techniques has
made possible not only the creation of more sophisticated 
devices but also greater control over their properties. 
Electron systems confined to small space regions, quantum dots,
and especially their transport properties have been studied
extensively for the last decade \cite{Del01,Kou97}.
In particular, an individual ultra-small metallic grain of
radius less than 5~nm was attached to two leads via
oxide tunnel barriers, thus forming a single electron
transistor (SET) \cite{Del01}.
Applying bias voltage, $V$, between two leads allows one
to study transport properties of the system, Fig.~\ref{fig:ch2_Fig1}.
Alternatively, a SET can be formed by depleting two-dimensional
electron gas at the interface of GaAs/AlGaAs heterostructure
by applying negative voltages to the metallic surface gates \cite{Kou97}.

In this chapter we will assume that the bias voltage is infinitesimally
small, $V\to 0$. This corresponds to the linear response regime.
In order to tunnel onto the quantum dot, an electron in the left
lead has to overcome a charging energy, $E_C=e^2/2C$, where $e>0$ is 
the elementary charge; $C$ is the capacitance of the quantum dot.
If $T\ll E_C$ then conductance through the system is exponentially 
suppressed. This phenomenon is called the Coulomb blockade.
However if we apply a voltage, $V_g$, to the additional gate 
capacitively coupled to the dot, the Coulomb blockade can be lifted.
Indeed, changing $V_g$ one can shift the position of energy
minimum so that energies of the quantum dot with $N_e$ and $N_e+1$
electrons will become equal and an electron can freely jump 
from the left lead onto the dot and then jump out into the other lead. 
Thus, current event has occurred and a peak in the conductance, $G$,
corresponding to this gate voltage is observed. By changing the gate 
voltage one can observe an oscillation of the conductance or Coulomb 
blockade oscillations.

One-particle energy levels in the quantum dot, $\{E_i\}$, are given
by the solution of the Schr\"odinger equation in the quantum dot's
potential. The mean spacing between these energy levels is $\delta E$.
The conventional assumption that $E_C\gg\delta E$ 
is not valid in the case of sufficiently small dots.
In fact, in the recent experiments \cite{Por97a,Por97b,Par00},
where a $C_{60}$ molecule has acted as a quantum dot,
the level spacing is of order charging energy. 
Experiment \cite{Por97a,Por97b}
was performed at $T=4.2 K$ as well as at room temperature.
In other experiments \cite{Ji00,Boc97} with quantum
dot formed by depleting 2DEG \cite{Ji00} and ropes of
carbon nanotubes acting as a quantum dot \cite{Boc97},
charging energy is only three times larger than the spacing $\delta E$.

Though Coulomb blockade oscillations have been studied in a
number of important limiting cases \cite{Gla89,Gla88,Mat96,Bee91}, 
the problem in the case when values of $E_C$, $\delta E$, and $T$ are 
all of the same order has not been theoretically addressed.
Let us note that energy levels of the quantum dots are random
and obey Wigner-Dyson statistics with the fluctuation of order
of their mean \cite{Meh91}.
Nonetheless to go as far as possible in the analytical treatment 
of the problem we have to assume that energy levels in the quantum 
dot are equidistant. In this chapter we derive an analytical expression 
for the linear conductance, $G=I/V|_{V\to 0}$, in the case of spinless 
as well as spin-$\frac{1}{2}$ fermions.

In Section~\ref{sec:ch2_model} we describe our model and the assumptions 
involved. We write the model assuming spin-$\frac{1}{2}$ fermions. 
In Section~\ref{sec:ch2_spinless} we consider the linear conductance 
in the case of spinless fermions. We obtain an analytical expression 
for the conductance and analyze its limiting cases. 
In Section~\ref{sec:ch2_edge_states} we consider one possible application 
of the Section~\ref{sec:ch2_spinless} results, namely tunneling through 
the edge states in a quantum dot placed into a strong magnetic field. 
In Section~\ref{sec:ch2_spinhalf} the linear conductance as well as 
its properties in the case of spin-$\frac{1}{2}$ fermions is considered.
In Section~\ref{sec:ch2_conclusions} we summarize our findings.

\section{The Model}
\label{sec:ch2_model}

\begin{figure}
\begin{center}
\includegraphics[width=10.0cm]{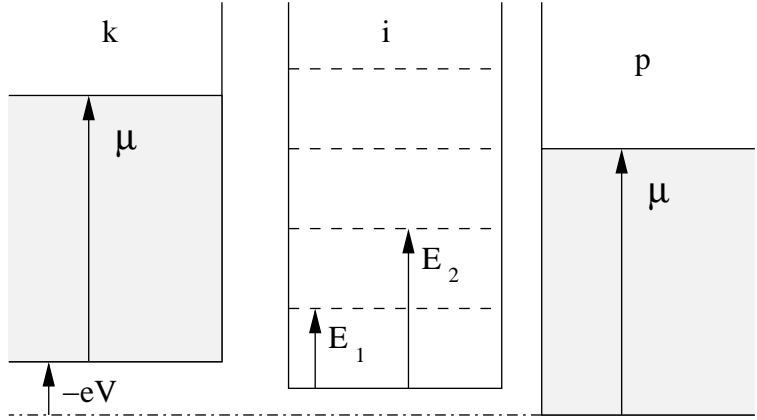}
\caption{Electrostatic potential energy along 
a line through the tunnel junctions.}
\label{fig:ch2_Fig2}
\end{center}
\end{figure}

Hamiltonian of the system in question is
\begin{equation}
{\hat H}={\hat H_{l}}+{\hat H_{d}}+{\hat T}.
\label{eq:ch2_hamilt}
\end{equation}
Here, the first term is the Hamiltonian of noninteracting electrons 
in the left and right leads:
\begin{eqnarray}
{\hat H_{l}} = 
\sum_{k \sigma}E_{k}c_{k\sigma}^{\dagger}c_{k\sigma}+ 
\sum_{p \sigma}E_{p}c_{p\sigma}^{\dagger}c_{p\sigma},
\end{eqnarray}
where a continuum of states in each lead, 
$\left| k \sigma \right>$, $\left| p \sigma \right>$
is assumed;
$E_k$, $E_p$ and $c_{k\sigma}$, $c_{p\sigma}$ are the energies and 
electron annihilation operators in the left and right leads, 
respectively; $\sigma$ stands for the $z$-component of spin.
The chemical potentials of the leads, $\mu\gg E_C,\delta E,T$, 
are shifted according to the bias voltage, $V$, 
applied, Fig.~\ref{fig:ch2_Fig2}. 
We will assume that leads are in thermal equilibrium at temperature 
$T$ and, thus, occupied according to the Fermi-Dirac distribution.

The second term in Eq.~(\ref{eq:ch2_hamilt}) is the Hamiltonian 
of the quantum dot:
\begin{eqnarray}
{\hat H_{d}}=\sum_{i\sigma}E_{i}
c_{i\sigma}^{\dagger}c_{i\sigma}
+{\hat U},
\end{eqnarray}
where first term is the kinetic energy of electrons in the 
quantum dot: $\{E_{i}\}$ is a discrete set of the quantum 
dot's energy levels; $c_{i\sigma}$'s are the annihilation 
operators. The second term, ${\hat U}$ describes the 
electron-electron interaction in the quantum dot.
We adopt the simplest model for the interaction, namely, 
the constant interaction model. In this model the Coulomb 
interaction of the electrons depends only 
on the total number of electrons in the quantum dot:
\begin{eqnarray}
U({\hat N})=E_{C}{\hat N}^{2}-eV_{e}{\hat N},
\end{eqnarray}
where 
$N=\sum_{i\sigma}c_{i\sigma}^{\dagger}c_{i\sigma}-N_{i}$
is the total number of excess electrons;
$N_i$ is the total number of positively charged ions.
The second term is the contribution from external charges.
They are supplied by the ionized donors and the gate:
$V_{e} = V_{d} + a V_{g}$, where $a$ is a function of the 
capacitance matrix elements of the system. Thus, $V_{e}$ can 
be varied continuously by changing gate voltage, $V_{g}$.
$U(N)$ can be rewritten as 
\begin{eqnarray}
U(N)=E_C(N-N_g)^2+\mbox{Const},
\end{eqnarray}
where $N_g=eV_{e}/2E_C$ is the dimensionless gate voltage.

The third term in Eq.~(\ref{eq:ch2_hamilt}) is the tunneling Hamiltonian:
\begin{eqnarray}
{\hat T}=
\sum_{ki\sigma}\left( t_{ki}{c}_{k\sigma}^{\dagger}
{c}_{i\sigma}+h.c.\right)+
\sum_{pi\sigma}\left( t_{pi}{c}_{p\sigma}^{\dagger}
{c}_{i\sigma}+h.c.\right),
\end{eqnarray}
where $t_{ki}$ and $t_{pi}$ are matrix elements of tunneling into
the left and right leads, respectively.

We assume that the dot is weakly coupled to the leads;
that is, the conductances of the dot-lead junctions
are small: $G^{l,r} \ll e^2/h$, where $h$ is Planck's constant.
Equivalently, the widths of the quantum dot's energy levels 
contributing to the conductance, $\Gamma_i=\Gamma_i^l+\Gamma_i^r$,
must be small compared to spacing between them: $\Gamma_i\ll\delta E$.
This, together with $\Gamma_i \ll T$ assumption, allows us to 
characterize the state of the dot by a set of occupation 
numbers, $\{ n_{i\sigma}\}$ \cite{Bee91}.

\section{Linear Conductance in the Spinless Case}
\label{sec:ch2_spinless}

The model formulated above has been studied by Beenakker 
in the sequential tunneling approximation \cite{Bee91};
that is, conservation of energy was assumed in each tunneling 
process, and cotunneling was neglected. Therefore, to find the 
stationary current, kinetic equation considerations can be applied.
In the linear response regime an analytical formula for the conductance 
has been obtained. In the case of spinless fermions \cite{Bee91}:
\begin{eqnarray}
G &=& \frac{e^2}{h T} \sum_{i=1}^{\infty} 
\frac{\Gamma_i^l \Gamma_i^r}{\Gamma_i^l + \Gamma_i^r}
\sum_{N_e=1}^{\infty} P_{eq}(N_e) F_{eq}(E_i|N_e)
\nonumber \\
&& \times [1-n_F(E_i - \mu + U(N)- U(N-1)],
\label{eq:ch2_spinlessg}
\end{eqnarray}
where
$$
\Gamma_i^l= 2\pi \sum_k |t_{ki}|^{2} 
\delta\left[ E_i-E_k+U(N)-U(N-1)\right]
$$
and
$$
\Gamma_i^r= 2\pi \sum_p |t_{pi}|^{2} 
\delta\left[ E_i-E_p+U(N)-U(N-1)\right]
$$
are widths of the quantum dot's level $i$ associated with
tunneling into the left and right leads, respectively;
$P_{eq}(N_e)$ is the equilibrium probability that the 
quantum dot contains $N_e$ electrons; 
$F_{eq}(E_i|N_e)$ is the occupation number of level $i$ 
given that the dot contains $N_e$ electrons; 
$n_F(E)$ is the Fermi-Dirac distribution; 
and $\mu$ is the chemical potential in the leads.

\begin{figure}
\begin{center}
\includegraphics[width=10.0cm]{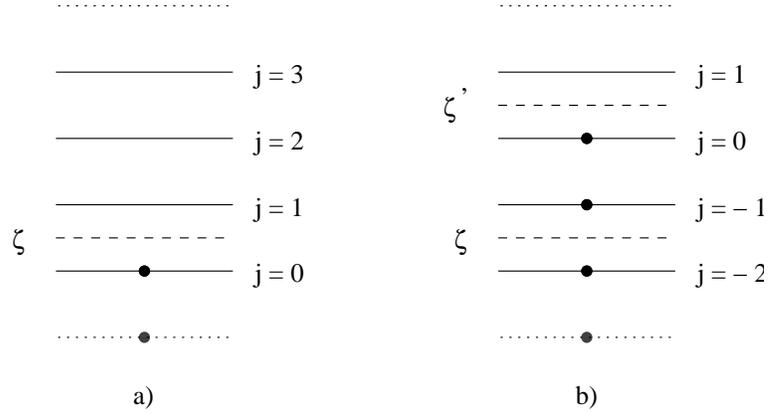}
\caption{a) occupation numbers in the canonical ensemble 
for the equidistant energy levels at $T=0$; b) mapping sum 
over $i$ onto sum over $j$ ($N=2$ case is shown).}
\label{fig:ch2_Fig3}
\end{center}
\end{figure}

The quantity $F_{eq}(E_i|N_e)$ in~(\ref{eq:ch2_spinlessg}) 
is the most non-trivial one to calculate. It is the occupation 
number of the level $i$ in the canonical ensemble ($N_e$ is fixed). 
In the limit $\delta E/T\to 0$, $F_{eq}(E_i|N_e)$ 
becomes a Fermi-Dirac distribution with the appropriately 
chosen chemical potential: $\tilde\mu=(E_{0}+E_{1})/2$,
where $E_{0}$ corresponds to the energy of the last 
occupied energy level at $T=0$, Fig.~\ref{fig:ch2_Fig3}(a); 
$E_{1}$ corresponds to the energy of the first empty energy 
level at $T=0$. In the opposite limit $\delta E/T\to\infty$, 
the Fermi-Dirac distribution with $\tilde\mu =(E_{0}+E_{1})/2$ 
apparently breaks down: the occupation number of level $j=1$, 
for example, see Fig.~\ref{fig:ch2_Fig3}a, is $n_1=e^{-\delta E/T}$, 
not $e^{-\delta E/2T}$ as the Fermi-Dirac distribution would 
predict \cite{Bee91}.

The occupation number in question is~\cite{Bee91}
\begin{eqnarray}
F_{eq}(E_i|N_e)&=&\frac{1}{P_{eq}(N_e)}\sum_{\{ n_l\}} P_{eq}(\{ n_l\})
\delta_{n_i,1} \delta_{N_e,\sum n_l}
\nonumber \\
&=&e^{\beta F(N_e)} \sum_{\{ n_l\}}e^{-\beta \sum E_l n_l}
\delta_{n_i,1}\delta_{N_e,\sum n_l},
\label{eq:ch2_tough}
\end{eqnarray}
where $P_{eq}(\{ n_l\})$ is the equilibrium probability of the
$\left| \{ n_l\}\right>$ state of the quantum dot; $\beta =1/T$;
and the detailed definition of $F(N_e)$ will follow. 
The reason for writing this equation is to show that analytical 
calculation of the occupation numbers is hardly possible for 
arbitrary quantum dot's energy level structure, $\{ E_i\}$. 

The only way to overcome this difficulty is to assume 
that energy levels in the quantum dot are equidistant.
Then one can use the bosonization technique \cite{Hal81}
(see Appendix~\ref{ap:ch2}) to find the exact analytical 
expression for the occupation numbers in the canonical ensemble.
It was done by Denton, Muhlschlegel, and Scalapino \cite{Den73}:
\begin{equation}
F_{eq}(E_i | N_e) \equiv n_j = \sum_{m=1}^{\infty} (-1)^{m-1} 
e^{- \frac{1}{2}[m^2 + (2j-1)m] \frac{\delta E}{T}},
\label{eq:ch2_dent}
\end{equation}
where
\begin{eqnarray}
j =\frac{E_i-\zeta}{\delta E}+\frac{1}{2}=\mbox{integer},
\label{eq:ch2_nzero}
\end{eqnarray}
$\delta E/T\equiv\delta =\mbox{const}$, $\zeta$ is the 
energy corresponding to the highest occupied energy level 
at $T=0$ plus $\delta E/2$, Fig.~\ref{fig:ch2_Fig3}(a). This 
quantity $\zeta$ is somewhat similar to the chemical potential 
of a dot, though, strictly speaking, the chemical potential 
is not well-defined for a dot in the canonical ensemble.
The difference, $\zeta -\mu$, is a linear function
of the gate voltage. Therefore, by properly adjusting 
``zero'' value of the gate voltage it can be put to zero.
Hereinafter, we assume that $\zeta -\mu=0$.

To calculate the conductance we also need to know $P_{eq}(N_e)$, 
the probability that the dot, in thermodynamic equilibrium with 
the reservoirs, contains $N_e$ electrons. It can be calculated 
in the grand canonical ensemble:
\begin{equation}
P_{eq}(N_e) = \frac{1}{Z_{\mu}} e^{- \beta \varphi (N_e)},
\label{eq:ch2_peq}
\end{equation}
where $Z_{\mu} = \sum_{N_e = 0}^{\infty} e^{- \beta \varphi (N_e)}$ 
is the grand  partition function; $\varphi (N_e)$ is the 
thermodynamic potential of the quantum dot. It can be expressed
via free energy of the dot's internal degrees of freedom, $F(N_e)$:
\begin{eqnarray}
\varphi (N_e) = F(N_e) - \mu N_e + U(N),
\end{eqnarray}
hence,
\begin{equation}
P_{eq}(N_e)=\frac{1}{Z_{\mu}}e^{- \beta U(N)}Z(N_e),
\label{eq:ch2_peq2}
\end{equation}
where
$$
Z(N_e)=e^{-\beta [F(N_e)-\mu N_e]}
=\sum\limits_{\{ n_{i}\}}e^{-\beta\sum (E_i-\mu)n_{i}}\delta_{N_e,\sum n_{i}}
$$
is the partition function in the canonical ensemble. In the 
last expression the sum is taken over all possible states, 
$\left| \{ n_{i}\}\right>$ of the quantum dot.
To calculate $Z(N_e)$ explicitly we need to assume 
that energy levels of the dot are equidistant:
\begin{equation}
Z(N_e) = e^{-(N_e - N_{i})^2 \delta /2} Z_{exc},
\label{eq:ch2_zne}
\end{equation}
where $N_e=N_{i}$ corresponds to the equilibrium number 
of the excess electrons $(\zeta =\mu)$; 
$Z_{exc}$ is the partition function of the thermal excitations.
Now, let us substitute Eq.~(\ref{eq:ch2_zne}) in Eq.~(\ref{eq:ch2_peq2}):
\begin{eqnarray}
P_{eq}(N_e) &=& \frac{1}{Z_{\mu}} e^{- \beta U(N)}
e^{-(N_e-N_{i})^2 \delta /2} Z_{exc} 
\nonumber \\
&=& \frac{1}{D'_{0}} e^{-\beta [U(N)+N^2\delta E/2]},
\label{eq:ch2_pofne}
\end{eqnarray}
where
\begin{eqnarray}
D'_{0}=\sum_{N=-\infty}^{\infty}e^{-\beta [U(N)+N^2\delta E/2]}.
\end{eqnarray}
Here, we have extended one limit of the sum to infinity since 
the Fermi energy is the largest energy scale of the problem.

Thus, we are well-equipped to calculate the conductance 
in the case of equidistant energy levels in the quantum 
dot at arbitrary ratio $\delta$.
The widths of energy levels, $\Gamma_{i}^{l}$ and $\Gamma_{i}^{r}$,
in the quantum dot are energy dependent, random 
quantities. Let us assume that quantum dot is weakly coupled 
to the leads via multichannel tunnel junctions:
$G^{l,r}=G^{l,r}_{1}N_{ch}\ll e^2/h$,
where $N_{ch}$ is the number of channels;
$G^{l,r}_{1}$ is the conductance of one channel.
Experimentally, this situation corresponds to
the metallic grain coupled to the leads via oxide tunnel
barriers \cite{Del01}. This setup allows one to 
decrease fluctuations of the energy levels' widths,
$\Gamma_{i}^{l}$ and $\Gamma_{i}^{r}$, by a factor
of $\sqrt{N_{ch}}$.
We also assume that the widths are slowly changing
functions of the energy, $E_{i}$.
Then, $\Gamma_{i}^{l}$ and $\Gamma_{i}^{r}$ can be
considered constants and evaluated at the chemical potential:
$\Gamma_i^l \approx \Gamma_{\mu}^l \equiv \Gamma^l$; 
$\Gamma_i^r \approx \Gamma_{\mu}^r \equiv \Gamma^r$.
There is a simple relation between these widths and conductances
of the corresponding junctions. In the case of spinless fermions:
\begin{equation}
   \Gamma^l=\frac{h G^l}{e^2}\delta E,
~~~\Gamma^r=\frac{h G^r}{e^2}\delta E.
\label{eq:ch2_glgr0}
\end{equation}
Let us substitute Eq.~(\ref{eq:ch2_pofne}) in Eq.~(\ref{eq:ch2_spinlessg}):
\begin{eqnarray}
G&=&\frac{G^lG^r}{G^l+G^r}\frac{\delta}{D'_{0}}
\sum_N e^{-\beta [U(N)+N^2\delta E/2]}
\sum_i F_{eq}(E_i|N_e)
\nonumber \\
&&\times \left\{ 1-n_F \left[ E_i-\mu +U(N)-U(N-1) \right] \right\}.
\label{eq:ch2_interm}
\end{eqnarray}
To take advantage of the expression for occupation numbers, 
Eq.~(\ref{eq:ch2_dent}), we need to map the sum over $i$ 
onto the sum over $j$. As illustrated in Fig.~\ref{fig:ch2_Fig3}, 
the mapping rule depends on the total number of excess electrons, $N$ 
(compare with Eq.~(\ref{eq:ch2_nzero}) written for $N=0$):
\begin{eqnarray}
F_{eq}(E_i|N_e) &=& n_j,\nonumber
\\
E_i-\mu =E_i-\zeta'+(\zeta'-\zeta)
&=&\left( j-\frac{1}{2}\right)\delta E+N\delta E,\nonumber
\end{eqnarray}
where $\zeta'=\zeta'(N)$ is the energy of the highest
occupied energy level in the dot with $N$ excess
electrons at $T=0$ plus $\delta E/2$.
We will also use the following identities:
\begin{eqnarray}
U(N)+\frac{\delta E}{2}N^2
=\left( E_C+\frac{\delta E}{2}\right) N^{2}-eV_{e}N
\nonumber \\
=\left( E_C+\frac{\delta E}{2}\right)
\left( N-\Delta_0 +\frac{1}{2}\right)^{2}+{\mathcal C}_{1},
\end{eqnarray}
where
${\mathcal C}_{1}=-\left(eV_{e}\right)^{2}/2(2E_C+\delta E)$;
\begin{eqnarray}
\Delta_0 \equiv \frac{eV_{e}}{2E_C+\delta E}+\frac{1}{2}
\end{eqnarray}
has been chosen so that $\Delta_0 =0$ corresponds to the 
maximum of the conductance peak; and
\begin{eqnarray}
U(N)-U(N-1)=(2N-1)E_C-eV_{e}
\nonumber \\
=(2N-1)E_C-(2E_C+\delta E)(\Delta_0-1/2).
\end{eqnarray}
Substituting these results in Eq.~(\ref{eq:ch2_interm}), we obtain
\begin{equation}
\frac{G(\Delta_0)}{G_{\infty}}=
\frac{\delta}{D_0}
\sum_N e^{-\varepsilon_0\left( N-\Delta_0+\frac{1}{2}\right)^2}
\sum_j\frac{n_j}{e^{-j\delta -2(N-\Delta_0)\varepsilon_0}+1},
\label{eq:ch2_g0}
\end{equation}
where
\begin{eqnarray}
D_0 = \exp \left(\beta{\mathcal C}_{1}\right) D'_0
=\sum_N e^{-\varepsilon_0\left( N-\Delta_0+\frac{1}{2}\right)^2};
\end{eqnarray}
$\varepsilon_0 \equiv \beta (E_C + \delta E/2)$; 
$G_{\infty} \equiv G^l G^r/(G^l + G^r)$ is the classical,
$E_C, \delta E \ll T \ll \mu$, limit of the conductance. We have 
also used the identity: 
$1-n_F(E) = (1+e^{-\beta E})^{-1}$. 
Eq.~(\ref{eq:ch2_g0}) is the general expression for the linear 
conductance in the spinless case for equidistant energy levels 
in the quantum dot at arbitrary values of $E_C$, $\delta E$ and $T$. 

One can immediately prove the following properties of the 
conductance, Eq.~(\ref{eq:ch2_g0}). First of all, 
$G(\Delta_0)=G(\Delta_0+M)$, where $M$ is an integer. 
In the gate voltage units, $\Delta V_e=(2E_C+\delta E)/e$ 
is a period of the conductance oscillations. This property reflects 
symmetry with respect to adding (removing) an electron to the quantum 
dot. Secondly, due to the electron-hole symmetry, conductance 
is an even function of $\Delta_0$: $G(\Delta_0)=G(-\Delta_0)$.
\begin{figure}
\begin{center}
\includegraphics[width=10.0cm]{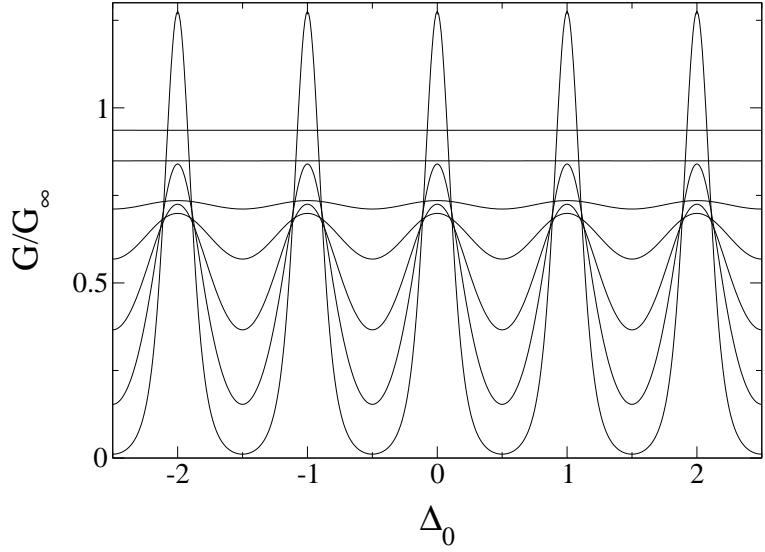}
\caption{Coulomb blockade oscillations of the conductance 
as a function of the dimensionless gate voltage, $\Delta_0$ 
at $\delta E=E_C$. Curves are plotted for different temperatures:
$T/E_C=T/\delta E\equiv 1/\delta =0.2, 0.35, 0.5, 0.7, 1, 2, 5$.}
\label{fig:ch2_gd0many}
\end{center}
\end{figure}

The linear conductance, Eq.~(\ref{eq:ch2_g0}), as a function 
of the dimensionless gate voltage $\Delta_0$ at $\delta E=E_C$ 
is plotted in Fig.~\ref{fig:ch2_gd0many} for different temperatures.
At low temperatures there are sharp Coulomb blockade peaks.
At high temperatures, $T\gg E_C,\delta E$, Coulomb blockade is 
lifted and small oscillations of the conductance can be observed. 
These oscillations are slightly non-sinusoidal and given 
by the following asymptotic formula:
\begin{eqnarray}
\frac{G(\Delta_0)-\overline{G(\Delta_0)}}{G_{\infty}}=
2\pi^{3/2}\frac{e^{-\varepsilon_0 /4}}{\sqrt{\varepsilon_0}}
\left[
e^{-\pi^2/\varepsilon_0}\cos (2\pi\Delta_0)
\right.
\nonumber \\
+\left.
e^{-2\pi^2/\varepsilon_0}\cos (4\pi\Delta_0)
+O(e^{-3\pi^2/\varepsilon_0})
\right],
\label{eq:ch2_hightexp}
\end{eqnarray}
where $\overline{G(\Delta_0)}$ is the average value of the 
conductance. The second term in Eq.~(\ref{eq:ch2_hightexp}) 
is due to inherently non-sinusoidal nature of the conductance 
oscillations, see Fig.~\ref{fig:ch2_gd0many}. To derive this 
expression one can use Poisson's summation formula.
\begin{figure}
\begin{center}
\includegraphics[width=10.0cm]{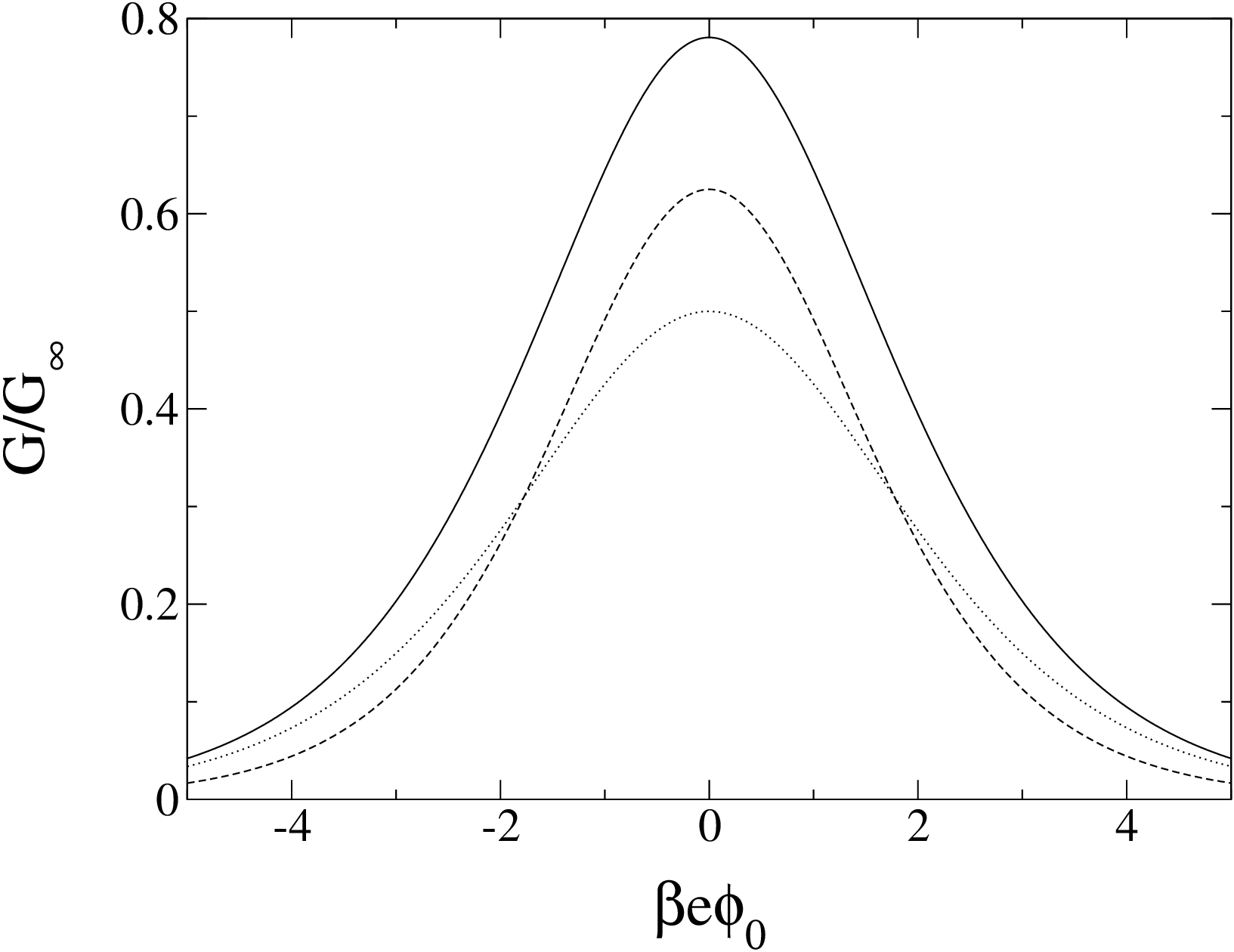}
\caption{The exact line shape of the conductance 
peak, Eq.~(\ref{eq:ch2_onepeak}), plotted at 
$1/\delta\equiv T/\delta E=0.4$ (solid curve). 
Dotted curve corresponds to blindly applying classical regime 
formula (\ref{eq:ch2_classical}) at $T/\delta E=0.4$.
Dashed curve corresponds to blindly applying 
formula (\ref{eq:ch2_resonant}) at $T/\delta E=0.4$.
The argument of the plot is linear function 
of the gate voltage, $\phi_0$.}
\label{fig:ch2_gd0peak}
\end{center}
\end{figure}

To study the line shape of a separate peak let us consider
the limit of large charging energy: $E_C\gg T,\delta E$
or, equivalently, $\varepsilon_0\gg 1,\delta$ in the
dimensionless units. In Eq.~(\ref{eq:ch2_g0}) only $N=-1,0$ 
terms in the sum over $N$ give substantial contribution to the
conductance near $\Delta_0=0$; all other terms are exponentially 
suppressed. Besides, the sum over $j$ at the $N=-1$ is 
$O(e^{-2\varepsilon_0})$ and, therefore, can also be neglected.
Hence, line shape of the conductance peak at $\Delta_0=0$ 
is given by
\begin{equation}
\frac{G(\Delta_0)}{G_{\infty}}=
\frac{\delta}{1+e^{-2\varepsilon_0\Delta_0}}
\sum_j\frac{n_j}{1+e^{-j\delta +2\varepsilon_0\Delta_0}}.
\end{equation}
It is more instructive to rewrite this equation as follows:
\begin{equation}
\frac{G(\phi_0)}{G_{\infty}}=
\frac{\delta}{1+e^{-\beta e\phi_0}}
\sum_j\frac{n_j}{1+e^{-j\delta +\beta e\phi_0}},
\label{eq:ch2_onepeak}
\end{equation}
where $\phi_0 =V_{e}-V_{e}^{(0)}$; $V_{e}^{(0)}$ is chosen so
that $\phi_0 =0$ corresponds to center of the conductance peak.
In the classical regime \cite{Gla89}, $T\gg\delta E$, line shape 
of the conductance peak is given by
\begin{equation}
\frac{G(\phi_0)}{G_{\infty}}=
\frac{\beta e\phi_0}{2\sinh (\beta e\phi_0)}.
\label{eq:ch2_classical}
\end{equation}
In the opposite limit of $\delta E\gg T$:
\begin{equation}
\frac{G(\phi_0)}{G_{\infty}}=
\frac{\delta}{2[1+\cosh (\beta e\phi_0)]}.
\label{eq:ch2_resonant}
\end{equation}
The exact line shape of the conductance peak, Eq.~(\ref{eq:ch2_onepeak}),
at $T/\delta E=0.4$ is shown in Fig.~\ref{fig:ch2_gd0peak}.
On the same figure we also plotted two conductance peaks
in the limiting cases, Eqs.~(\ref{eq:ch2_classical}) and (\ref{eq:ch2_resonant}),
out of their validity region at $T/\delta E=0.4$.
Nevertheless, it is interesting that the exact conductance 
peak is higher than both of the limiting cases peaks.
The peak's height is given by
\begin{equation}
\frac{G(0)}{G_{\infty}}
=\frac{\delta}{2}
\sum_j\frac{n_j}{1+e^{-j\delta}}=
\left\{
\begin{array}{l}
1/2,~~T\gg\delta E \\
\delta /4,~~\delta E\gg T\gg\Gamma_i
\end{array}
\right. .
\label{eq:ch2_eqheight}
\end{equation}
Temperature dependence of the conductance peak's height
was numerically calculated in the Ref.~\cite{Bee91}, 
see Fig.~2 there.

\section{Application to Tunneling Through Quantum
Hall Edge States in a Quantum Dot}
\label{sec:ch2_edge_states}

Formulas for the linear conductance in the case of equidistant
energy levels in a dot and spinless fermions derived 
in Section~\ref{sec:ch2_spinless} can be applied to 
a number of physical problems.

\begin{figure}
\begin{center}
\includegraphics[width=10.0cm]{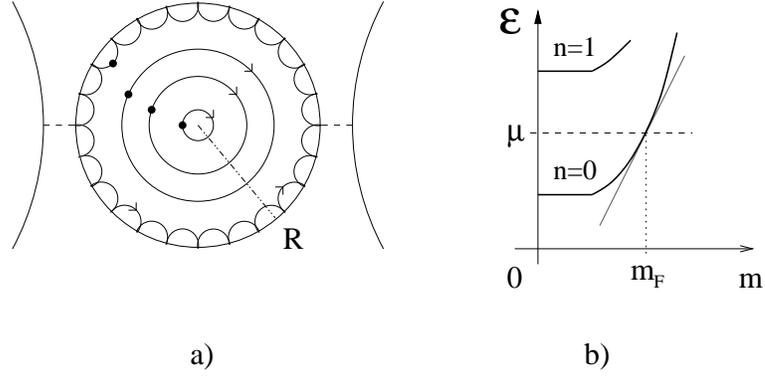}
\caption{Geometry and spectrum of the quantum dot states:
a) symmetric gauge eigenstates, $\left| m\right>$, including 
edge state are shown schematically; b) energy spectrum of the
eigenstates with different angular momenta,
two lowest Landau levels are shown.}
\label{fig:ch2_qhe}
\end{center}
\end{figure}

Let us consider, for example, a quantum dot formed by confining
a two-dimensional electron gas by a circularly symmetric
electrostatic potential, $U(r)$. We assume that $U(r)$ 
is zero at the origin and takes large value at $r=R$, 
where $R$ is the radius of the dot, Fig.~\ref{fig:ch2_qhe}(a). 
Let us apply a strong magnetic field, $B$, perpendicular 
to the plane of the dot. This situation corresponds to the 
quantum Hall regime and was reviewed in Ref.~\cite{Mac96}.

To solve the one-electron Schr\"odinger equation in this geometry
it is convenient to choose the symmetric gauge. Then, angular
momentum is, clearly, an integral of motion. In each Landau level, 
$n$, states with larger angular momentum, $\left| m\right>$, are 
localized further from the origin, near a circle with 
radius $R_m=l_H\sqrt{2(m+1)}$, where $l_H=\sqrt{\hbar c/eB}$ 
is the magnetic length, $c$ is the speed of light. 
The presence of the confinement potential leads to an increase 
in energy for the symmetric gauge eigenstates with $R_m$ 
of order or larger than $R$ (or, $m$ of order or larger than $m_F$, 
see Fig.~\ref{fig:ch2_qhe}(b)). 
For the states with $R_m\sim R$ an electron is influenced by
both the electric field of the boundary, $E(R)=U'(R)/e$, and 
strong, perpendicular to the electric, magnetic field. 
Thus, near the edge electron executes rapid cyclotron 
orbits centered on a point that slowly drifts in the direction 
of ${\bf E}\times{\bf B}$, that is, along the boundary. 
Thus, Quantum Hall edge states are formed, Fig.~\ref{fig:ch2_qhe}(a). 
It is important to notice that in this closed geometry
electron system has only one edge.
In this consideration we also assume that $l_H\ll R$.

For simplicity let us consider the case when only the zeroth
Landau level crosses the chemical potential, that is,
there is only one type of edge states. This corresponds
to a sufficiently strong magnetic field so that filling 
factor, $\nu$ is equal to $1$, Fig.~\ref{fig:ch2_qhe}(b).

Now, we are ready to consider transport through this
type of quantum dot in the strong magnetic field.
Let us weakly couple it to two leads and apply
an infinitesimally small bias voltage between them. 
An electron from the left lead can now tunnel into dot's 
edge state and then tunnel into the right lead as
illustrated by dashed lines in Fig.~\ref{fig:ch2_qhe}(a).

The energy spectrum of edge states can be linearized as follows
\begin{eqnarray}
{\mathcal E}_{m}=\mu +\left. \frac{\partial {\mathcal E}}{\partial m}
\right|_{m_F} \left( m-m_F\right).
\end{eqnarray}
Thus, energy levels of the edge states are equally spaced
with the spacing~\cite{Mac96}
\begin{eqnarray}
\delta E=
\left.\frac{\partial {\mathcal E}_{m}}{\partial m}\right|_{m_F}=
\left.\frac{\partial {\mathcal E}_{m}}{\partial R_m}\right|_{R}
\left.\frac{\partial R_m}{\partial m}\right|_{m_F}=
eE(R) \frac{l_H^2}{R}.
\label{eq:ch2_qhespacing}
\end{eqnarray}
This fact makes formulas derived in Section~\ref{sec:ch2_spinless}
applicable to this problem. Essential assumption here is that 
dispersion curve, Fig.~\ref{fig:ch2_qhe}(b) is almost
linear in the range of angular momentums:
$|m-m_{F}|\lesssim\Delta m$, 
where $\Delta m=\mbox{max}(1,T/\delta E)$.

In the case at hand, spacing, $\delta E$ is inversely 
proportional to the size of a dot just like charging
energy, $E_C$. Hence, their ratio does not depend on
the size of a dot and is given by
\begin{eqnarray}
\frac{\delta E}{E_C}\sim\frac{\epsilon}{\alpha}\frac{E(R)}{B},
\end{eqnarray}
where $\epsilon$ is the dielectric constant of the media 
around the interface;
$\alpha =e^{2}/\hbar c$ is the fine structure constant.
Therefore, in this case oscillations of the conductance
given by Eq.~(\ref{eq:ch2_g0}) are determined by only one
parameter $\delta =\delta E/T$.

In conclusion, let us consider the case
of an arbitrary shaped quantum dot.
In this case, $m$ is just the index of an edge state
and no longer associated with the angular momentum. 
The phase along the boundary 
for the $m$-th edge state is
\begin{eqnarray}
\theta_{m}=\int_{0}^{L}dx~k_{m}(x),
\end{eqnarray}
where $k_{m}(x)$ is the corresponding wave vector,
$x$ parametrizes the boundary, and $L$ is its length.
The phase difference between two consecutive edge states is
\begin{eqnarray}
\theta_{m+1}-\theta_{m}=2\pi
=\int_{0}^{L}dx\left[k_{m+1}(x)-k_{m}(x)\right],
\end{eqnarray}
where
$k_{m+1}(x)-k_{m}(x)=\left({\mathcal E}_{m+1}-{\mathcal E}_{m}\right)/\hbar v(x)$,
$v(x)=cE(x)/B$ is a drift speed along the boundary.
Then, the spacing between edge states' energy levels is
\begin{eqnarray}
\delta E=2\pi\hbar
\left[\int_{0}^{L}\frac{dx}{v(x)}\right]^{-1}
=2\pi el_H^2\left[\int_{0}^{L}\frac{dx}{E(x)}\right]^{-1}.
\label{eq:ch2_qhespacingrealdot}
\end{eqnarray}
Though the electric field $E(x)$ at the boundary
slightly changes as one goes from one edge state to the other,
this effect is small and we neglect it.
Therefore, the energy levels of the edge states 
are equidistant with the spacing given by
Eq.~(\ref{eq:ch2_qhespacingrealdot}).

In the case of the circularly symmetric quantum dot,
$E(x)$ is constant, and one can easily perform
the integration in Eq.~(\ref{eq:ch2_qhespacingrealdot}).
This leads to the previously obtained expression
for the level spacing, Eq.~(\ref{eq:ch2_qhespacing}).

\section{Linear Conductance in the Spin-$\frac{1}{2}$ Case}
\label{sec:ch2_spinhalf}

Formula (\ref{eq:ch2_spinlessg}) for the linear conductance 
in the spinless case can be easily generalized to the 
spin-$\frac{1}{2}$ case by counting each energy level 
twice~\cite{Bee91}:
\begin{eqnarray}
G &=& 2 \frac{e^2}{hT} \sum_{i=1}^{\infty}
\frac{\Gamma_i^l \Gamma_i^r}{\Gamma_i^l + \Gamma_i^r}
\sum_{N_e =1}^{\infty}P_{eq}(N_e)F_{eq}(E_{i\uparrow}|N_e)
\nonumber \\
&& \times [1 - n_F(E_i - \mu + U(N) - U(N-1))],
\label{eq:ch2_beespin}
\end{eqnarray}
where $F_{eq}(E_{i\uparrow}|N_e)$ is the occupation number 
of the quantum dot's energy level $i$ with a spin-up electron, 
($i,\uparrow$) in the canonical ensemble: 
number of electrons in the dot, $N_e$, is fixed.

As in the spinless case, to carry out analytical consideration 
we have to assume that energy levels in the quantum dot are 
equally spaced. Presence of the spin degeneracy makes the 
calculations more complicated.

First of all, let us find the occupation number 
$F_{eq}(E_{i\uparrow}|N_e)$.
Let us consider spin-up and spin-down electron subsystems. 
Ground state energy of the system is
$$
E_{g}=\left(N^{2}_{\uparrow}+N^{2}_{\downarrow}\right)
\frac{\delta E}{2}
=\left[
\left(N_{\uparrow}+N_{\downarrow}\right)^{2}
+\left(N_{\uparrow}-N_{\downarrow}\right)^{2}
\right]
\frac{\delta E}{4},
$$
where $N_{\sigma}=(N_e)_{\sigma}-N_{i}/2$ is the number 
of excess electrons in the spin-$\sigma$ subsystem; 
$N_{i}$ is chosen even.
$N_{\uparrow}+N_{\downarrow}\equiv N$ is the total 
number of excess electrons in the quantum dot; 
$(N_{\uparrow}-N_{\downarrow})/2\equiv S_z$
is $z$-component of the total electron spin.
Using these identities, one can find that
\begin{eqnarray}
E_{g}=\frac{1}{4}N^{2}\delta E
+S_{z}^{2}\delta E.
\end{eqnarray}
While $S_{z}$ is subjected to the thermodynamic
fluctuations, $N$ is fixed.
\begin{figure}
\begin{center}
\includegraphics[width=10.0cm]{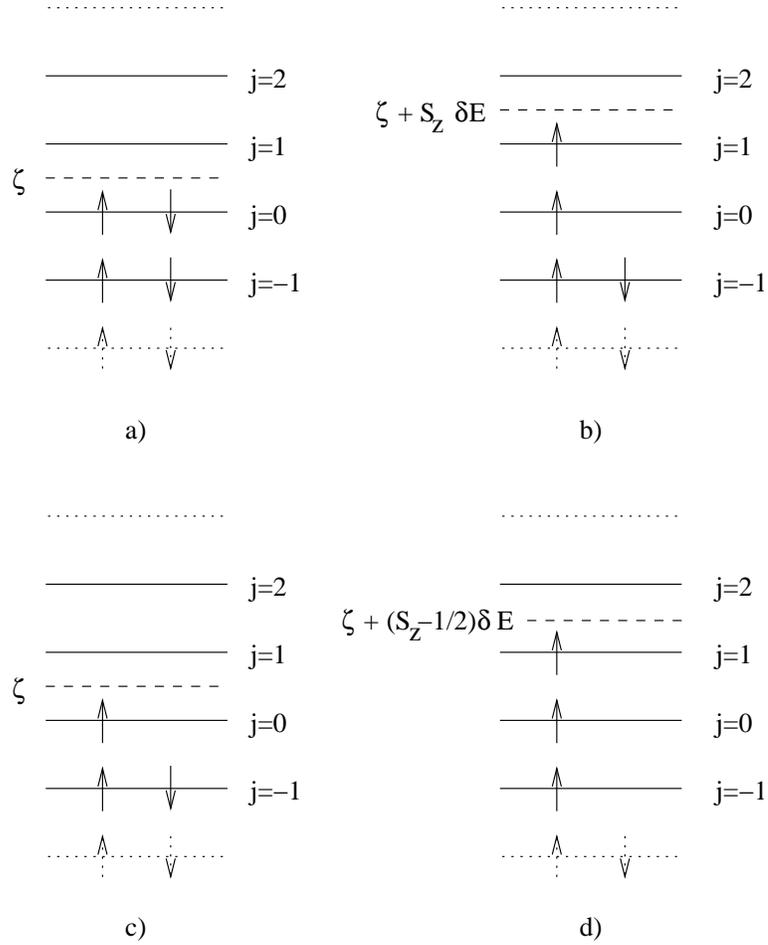}
\caption{Parameter $\zeta'$ of the spin-up electron 
subsystem for even number of electrons: a) at $S_z = 0$, 
b) at arbitrary integer $S_z$ ($S_z=1$ case is shown); 
for odd number of electrons: c) at $S_z = 1/2$, 
d) at arbitrary half-integer $S_z$ ($S_z=3/2$ case is shown).}
\label{fig:ch2_Fig4}
\end{center}
\end{figure}

The occupation number in question, 
$F_{eq}(E_{i\uparrow}|N_e)\equiv n_{j\uparrow}$,
is known if, in addition to $N$, 
the $z$-component of the total spin, $S_{z}$, is fixed. 
In the case of an even number of electrons, 
parameter $\zeta'$ for the spin-up electron subsystem 
is equal to $\zeta +S_{z}\delta E$, given $S_{z}$, 
see Figs.~\ref{fig:ch2_Fig4}(a) and \ref{fig:ch2_Fig4}(b). 
Occupation numbers in the spin-up subsystem at fixed $S_{z}$ 
are given by Eq.~(\ref{eq:ch2_dent}) 
with the appropriately chosen parameter $\zeta'$: $n_{j-S_{z}}$. 

However, $z$-component of the total spin, $S_{z}$, 
is not fixed but subjected to the thermodynamic fluctuations. 
Therefore, to find the occupation numbers, 
$n_{j\uparrow}^{ev}$, we have to account for all
possible values of $S_{z}$:
\begin{eqnarray}
n_{j\uparrow}^{ev}
=\sum_{S_z=-N_e/2}^{N_e/2}P(S_{z})~n_{j-S_{z}},
\label{eq:ch2_nevmod}
\end{eqnarray}
where
\begin{eqnarray}
P(S_{z})=\frac{e^{-S_{z}^{2}\delta}}
{\sum\limits_{S_z=-N_e/2}^{N_e/2}e^{-S_{z}^{2}\delta}}
\label{eq:ch2_pszmod}
\end{eqnarray}
is the probability that $z$-component of the total
spin of the quantum dot is equal to $S_{z}$.
Substituting Eqs.~(\ref{eq:ch2_pszmod}) and (\ref{eq:ch2_dent})
into Eq.~(\ref{eq:ch2_nevmod}) we obtain
$$
n_{j\uparrow}^{ev}
=\sum\limits_{m=1}^{\infty}(-1)^{m-1}
e^{-\left[ m^{2}+(2j-1)m\right]\frac{\delta}{2}}~
\frac{\sum\limits_{S_z=-N_e/2}^{N_e/2}e^{-\left(S_{z}^{2}-mS_{z}\right)\delta}}
{\sum\limits_{S_z=-N_e/2}^{N_e/2}e^{-S_{z}^{2}\delta}}.
$$
Since number of electrons in the quantum dot is even,
$S_{z}$ may take only integer values. Therefore,
$$
n_{j \uparrow}^{ev}=\sum\limits_{m=1}^{\infty}(-1)^{m-1}
e^{-\left[ \frac{m^2}{4}+(j-\frac{1}{2})m\right] \delta}~
\frac{\sum\limits_{S_z = -\infty}^{\infty} 
e^{-(S_z -\frac{m}{2})^2\delta}}
{\sum\limits_{S_z = -\infty}^{\infty} e^{-S_z^2 \delta}},
$$
where we extended limits of the sum over $S_{z}$ to
infinities since the Fermi energy is the largest energy
scale in the problem.
Separating $m=2r-1$ and $m=2r$ parts of the sum, 
where $r$ is a positive integer, we obtain final expression 
for the occupation numbers in the case of even number of electrons:
\begin{equation}
n_{j \uparrow}^{ev} 
= A (\delta) \sum_{r=1}^{\infty} 
e^{-(r-\frac{1}{2})(r+2j-\frac{3}{2})\delta} -
\sum_{r=1}^{\infty} e^{-r(r+2j-1)\delta},
\label{eq:ch2_nevanalytic}
\end{equation}
where
\begin{eqnarray}
A(\delta )
&=& \frac{\sum\limits_{s=-\infty}^{\infty} 
e^{-\left( s-\frac{1}{2}\right)^2 \delta}}
{\sum\limits_{s=-\infty}^{\infty} e^{- s^2 \delta}}.
\end{eqnarray}
In two limiting cases
\begin{eqnarray}
A(\delta )
&=& \left\{
\begin{array}{l}
2e^{-\delta /4}
\left[ 1+O\left( e^{-\delta}\right)\right],~~\delta E\gg T  \\
\\
1-4e^{-\pi^2/\delta}+O\left( e^{-2\pi^2/\delta }\right),~~T\gg\delta E
\nonumber
\end{array}
\right. .
\end{eqnarray}
Analytical expression for the high-temperature limit of 
$A(\delta )$ can be obtained using Poisson's summation formula.

One can easily prove the following properties 
of the occupation numbers $n_{j \uparrow}^{ev}$
valid at arbitrary temperature:
\begin{eqnarray}
n_{j \uparrow}^{ev} &=& 1 - n_{1-j,\uparrow}^{ev},
\\
e^{j \delta} n_{j \uparrow}^{ev} + e^{-j \delta} n_{-j \uparrow}^{ev}
&=& A( \delta ) e^{\delta /4}.
\end{eqnarray}
They are valid due to the electron-hole symmetry and 
similar to the following properties of the Fermi-Dirac 
distribution: $n_F(E)=1-n_F(-E)$ and 
$e^{\beta E}n_F(E)+e^{-\beta E}n_F(-E)$ $=1$.

Similarly, one can find occupation numbers in the case 
of odd number of electrons in the quantum dot, $N_{e}$.
Energy level which contains one electron at $T = 0$ 
will be referred to as $j=0$ level. In this case 
electron-hole symmetry corresponds to $j \to -j$ transformation. 
Parameter $\zeta'$ of the spin-up electron subsystem 
at a given $S_z$ is equal to $\zeta +(S_z -1/2)\delta E$, 
see Figs.~\ref{fig:ch2_Fig4}(c) and \ref{fig:ch2_Fig4}(d).
Therefore,
\begin{eqnarray}
n_{j \uparrow}^{od}
=\sum\limits_{S_{z}=-N_{e}/2}^{N_{e}/2}
P\left( S_{z}\right)~n_{j-\left( S_{z}-\frac{1}{2}\right)}
\nonumber \\
=\sum\limits_{m=1}^{\infty} (-1)^{m-1} 
e^{-\left( \frac{m^2}{4}+jm\right) \delta}
~\frac{\sum\limits_{S_{z}=-N_{e}/2}^{N_{e}/2} 
e^{-\left( S_z -\frac{m}{2}\right)^2 \delta}}
{\sum\limits_{S_{z}=-N_{e}/2}^{N_{e}/2} e^{-S_z^2 \delta}}.
\end{eqnarray}
Since number of electrons in the quantum dot is odd,
$S_{z}$ may take only half-integer values. Separating 
odd and even parts of the sum over $m$, we obtain:
\begin{equation}
n_{j \uparrow}^{od}=\frac{1}{A( \delta )}\sum_{r=1}^{\infty}
e^{-\left( r-\frac{1}{2}\right)\left( r+2j-\frac{1}{2}\right)\delta}-
\sum_{r=1}^{\infty} e^{-r(r+2j) \delta}.
\label{eq:ch2_nodanalytic}
\end{equation}
Property of the electron-hole symmetry reads as follows:
$n_{j\uparrow}^{od}=1-n_{-j\uparrow}^{od}$.

It turns out that there exists simple relation between 
$n^{ev}$ and $n^{od}$ occupation numbers:
\begin{equation}
n_{j \uparrow}^{ev} =A(\delta ) 
e^{-\left( j-\frac{1}{4}\right)\delta} n_{-j \uparrow}^{od}.
\label{eq:ch2_evod}
\end{equation}
This property is the analog of $n_F(E)=e^{-\beta E}n_F(-E)$ 
one of the Fermi-Dirac distribution. It will allow us to get rid of
$n^{od}$ occupation numbers in the final expression for the conductance.

Now we are in a position to find the probability that a dot, 
in thermodynamic equilibrium with the reservoirs, 
contains $N_e$ electrons, $P_{eq}(N_e)$.
Eq.~(\ref{eq:ch2_peq2}) written for the spinless case is still 
applicable if we keep in mind that energy levels in the 
quantum dot are doubly degenerate. Partition function
of the dot's internal degrees of freedom in the canonical
ensemble is
\begin{eqnarray}
{\mathcal Z}(N_e) &=& \sum\limits_{\{ n_{i\sigma}\}}
e^{- \beta \sum (E_i - \mu)n_{i\sigma}} \delta_{N_e, \sum n_{i\sigma}}
\nonumber \\
&=& \sum\limits_{N_e^{\uparrow}=0}^{N_e} 
\sum\limits_{N_e^{\downarrow}=0}^{N_e}
Z(N_e^{\uparrow}) Z(N_e^{\downarrow})
\delta_{N_e, N_e^{\uparrow}+N_{e}^{\downarrow}},
\label{eq:ch2_eq9}
\end{eqnarray}
where
\begin{eqnarray}
Z(N_e^{\uparrow})=\sum_{\{n_i\}}e^{-\beta\sum (E_i - \mu)n_i}
\delta_{N_e^{\uparrow}, \sum n_i}
\end{eqnarray}
is the partition function of the spin-up electron subsystem in 
the canonical ensemble. Mathematically, expression for 
$Z(N_e^{\uparrow})$ is identical to the one for $Z(N_e)$ 
in the spinless case. Thus, one can directly apply the result
obtained previously, Eq.~(\ref{eq:ch2_zne}):
\begin{eqnarray}
Z(N_e^{\uparrow})=e^{-(N_e^{\uparrow}-N_{i}/2)^2\delta /2}Z_{exc},
\end{eqnarray}
where $N_e^{\uparrow}=N_{i}/2$ is the equilibrium number of 
electrons in the spin-up subsystem. Similar result is valid 
for the partition function of the spin-down electron subsystem, 
$Z(N_e^{\downarrow})$. 
Substituting these results in Eq.~(\ref{eq:ch2_eq9}) we obtain:
\begin{eqnarray}
{\mathcal Z}(N_e)=Z_{exc}^2 
\, \sum_{N_e^{\uparrow}=0}^{N_e} 
\sum_{N_e^{\downarrow}=0}^{N_e}
e^{-\left[ (N_e^{\uparrow} - N_{i}/2)^2 + 
(N_e^{\downarrow} - N_{i}/2)^2\right]
\delta /2} \delta_{N_e^{\uparrow} + N_e^{\downarrow}, N_e}
\nonumber .
\end{eqnarray}
The exponent can be simplified as follows
$$
\left( N_e^{\uparrow}-\frac{N_{i}}{2}\right)^2 
+\left( N_e^{\downarrow}-\frac{N_{i}}{2}\right)^2
=\frac{N^2}{2}+\frac{\left( N_e^{\uparrow} - 
N_e^{\downarrow}\right)^2}{2},
$$
hence,
\begin{eqnarray}
\frac{{\mathcal Z}(N_e)}{Z_{exc}^2} &=& e^{-N^2\delta /4}
\sum_{N_e^{\uparrow}=0}^{N_e}\sum_{N_e^{\downarrow}=0}^{N_e}
e^{-(N_e^{\uparrow}-N_e^{\downarrow})^2\delta /4} 
\delta_{N_e^{\uparrow}+N_e^{\downarrow}, N_e}
\nonumber \\
&=& e^{-N^2 \delta /4}\sum_{N_e^{\uparrow}=0}^{N_e}
e^{- (N_e^{\uparrow}-N_e /2)^2\delta}
= e^{-N^2 \delta /4} \sum_{s = - N_e /2}^{N_e /2}
e^{-s^2 \delta},
\nonumber
\end{eqnarray}
where in the second equality we took advantage of the delta 
symbol. Sum in the last line is taken over integer values
of $s$ if $N_e$ is even or half-integer values of $s$ if 
$N_e$ is odd. Limits of the sum over $s$ can be
extended to infinities since we assume that $\mu\gg T$.
According to Eq.~(\ref{eq:ch2_peq2}) probability that quantum dot, 
in thermodynamic equilibrium with the reservoirs, contains 
$N_e$ electrons is
\begin{eqnarray}
P_{eq}(N_e) &=& \frac{{\mathcal Z}(N_e)}{Z_{\mu}}e^{-\beta U(N)}
=\frac{Z_{exc}^2}{Z_{\mu}}
e^{-\beta {\tilde U}(N)}
\sum_{s=-N_e/2}^{N_e/2}e^{-s^2\delta}
\nonumber
\\
&=& \frac{e^{-\beta {\tilde U}(N)}
\sum\limits_{s=-N_e/2}^{N_e/2}e^{-s^2\delta}}
{\sum\limits_{N=ev}
e^{-\beta {\tilde U}(N)}
\sum\limits_{s} e^{-s^2 \delta}
+\sum\limits_{N=od}
e^{-\beta {\tilde U}(N)} 
\sum\limits_{s}
e^{-\left( s-\frac{1}{2}\right)^2\delta}},
\nonumber
\end{eqnarray}
where ${\tilde U}(N)\equiv U(N)+N^2\delta E/4$;
and we used the fact that $N$ and $N_e$ have the same parity 
since $N_{i}$ is chosen even. 
Therefore, sums over $N=ev$ and $N=od$ are taken over 
$N=0,\pm 2,\pm 4,\ldots$ and $N=\pm 1,\pm 3,\ldots$ values, 
respectively. At this point in the calculation we need to 
specify whether the total number of electrons in the dot 
is even or odd:
\begin{equation}
P_{eq}(N_e)=
\left\{
\begin{array}{l}
\left( D'\right)^{-1}
e^{-\beta {\tilde U}(N)},~~N~\mbox{is~even}
\\
\left( D'\right)^{-1} A(\delta )
e^{-\beta {\tilde U}(N)},~~N~\mbox{is~odd}
\end{array}
\right.
,
\label{eq:ch2_pnespin}
\end{equation}
where
\begin{eqnarray}
D'\equiv\sum_{N=even}e^{-\beta {\tilde U}(N)}
+A(\delta )\sum_{N=odd}e^{-\beta {\tilde U}(N)}.
\end{eqnarray}

Now we are prepared to calculate the conductance, 
Eq.~(\ref{eq:ch2_beespin}), in the case of the 
equidistant double degenerate energy levels 
in the dot at an arbitrary $\delta E/T$ and
$E_{C}/\delta E$ ratios.
Similarly to the consideration in the spinless case
we assume that quantum dot is weakly coupled to the
leads via multichannel tunnel junctions, and 
tunneling widths of the energy levels in the quantum dot,
$\Gamma_{i}^{l}$ and $\Gamma_{i}^{r}$, are slowly changing
functions of the energy, $E_i$.
\begin{figure}
\begin{center}
\includegraphics[width=14.0cm]{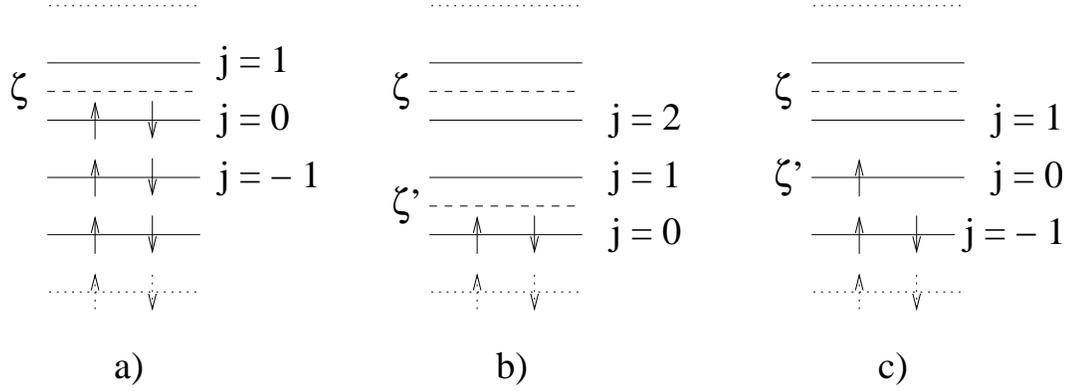}
\caption{Mapping of the sum over $i$ onto the sum over $j$ for 
a) $N=0$; 
b) even $N$ ($N=-4$ case is shown); 
c) odd $N$ ($N=-3$ case is shown).}
\label{fig:ch2_Fig6}
\end{center}
\end{figure}
Then, these tunneling widths can be considered 
constants and evaluated at the chemical potential:
$\Gamma_i^l\approx\Gamma^l_{\mu}\equiv \Gamma^l$; 
$\Gamma_i^r\approx\Gamma^r_{\mu}\equiv \Gamma^r$.
Furthermore, they can be expressed via
conductances of the corresponding junctions:
\begin{equation}
\Gamma^l=\frac{hG^l}{e^2}\frac{\delta E}{2},
~~~\Gamma^r=\frac{hG^r}{e^2}\frac{\delta E}{2}.
\label{eq:ch2_glgr}
\end{equation}
There is an additional factor of $1/2$ here compared to the 
spinless case, Eq.~(\ref{eq:ch2_glgr0}), due to the double 
degeneracy of each energy level in the quantum dot.
First of all, let us break the sum over $N_e$ in 
Eq.~(\ref{eq:ch2_beespin}) in two parts: $N=even$ and 
$N=odd$, and apply Eqs.~(\ref{eq:ch2_glgr}) and (\ref{eq:ch2_pnespin}):
\begin{eqnarray}
G = \frac{G^lG^r}{G^l+G^r}\frac{\delta}{D'}
\nonumber \\
\times
\left( 
\sum_{N=even}e^{-\beta {\tilde U}(N)}
\sum_{i=1}^{\infty} F_{eq}(E_{i\uparrow}|N_e)
\left\{ 1-n_F\left[ E_i-\mu +U(N)-U(N-1)\right]\right\}
\right.
\nonumber \\
\left.
+A(\delta )\sum_{N=odd}e^{- \beta {\tilde U}(N)}
\sum_{i=1}^{\infty}F_{eq}(E_{i\uparrow}|N_e)
\left\{ 1-n_F\left[ E_i-\mu +U(N)-U(N-1)\right]\right\}
\right).
\label{eq:ch2_beforefinal}
\end{eqnarray}
To take advantage of the occupation numbers we derived, 
Eqs.~(\ref{eq:ch2_nevanalytic}) and (\ref{eq:ch2_nodanalytic}),
we need to map each of the sums over $i$ onto the sum 
over $j$. The first sum over $i$ in Eq.~(\ref{eq:ch2_beforefinal}) 
is taken at even number of excess electrons, see 
Figs.~\ref{fig:ch2_Fig6}(a) and \ref{fig:ch2_Fig6}(b), hence
\begin{eqnarray}
F_{eq}\left( E_{i\uparrow}|N_e\right) &=&n_{j\uparrow}^{ev},
\nonumber \\
E_i-\mu =E_i-\zeta' +\left( \zeta' -\zeta\right) &=&
\left( j-\frac{1}{2}\right)\delta E+\frac{1}{2}N\delta E.
\nonumber
\end{eqnarray}
Remember that by properly choosing ``zero'' 
of the gate voltage we put $\zeta =\mu$.
The second sum over $i$ is taken at odd number of excess electrons,
see Figs.~\ref{fig:ch2_Fig6}(a) and \ref{fig:ch2_Fig6}(c), therefore
\begin{eqnarray}
F_{eq}\left( E_{i\uparrow}|N_e\right) &=& n_{j\uparrow}^{od},
\nonumber \\
E_i-\mu =E_i -\zeta' +\left(\zeta' -\zeta\right) &=&
j\delta E +\frac{1}{2}N\delta E.
\nonumber
\end{eqnarray}
We will also use the following identities:
\begin{eqnarray}
{\tilde U}(N)=\left( E_C +\frac{\delta E}{4}\right) N^2 -eV_{e}N 
=\left( E_C +\frac{\delta E}{4}\right)
\left( N -\Delta +\frac{1}{2}\right)^2+{\mathcal C}_{2},
\nonumber
\end{eqnarray}
where ${\mathcal C}_{2}=-(eV_{e})^{2}/(4E_C+\delta E)$;
\begin{eqnarray}
\Delta\equiv\frac{eV_{e}}{2(E_{C}+\delta E/4)}+\frac{1}{2}
\end{eqnarray}
is the dimensionless gate voltage,
$\Delta =0$ corresponds to a position of the conductance peak; and
\begin{eqnarray}
U(N)-U(N-1)=(2N-1)E_C-eV_{e}
=(2N-1)E_C
-2(E_C+\frac{\delta E}{4})(\Delta -\frac{1}{2}).
\nonumber
\end{eqnarray}
Substituting these results in Eq.~(\ref{eq:ch2_beforefinal}) we obtain
\begin{eqnarray}
\frac{G(\Delta )}{G_{\infty}}&=&
\frac{\delta}{D}
\sum_{N=even} 
e^{-\varepsilon\left( N-\Delta +1/2\right)^2}
\nonumber \\
&\times &
\sum_{j=-\infty}^{\infty}n_{j\uparrow}^{ev}
\left[
\frac{1}{e^{-(j-1/4)\delta-2(N-\Delta)\varepsilon}+1}
+\frac{1}{e^{-(j-1/4)\delta+2(N+1-\Delta)\varepsilon}+1}
\right]
,
\label{eq:ch2_spinhalfcond}
\end{eqnarray}
where
\begin{eqnarray}
D = \exp \left(\beta{\mathcal C}_{2}\right) D' 
= \sum_{N=even} e^{-\varepsilon (N-\Delta +1/2)^2}
+ A(\delta ) \sum_{N=odd} e^{-\varepsilon (N-\Delta +1/2)^2};
\end{eqnarray}
\begin{eqnarray}
\varepsilon\equiv\beta\left( E_C+\frac{\delta E}{4}\right);
\end{eqnarray}
$G_{\infty}\equiv G^lG^r/(G^l+G^r)$
is the high-temperature, $E_C, \delta E\ll T\ll\mu$, 
limit of the conductance. 
To eliminate $n^{od}$ from the final expression 
we used useful property of the occupation numbers 
given by Eq.~(\ref{eq:ch2_evod}).

\begin{figure}
\begin{center}
\includegraphics[width=10.0cm]{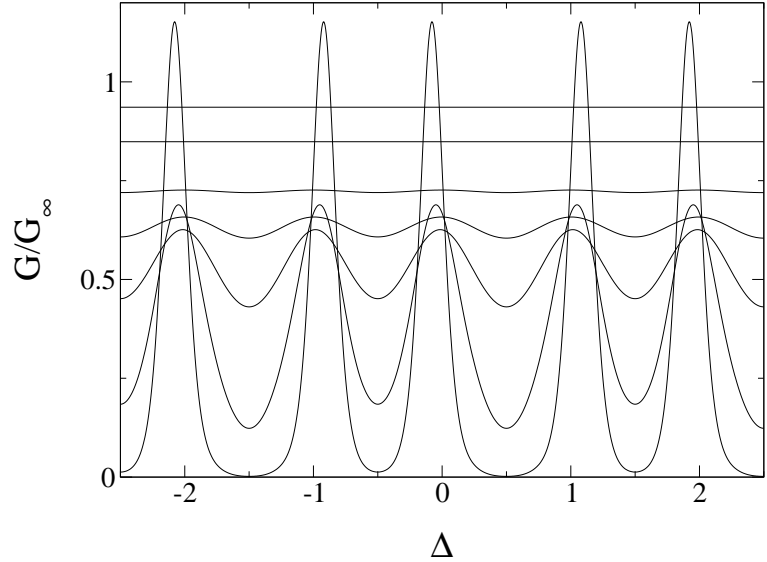}
\caption{Linear conductance oscillations as a function of 
the dimensionless gate voltage, $\Delta$, at $\delta E=E_C$.
Curves are plotted for different temperatures:
$T/E_C=T/\delta E\equiv 1/\delta =0.15, 0.3, 0.5, 0.7, 1, 2, 5$
using Eq.~(\ref{eq:ch2_spinhalfcond}).}
\label{fig:ch2_spinhalfplot}
\end{center}
\end{figure}

Formula (\ref{eq:ch2_spinhalfcond}) is the main result of 
this chapter. It is the analytical expression for the linear 
conductance in the spin-$\frac{1}{2}$ case for equidistant 
energy levels in the quantum dot. One can use 
Eq.~(\ref{eq:ch2_spinhalfcond}) to plot Coulomb blockade 
oscillations of the conductance as a function of the 
dimensionless gate voltage, $\Delta$, at arbitrary values 
of $E_C$, $\delta E$ and $T$. Particularly, when all energy 
scales are of the same order: $E_C \sim \delta E \sim T$, 
numerical calculation is a breeze, Fig.~\ref{fig:ch2_spinhalfplot}.

One can immediately notice the following properties of the 
linear conductance. First of all, $G(\Delta )=G(\Delta +2M)$, 
where $M$ is an integer. In other words, $(4E_C+\delta E)/e$ 
is the conductance period in the gate voltage units. 
This property reflects symmetry with respect to
adding (removing) two electrons to (from) the quantum dot. 
Secondly, conductance is a symmetric function with respect to 
the center of a valley, $\Delta'=M+1/2$, where $M$ is an integer.
That is, $G(\Delta -\Delta')=G(\Delta'-\Delta)$.
This is a reflection of the electron-hole symmetry.
These properties of the conductance oscillations
are not generic. They are valid due to the assumption 
of equally spaced energy levels in a quantum dot.

At high temperatures, $T\gg E_C, \delta E$, conductance 
peaks overlap and their maximums become almost 
equidistant, Fig.~\ref{fig:ch2_spinhalfplot}. As a result, instead 
of separate peaks, the conductance in this limit has 
oscillatory behavior.
\begin{figure}
\begin{center}
\includegraphics[width=10.0cm]{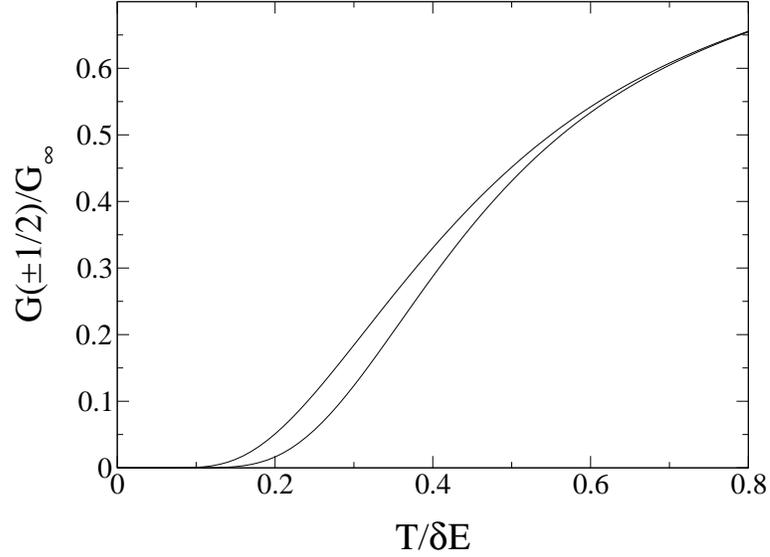}
\caption{Temperature dependence of the conductance in the odd, 
$G(-1/2)/G_{\infty}$ and even, $G(1/2)/G_{\infty}$ valleys 
at $\delta E=E_C$. Upper curve corresponds to the conductance 
in the odd valley.}
\label{fig:ch2_Fig8}
\end{center}
\end{figure}

Let us find temperature dependence of the conductance in 
the valleys. In the sequential tunneling approximation 
the conductance in the valleys decays exponentially as
$T\to 0$, Fig.~\ref{fig:ch2_Fig8}. At low temperature number of 
electrons in a dot in the valleys is almost quantized. 
We will call the valley ``odd'' (``even'') if it corresponds 
to odd (even) number of electrons in the dot. We find that 
at any temperature the conductance in the odd valley 
is larger than that in the even one, Fig.~\ref{fig:ch2_Fig8}. 
This feature is robust with respect to the distribution of 
energy levels in a quantum dot.

However, it is important to mention that at low temperatures,
$T<T_{in}$, where
\begin{eqnarray}
T_{in}\simeq\frac{E_C}{\ln
\left(
\frac{e^2/\hbar}{G^l+G^r}
\right)}
\end{eqnarray}
cotunneling \cite{Gla00} will dominate sequential
tunneling contribution to the conductance in the valleys.
Therefore, temperature dependence of the conductance in 
the valleys, Fig.~\ref{fig:ch2_Fig8}, is valid only for
the temperatures $T>T_{in}$.

Let us analyze the limit of large charging energy: 
$E_C\gg T, \delta E$ or, equivalently, 
$\varepsilon\gg 1, \delta$ in the dimensionless units.
In this limit, two adjacent peaks in the conductance have 
exponentially small, $\sim e^{-E_C/T}$, overlap with each other.
Thus, it makes perfect sense to study the line shape of a 
separate peak. Let us determine line shape of the conductance
peak near $\Delta =0$. In the numerator of Eq.~(\ref{eq:ch2_spinhalfcond}) 
only the $N=0$ term in the sum over $N$ survives;
moreover, at $N=0$ second term in square brackets 
is $O(e^{-2\varepsilon})$. 
In the denominator, only the $N=-1,0$ terms matter.
Hence, the line shape of the conductance peak near $\Delta =0$ at 
arbitrary $\delta E/T$ ratio is given by
\begin{equation}
\frac{G(\phi )}{G_{\infty}}=
\frac{\delta}{1+A(\delta )e^{-\beta e\phi}}
\sum_{j=-\infty}^{\infty}
\frac{n_{j\uparrow}^{ev}}{1+e^{-(j-1/4)\delta +\beta e\phi}},
\label{eq:ch2_peakspinexact}
\end{equation}
where we used the following identity:
$$
2\varepsilon\Delta 
= \beta e\left( V_{e}-V_{e}^{(0)}\right)
= \beta e\phi .
$$
Clearly, $\phi =0$ corresponds to $\Delta =0$.
In the classical regime, $T\gg \delta E$, the line shape 
of the conductance peak is given by~\cite{Gla89}
\begin{equation}
\frac{G(\phi )}{G_{\infty}}=
\frac{\beta e\phi}{2\sinh (\beta e\phi )}.
\label{eq:ch2_spinclassical}
\end{equation}
Formally, this equation is identical to that of the spinless case, 
Eq.~(\ref{eq:ch2_classical}). Nonetheless, the values of $G_{\infty}$ 
are different in these two cases by a factor of $2$. 
This is due to spin degeneracy of each energy level 
in the spin-$\frac{1}{2}$ case, 
compare Eqs.~(\ref{eq:ch2_glgr0}) and (\ref{eq:ch2_glgr}). 
In the limit of $\delta E\gg T$~\cite{Gla88}:
\begin{equation}
\frac{G(\phi )}{G_{\infty}}=
\frac{\delta}{3+2\sqrt{2}\cosh 
\left(
\beta e\phi +\frac{1}{4}\delta -\frac{1}{2}\ln 2 
\right)}.
\label{eq:ch2_eq12}
\end{equation}
\begin{figure}
\begin{center}
\includegraphics[width=10.0cm]{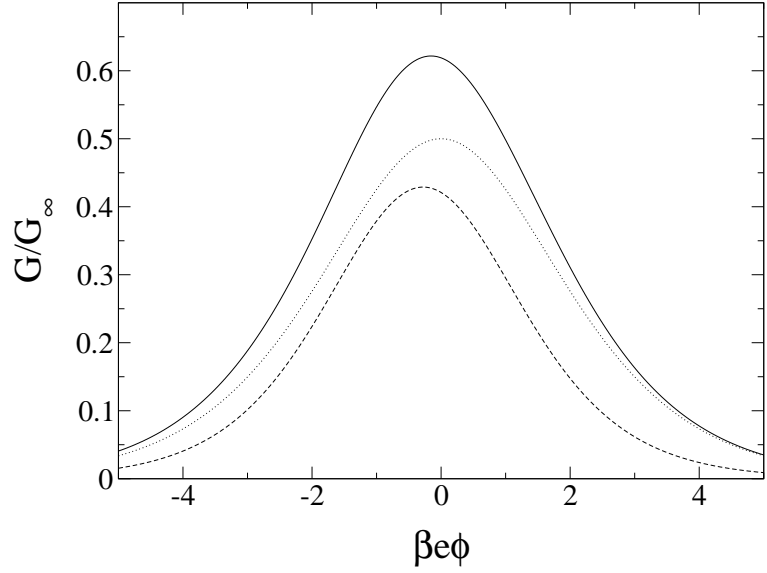}
\caption{The exact line shape of the conductance peak, 
Eq.~(\ref{eq:ch2_peakspinexact}), plotted at 
$1/\delta\equiv T/\delta E =0.4$ (solid curve).
Dotted curve corresponds to blindly applying 
classical regime formula, Eq.~(\ref{eq:ch2_spinclassical}),
at $T/\delta E =0.4$.
Dashed curve corresponds to blindly applying 
formula (\ref{eq:ch2_eq12}) at $T/\delta E=0.4$. 
The argument of the plot is linear function of 
the gate voltage, $\phi$.}
\label{fig:ch2_Fig9}
\end{center}
\end{figure}
This peak has its maximum at 
\begin{equation}
e\phi_{LT}\equiv
e\phi_{m}\left( T\ll\delta E\right)
=-\frac{1}{4}\delta E+\frac{\ln 2}{2}T,
\label{eq:ch2_phirt}
\end{equation}
and is symmetric with respect to this value:
$G(\phi -\phi_{LT})=G(\phi_{LT}-\phi )$.
The exact line shape of the conductance peak, 
Eq.~(\ref{eq:ch2_peakspinexact}), at $T/\delta E=0.4$ and
two limiting cases conductance peaks, 
Eqs.~(\ref{eq:ch2_spinclassical}) and (\ref{eq:ch2_eq12}), plotted
out of their validity region at $T/\delta E =0.4$
are shown in Fig.~\ref{fig:ch2_Fig9}.
As in the spinless case, the exact conductance
peak is higher than both of the limiting cases
peaks.

\begin{figure}
\begin{center}
\includegraphics[width=10.0cm]{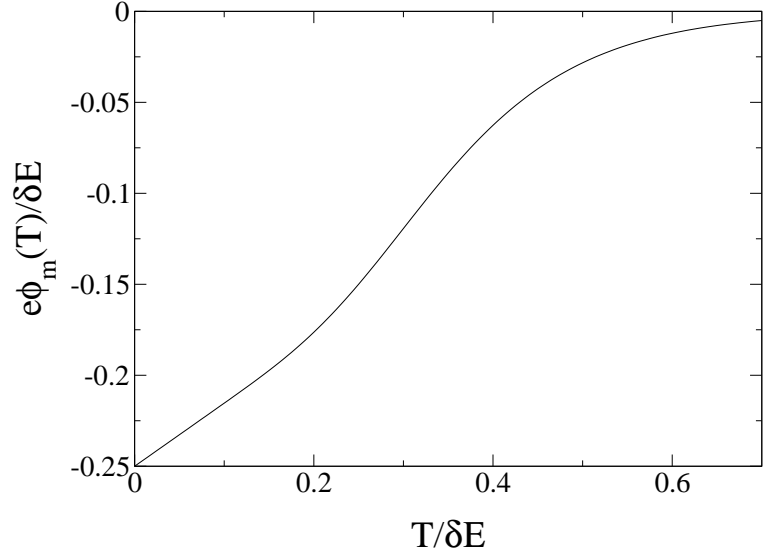}
\caption{Temperature dependence of the dimensionless
peak's maximum position, $e\phi_m(T)/\delta E$.
In the low temperature limit: $e\phi_m(0)/\delta E=-1/4$
according to Eq.~(\ref{eq:ch2_phirt}).}
\label{fig:ch2_Fig10}
\end{center}
\end{figure}

The position of the peak's maximum, $\phi_m =\phi_m (T)$,
is shifted to the left from its high temperature limit,
$\phi_{CL}\equiv\phi_m(T\gg\delta E)=0$.
It is determined by the equation for $\phi_m$:
$G'(\phi_m )=0$, where $G(\phi )$ is given by 
Eq.~(\ref{eq:ch2_peakspinexact}).
The dimensionless position of the peak's maximum, 
$e\phi_m(T)/\delta E$, as a function of the temperature, 
$T/\delta E$, is numerically plotted in Fig.~\ref{fig:ch2_Fig10}.

\begin{figure}
\begin{center}
\includegraphics[width=10.0cm]{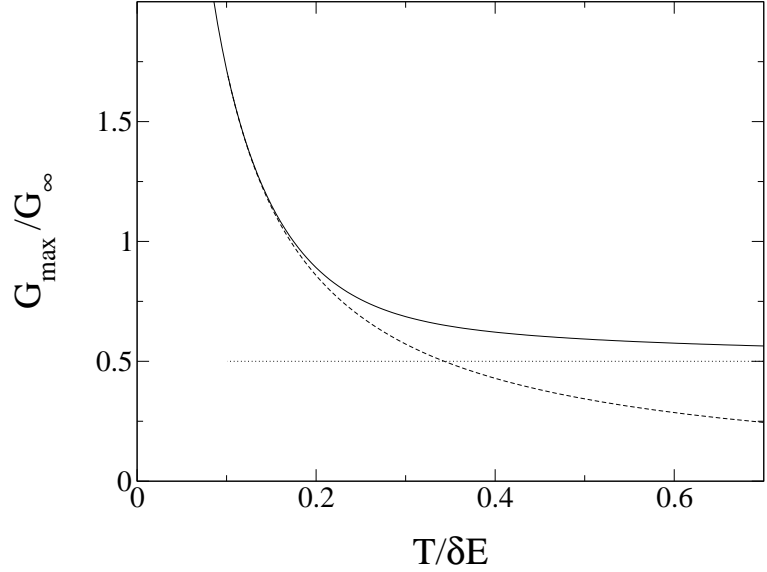}
\caption{Height of the conductance peak, $G_{max}/G_{\infty}$, 
as a function of the temperature, $T/\delta E$. 
Dotted and dashed curves correspond to two limiting cases 
peak heights, Eq.~(\ref{eq:ch2_spheight}).}
\label{fig:ch2_heightspin}
\end{center}
\end{figure}

The conductance peak height is $G_{max}=G(\phi_m)$. In the
limiting cases:
\begin{equation}
\frac{G_{max}}{G_{\infty}}=\frac{G(\phi_m)}{G_{\infty}}=
\left\{
\begin{array}{l}
1/2,~~T\gg\delta E \\
(3-2\sqrt{2}) \delta,~~\delta E\gg T\gg\Gamma_i
\end{array}
\right.
.
\label{eq:ch2_spheight}
\end{equation}
Peak's height as a function of the temperature, $T/\delta E$,
can be plotted numerically, Fig.~\ref{fig:ch2_heightspin}.

\section{Conclusions}
\label{sec:ch2_conclusions}

We have studied Coulomb blockade oscillations of the
linear conductance through a quantum dot weakly
coupled to the leads via multichannel tunnel
junctions in the sequential tunneling approximation. 
To obtain analytical results we have assumed that 
the energy levels in the dot are equally spaced.
The electron-electron interaction in a quantum dot 
has been described by the constant interaction model;
though, thermal excitations with all possible spins 
have been taken into account.

The linear conductance in the spinless case is given 
by Eq.~(\ref{eq:ch2_g0}). It is valid at arbitrary 
values of $E_C$, $\delta E$ and $T$.
The line shape of an individual conductance peak at arbitrary
ratio $\delta =\delta E/T$ is given by Eq.~(\ref{eq:ch2_onepeak}).
Exact conductance peak is higher than both of the 
limiting cases peaks at any gate voltage
as is illustrated in Fig~\ref{fig:ch2_gd0peak}.
An analytical expression for the height of the conductance
peak at any ratio $\delta$ is obtained, Eq.~(\ref{eq:ch2_eqheight}).

In Section~\ref{sec:ch2_edge_states} we applied the 
spinless case theory result to the problem of 
the transport via a dot in the quantum Hall regime.
Energy levels in a dot in this case are equidistant
with the spacing given by Eq.~(\ref{eq:ch2_qhespacingrealdot}).

Linear conductance in the case of spin-$\frac{1}{2}$ 
electrons at arbitrary values of $E_C$, $\delta E$, and $T$ 
is given by Eq.~(\ref{eq:ch2_spinhalfcond}).
In particular, this equation allows one to plot
the conductance oscillations in the regime when
the charging energy, level spacing in the dot, and
the temperature are all of the same order, Fig.~\ref{fig:ch2_spinhalfplot}.
We find that the period of Coulomb blockade
oscillations is doubled compared to the model with
a continuous electronic spectrum in the dot.
Equation~(\ref{eq:ch2_spinhalfcond}) is the main result of the chapter.

We also find that conductance in the odd valley is larger 
than that in the even one at any temperature, Fig.~\ref{fig:ch2_Fig8}.
The difference between conductances has the largest value at
$T\approx 0.3~\delta E$ (at $\delta E=E_C$). 
The sign of the difference is the same as for the quantum dot 
in the Kondo regime \cite{Gla00}.
Kondo effect takes place at very low temperatures,
$T\lesssim T_{K}\ll T_{in}$,
where $T_K$ is the Kondo temperature,
and leads to the logarithmic enhancement
of the conductance in the odd valleys \cite{Ale02}.
Our consideration shows that even-odd asymmetry
exists at much higher temperatures.

Line shape of the conductance peak is given by 
Eq.~(\ref{eq:ch2_peakspinexact}). As in the spinless case,
the conductance peak is higher than both of the 
limiting cases peaks at any gate voltage, Fig~\ref{fig:ch2_Fig9}.
As we increase the temperature peaks' maximums shift and
become more equidistant, Fig~\ref{fig:ch2_Fig10}.
The peak's height as a function of the temperature
is calculated numerically and plotted in Fig.~\ref{fig:ch2_heightspin}.

Though we have found physical system which has equidistant energy 
levels in the spinless case, see Section~\ref{sec:ch2_edge_states}, 
we are not aware of any such system in the spin-$\frac{1}{2}$ case. 
In the case of a chaotic quantum dot 
Wigner-Dyson model gives a fairly good 
approximation for the distribution of the energy levels of the dot. 
If we had assumed Wigner-Dyson distribution of the quantum dot's 
energy levels then we would have had to give up the hope of finding 
a solution. It goes back to the very difficult problem of finding 
occupation numbers of the dot's energy levels in the canonical ensemble. 
The only way to solve it is to assume that energy levels 
in the quantum dot are equally spaced. 
Then one can use the bosonization technique to find 
the occupation numbers. Assumption of the equidistant energy levels is in 
line with the level repulsion property of the Wigner-Dyson distribution. 
Therefore, the analytical consideration of 
this reasonably simplified model, in our opinion, 
is a significant step forward in the solution of the 
general problem.

Though we do not expect our quantitative results to precisely describe 
a quantum dot with random energy levels, they certainly give correct order 
of magnitude for the conductance oscillations and their generic features.

%% file: chapters/chapter3.tex
\setlength{\textheight}{8.0in}
\clearpage
\chapter{Coulomb Blockade Peak Spacings: 
Interplay of Spin and Dot-Lead Coupling}
\label{ch:ch3}
\thispagestyle{botcenter}
\setlength{\textheight}{8.4in}

\section{Overview}

For Coulomb blockade peaks in the linear conductance 
of a quantum dot, we study the correction to the spacing 
between the peaks due to dot-lead coupling. 
This coupling can affect measurements in which Coulomb 
blockade phenomena are used as a tool to probe the energy 
level structure of quantum dots.  The electron-electron 
interactions in the quantum dot are described by the 
constant exchange and interaction (CEI) model while 
the single-particle properties are described by random 
matrix theory.  We find analytic expressions for both 
the average and rms mesoscopic fluctuation of the correction.  
For a realistic value of the exchange interaction constant 
$J_{\rm s}$, the ensemble average correction to the peak 
spacing is two to three times smaller than that at 
$J_{\rm s} \!=\! 0$.  As a function of $J_{\rm s}$, the 
average correction to the peak spacing for an even valley 
decreases monotonically, nonetheless staying positive. 
The rms fluctuation is of the same order as the average and 
weakly depends on $J_{\rm s}$.  For a small fraction of quantum 
dots in the ensemble, therefore, the correction to the peak 
spacing for the even valley is negative. The correction to 
the spacing in the odd valleys is opposite in sign to that 
in the even valleys and equal in magnitude. These results are 
robust with respect to choice of the random matrix ensemble 
or change in parameters such as charging energy, mean level 
spacing, or temperature.  

The work in this chapter was done in collaboration with 
Harold~U. Baranger.

\section{Introduction}

\begin{figure}
\begin{center}
\includegraphics[width=11.2cm]{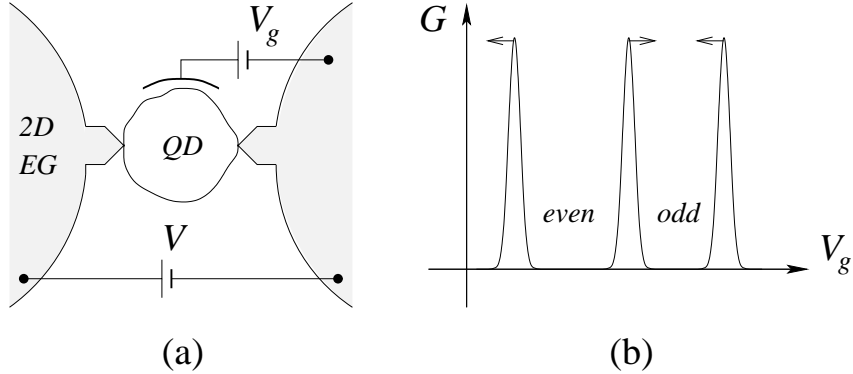}
\caption{(a) Scheme of the Coulomb blockade setup; 
(b) Oscillations of the SET linear conductance $G$ as the 
back-gate voltage $V_g$ is changed. ``Even'' (``odd'') 
corresponds to an even (odd) number of electrons in the valley.  
Arrows show average shifts in the positions of the peaks' maxima 
due to the finite dot-lead couplings.}
\label{fig:ch3_su}
\end{center}
\end{figure}

Progress in nanoscale fabrication techniques has made possible 
not only the creation of more sophisticated devices but also 
greater control over their properties.  Electron systems confined 
to small regions -- quantum dots (QD) -- and especially their 
transport properties have been studied extensively for the last 
decade~\cite{Kou97,Del01}.  One of the most popular devices is 
a lateral quantum dot, formed by depleting the two-dimensional 
electron gas (2DEG) at the interface of a semiconductor 
heterostructure. By appropriately tuning negative potentials 
on the metal surface gates, one can control the QD size, the 
number of electrons $n$ it contains, as well as the tunnel 
barrier heights between the QD and the large 2DEG regions, 
which act as leads.  Applying bias voltage $V$ between these 
leads allows one to study transport properties of a single 
electron transistor (SET), Fig.~\ref{fig:ch3_su}(a)~\cite{Kou97}.

We study properties of the conductance $G$ through a QD in the 
linear response regime. We assume that the dot is weakly coupled 
to the leads: $G_{L,R} \!\ll\! e^2/h$, where $G_{L,R}$ are the 
conductances of the dot-lead tunnel barriers, $e \!>\! 0$ is the 
elementary charge, and $h$ is Planck's constant.

To tunnel onto the quantum dot, an electron in the left lead 
has to overcome a charging energy $E_C \!=\! e^2/2C$, where 
$C$ is the capacitance of the QD, a phenomenon called the Coulomb 
blockade.  However, if we apply voltage $V_g$ to an additional 
back-gate capacitively coupled to the QD, see 
Fig.~\ref{fig:ch3_su}(a), the Coulomb blockade can be lifted.  
Indeed, by changing $V_g$ one can change the electrostatics so 
that energies of the quantum dot with $n$ and $n+1$ electrons 
become equal, and so an electron can freely jump from the left 
lead onto the QD and then out to the right lead.  Thus, a current 
event has occurred, and a peak in the conductance corresponding 
to that back-gate voltage, $V_{g,n+\frac{1}{2}}$, is observed.  
By sweeping the back-gate voltage, a series of peaks is observed, 
Fig.~\ref{fig:ch3_su}(b).

In this chapter we calculate the correction to the spacing 
between Coulomb blockade peaks due to finite dot-lead tunnel 
coupling.  In recent years, low-temperature Coulomb blockade 
experiments have been repeatedly used to probe the energy level 
structure of quantum dots~\cite{Kou97,Del01,Alh00}.  The dot-lead 
tunnel coupling discussed here may influence such 
a measurement -- the presence of leads may change what one 
sees -- and so an understanding of coupling effects is needed.  
One dramatic consequence is the Kondo effect in quantum 
dots~\cite{Gol98a,Gol98b,Gla00}. Here we assume that $T \!\gg\! T_{K}$, 
where $T_{K}$ is the Kondo temperature, and, therefore, do not 
consider Kondo physics, focusing instead on less dramatic effects 
that, however, survive to higher temperature.

We study an ensemble of chaotic ballistic (or chaotic disordered) 
quantum dots with large dimensionless conductance, $g \!\gg\! 1$.  
The dimensionless conductance is defined as the ratio of the 
Thouless energy $E_T$ to the mean level spacing $\Delta$: 
$g \!\equiv\! E_T/\Delta$~\cite{Alh00}.  For isolated quantum 
dots with large dimensionless conductance, the distribution of 
$g$ energy levels $\{\varepsilon_k\}$ near the Fermi level and 
the corresponding wave functions $\{\psi_{k}({\bf r})\}$ can be 
approximated by random matrix theory 
(RMT)~\cite{Bee97,Alh00,Meh91}.  As will be evident from what 
follows, the leading contribution to the results obtained here 
comes from $\xi \!\equiv\! 2E_C/\Delta$ energy levels near the 
Fermi level; thus, if $E_{C}\lesssim E_{T}$ the statistics of 
these $\xi$ levels can be described by RMT.  We furthermore 
neglect the spin-orbit interaction and, therefore, consider only 
the Gaussian orthogonal (GOE) and Gaussian unitary (GUE) 
ensembles of random matrices.

The microscopic theory of electron-electron interactions in a 
quantum dot with large dimensionless conductance brings about 
a remarkable result~\cite{Kur00,Ale02}.  To leading order, the 
interaction Hamiltonian depends only on the squares of the 
following two operators: 
(i) the total electron number operator 
$\hat{n} = \sum c_{k\sigma}^{\dagger} c_{k\sigma}$ where 
$\left\{ c_{k\sigma}\right\}$ are the electron annihilation 
operators and $\sigma$ labels spin, and (ii) the total spin operator 
$   {\hat {\bf S}}=\frac{1}{2}\sum
      c_{k\sigma_{1}}^{\dagger}
   \left<\sigma_{1}\right|{\hat {\bf \sigma}}\left|\sigma_{2}\right>
      c_{k\sigma_{2}}
$
where $\left\{ {\hat \sigma_{i}} \right\}$ are the Pauli matrices.  
The leading order part of the Hamiltonian reads~\cite{Kur00,Ale02}
 \begin{eqnarray}
   H_{\rm int} = E_C \, \hat{n}^{2} - J_{\rm s}\, {\hat {\bf S}}^2
         \label{eq:ch3_cei}
 \end{eqnarray}
where $E_{C}$ is the redefined value of the charging 
energy~\cite{Kur00,Usa02} and $J_{\rm s} \!>\! 0$ is the exchange 
interaction constant.  Higher order corrections are of order 
$\Delta/g$~\cite{Kur00,Ale02,Usa01,Usa02}.  The coupling constants 
in (\ref{eq:ch3_cei}) are invariant with respect to different 
realizations of the quantum dot potential.  This ``universal'' 
Hamiltonian is also invariant under arbitrary rotation of the 
basis and, therefore, compatible with RMT.  In principal, the 
operator of interactions in the Cooper channel can appear in the 
``universal'' Hamiltonian for the GOE case.  However, if the 
quantum dot is in the normal state at $T \!=\! 0$, then the 
corresponding coupling constant is positive and is renormalized 
to a very small value~\cite{Kur00,Alt85}. The ``universal'' part 
of the Hamiltonian given by Eq.~(\ref{eq:ch3_cei}) is called the 
constant exchange and interaction (CEI) model~\cite{Usa01,Usa02}.

The total Hamiltonian of the quantum dot in the $g \!\to\! \infty$ 
limit thus has two parts, the single-particle RMT Hamiltonian and 
the CEI model describing the interactions.  The capacitive 
QD-backgate coupling generates an additional term which is linear 
in the number of electrons:
 \begin{eqnarray}
   H_{\rm dot} = \sum_{k\sigma} \varepsilon_{k}
      \, c_{k\sigma}^{\dagger} c_{k\sigma}
         + E_C \left( \hat{n} - {\mathcal N} \right)^{2}
            - J_{\rm s}\, {\hat {\bf S}}^2
   \label{eq:ch3_dot}
 \end{eqnarray}
where ${\mathcal N} \!=\! C_{g} V_{g}/e$ is the dimensionless 
back-gate voltage and $C_{g}$ is the QD-backgate capacitance. 

The CEI model contains an additional exchange interaction term 
as compared to the conventional constant interaction model 
(CI model)~\cite{She72,Kul75,Kou97}.  Exchange is important as 
$J_{\rm s}$ is of the same order as $\Delta$, the mean 
single-particle level spacing.  Indeed, in the realistic case 
of a 2DEG in a GaAs/AlGaAs heterostructure with gas parameter 
$r_s \!=\! 1.5$, the static random phase approximation gives 
$J_{\rm s}\!\approx\! 0.31\Delta$~\cite{Ore02}.  Therefore, as 
we sweep the back-gate voltage, adding electrons to the quantum 
dot, the conventional up-down filling sequence may be 
violated~\cite{Bro99,Bar00}.  Indeed, energy level spacings do 
fluctuate: If for an even number of electrons $n$ in the QD 
the corresponding spacing, 
$\varepsilon_{\frac{n}{2}+1}-\varepsilon_{\frac{n}{2}}$, is 
less than $2J_{\rm s}$, then it becomes energetically favorable 
to promote an electron to the next orbital instead of putting 
it in the same one; thus, a triplet state ($S \!=\! 1$) is formed.  
Higher spin states are possible as well.  For $r_{s} \!=\! 1.5$ 
the probability of forming a higher spin ground state is 
$P_1(S\!>\!0)\approx 0.26$ and $P_2(S\!>\!0)\approx 0.19$ for 
the GOE and GUE, respectively. The lower the electron density 
in the QD, the larger $r_s$ and, consequently, the larger the 
exchange interaction constant $J_{\rm s}$.  

\begin{figure}
\begin{center}
\includegraphics[width=10.0cm]{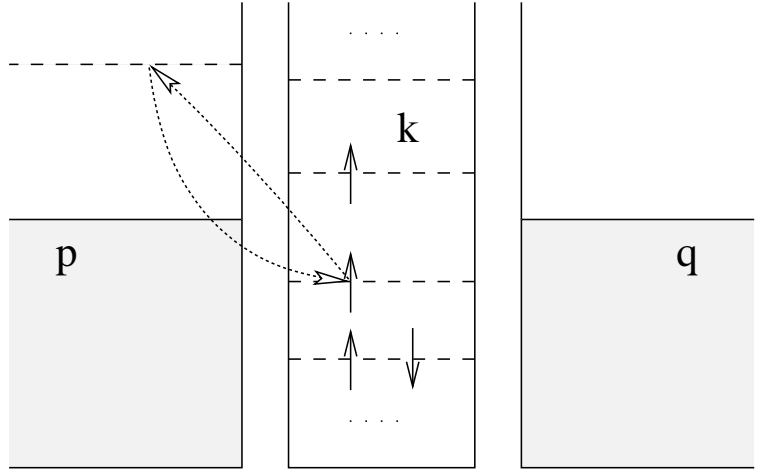}
\caption{One example of the virtual processes contributing 
to $E^{(2)}$.  This virtual process corresponds to an electron 
tunneling out of the quantum dot into the left lead and then 
tunneling back into the same level in the QD.}
\label{fig:ch3_vp}
\end{center}
\end{figure}

The back-gate voltage corresponding to the conductance peak 
maximum ${\mathcal N}_{n-\frac{1}{2}}$ is found by equating the 
energy for $n \!-\! 1$ electrons in the dot with that for $n$ 
electrons:\footnote{More precisely, the conductance 
peak has its maximum at the gate voltage corresponding to the 
maximum of the total amplitude of the tunneling through many 
energy levels in a QD: ${\mathcal A}=\sum_{k\sigma}{\mathcal A}_{k\sigma}$, 
see Ref.~\cite{Kam00}.  Therefore, in addition to the dominant 
resonant term which gives the charge degeneracy point result for 
the peak maximum [Eq.~(\ref{eq:ch3_energy})], the total tunneling 
amplitude includes elastic cotunneling terms (we assume that 
$T\ll\Delta$) with random phases.  However, one can show (see 
Ref.\,\cite{Kam00} for details) that these cotunneling terms 
give negligible fluctuating correction to the position of the 
peak maximum with 
$\mbox{mean}\big(\delta{\mathcal N}_{n-\frac{1}{2}}\big)\approx 0$ and 
$\mbox{rms}\big(\delta{\mathcal N}_{n-\frac{1}{2}}\big) \approx\left(g_L+g_R\right)^{3}/\left(4\pi\right)^{3}\xi$.  The coefficient in the 
last equation corresponds to the CI model.}
 \begin{eqnarray}
      E_{n-1} ( {\mathcal N}_{n - \frac{1}{2}} ) = 
         E_{n} ( {\mathcal N}_{n - \frac{1}{2}} ) \;.
   \label{eq:ch3_energy}
 \end{eqnarray}
As we are interested in the effect of dot-lead coupling on 
these peak positions, it is natural to expand the energies 
perturbatively in this coupling: 
$E = E^{(0)} + E^{(2)} + \dots\,\,$.  One possible virtual 
process contributing to $E^{(2)}$ is shown in 
Fig.~\ref{fig:ch3_vp}.  Electron occupations of the QD 
``to the left'' and ``to the right'' of the conductance peak 
[see Fig.~\ref{fig:ch3_su}(b)] are different; hence, the 
corrections $E^{(2)}$ to the energies are different. Therefore, 
the position of the peak maximum acquires corrections as well, 
${\mathcal N} = {\mathcal N}^{(0)} + {\mathcal N}^{(2)} + \dots\,\,$, as 
does the spacing 
between two adjacent peaks.

This physical scenario has been considered by Kaminski and 
Glazman with the interactions treated in the CI model, i.e. 
neglecting exchange~\cite{Kam00}.  The ensemble averaged 
change in the spacing and its rms due to mesoscopic fluctuations 
were calculated.  On average, ``even'' spacings (that is, 
spacings corresponding to an even number of electrons in the 
valley) increase, while ``odd'' spacings decrease (by the same 
amount)~\cite{Kam00}:
 \begin{eqnarray}
   2E_C~\overline{{\mathcal U}_{n}^{(2)}(J_{\rm s} \!=\! 0)} =
      \Delta\,\frac{g_L+g_R}{2\pi^2}\,\ln\frac{2E_C}{T},
 \end{eqnarray}
where ${\mathcal U}$ is the dimensionless spacing normalized by 
$2E_C$ and $g_{L,R} \!=\! G_{L,R}/(2e^2/h)$ are the 
dimensionless dot-lead conductances.

In this chapter we calculate the same quantities but with the 
electron-electron interactions in the QD described by the more 
realistic CEI model.  We find that the average change in the 
spacing between conductance peaks is significantly less than 
that predicted by the CI model.  However, the fluctuations are 
of the same order.  In contrast to the CI result~\cite{Kam00}, 
for large enough $J_{\rm s}$, we find that ``even'' spacings 
do not necessarily increase (likewise, ``odd'' spacings do not 
necessarily decrease).

The chapter is organized as follows.  In 
Section~\ref{sec:ch3_theham} we write down the total Hamiltonian 
of the system and find the condition for the tunneling Hamiltonian 
to be considered as a perturbation.  In 
Section~\ref{sec:ch3_approach} we describe the approach and 
make symmetry remarks.  In Section~\ref{sec:ch3_case} 
we perform a detailed calculation of the correction to the 
spacing between Coulomb blockade peaks for the 
$\frac{1}{2} \to 1 \to \frac{1}{2}$ spin sequence.  
In Section~\ref{sec:ch3_averaged} we find the ensemble averaged 
correction to the peak spacing.  The rms of the fluctuations 
of the correction to the peak spacing is calculated in 
Section~\ref{sec:ch3_fluctuations}.  
In Section~\ref{sec:ch3_conclusions} we summarize our findings 
and discuss their relevance to the available experimental 
data~\cite{Mau99,Jeo}.

\section{The Hamiltonian}
\label{sec:ch3_theham}

The Hamiltonian of the system in Fig.~\ref{fig:ch3_su}(a) 
consists of the QD Hamiltonian [Eq.~(\ref{eq:ch3_dot})], the 
leads Hamiltonian, and the tunneling Hamiltonian accounting 
for the dot-lead coupling:
 \begin{eqnarray}
   H = H_{\rm dot} + H_{\rm leads} + H_{\rm tun}.
      \label{eq:ch3_hamilt}
 \end{eqnarray}
The leads Hamiltonian can be written as follows
 \begin{eqnarray}
   H_{\rm leads} = \sum_{{\bf p}\sigma}\varepsilon_{{\bf p}}
      c_{{\bf p}\sigma}^{\dagger} c_{{\bf p}\sigma} +
   \sum_{{\bf q}\sigma} \varepsilon_{{\bf q}}
      c_{{\bf q}\sigma}^{\dagger} c_{{\bf q}\sigma},
 \end{eqnarray}
where $\{\varepsilon_{\bf p}\}$ and $\{\varepsilon_{\bf q}\}$ 
are the one-particle energies in the left and right leads, 
respectively, measured with respect to the chemical potential 
(see Fig.~\ref{fig:ch3_vp}). We assume that the leads are large; 
therefore we (i) neglect electron-electron interactions in the 
leads and (ii) assume a continuum of states in each lead.  The 
tunneling Hamiltonian is~\cite{Coh62}
 \begin{equation}
   H_{\rm tun} = \sum_{k{\bf p}\sigma}(t_{k{\bf p}} c_{k\sigma}^{\dagger}
         c_{{\bf p}\sigma} + {\rm h.c.}) +
      \sum_{k{\bf q}\sigma}(t_{k{\bf q}} c_{k\sigma}^{\dagger}
         c_{{\bf q}\sigma} + {\rm h.c.}),
   \label{eq:ch3_tunnelingh}
 \end{equation}
where $\{ t_{k{\bf p}}\}$ and $\{ t_{k{\bf q}}\}$ are the 
tunneling matrix elements.

We assume that $T \!\ll\! \Delta$ and, therefore, neglect excited 
states of the QD concentrating on ground state properties only. 
We also assume that the QD is weakly coupled to the leads, 
treating the tunneling Hamiltonian as a perturbation.  Corrections 
to the position of the peak maximum can be expressed in terms of 
corrections to the ground state energies of the QD via 
Eq.~(\ref{eq:ch3_energy}).  The perturbation series for these 
corrections contains only even powers as $H_{\rm tun}$ is 
off-diagonal in the eigenbasis of $H_{0}$.  The $2m^{\rm th}$ 
correction to the position of the peak is roughly
 \begin{eqnarray}
      2E_C~\overline{{\mathcal N}^{(2m)}}
   \approx \Delta\,\frac{g_L+g_R}{4\pi^2} \,\ln\frac{2E_C}{T}
   \left(\frac{g_L+g_R}{4\pi^2} \,\ln\frac{2E_C}{\Delta}
         \,\ln\frac{2E_C}{T} \right)^{m-1}.
 \end{eqnarray}
Thus, finite-order perturbation theory is applicable if~\cite{Kam00}
 \begin{eqnarray}
   \frac{g_L+g_R}{4\pi^2}~ \ln\left(\frac{2E_C}{\Delta}\right)
      \ln\left(\frac{2E_C}{T}\right) \ll 1.
 \end{eqnarray}
To loosen this restriction one should deploy a renormalization 
group technique which, however, is beyond the scope of this 
chapter~\cite{Kam00,Gla90,Mat91,And70}.

\section{Plan of the Calculation}
\label{sec:ch3_approach}

As the exchange interaction constant $J_{\rm s}$ becomes larger, 
more values of the QD spin $S$ become accessible.  The structure 
of the corrections to the ground state energies depends on the 
total QD spin $S$, and this structure becomes very complicated 
for large values of the spin $S$. Fortunately, for the realistic 
case $r_{s} = 1.5$, the probability of spin values higher than 
$\frac{1}{2}$ in an ``odd'' valley is small: 
$P_2(S \!>\! \frac{1}{2}) \approx 0.01$ for the GUE.  Hence, we 
can safely assume that in the ``odd'' valley the spin is always 
equal to $\frac{1}{2}$.  In the ``even'' valley, one has to allow 
both $S \!=\! 0$ and $S \!=\! 1$ states.

The structure of the expression for the spacing between peaks 
depends on the allowed spin sequences.  For an ``even'' valley 
there are only two possible spin sequences:
 \begin{eqnarray}
   {\textstyle \frac{1}{2}} \to 0 \to {\textstyle \frac{1}{2}}~~~~\mbox{and}~~~~
      {\textstyle \frac{1}{2}} \to 1 \to {\textstyle \frac{1}{2}}
         \label{eq:ch3_seqeven}
 \end{eqnarray}
where the number in the middle is the spin in the ``even'' 
valley, while the numbers to the left and right are spin values 
in the adjacent valleys, Fig.~\ref{fig:ch3_su}(b). For an ``odd'' 
valley there are four possibilities:
 \begin{eqnarray}
   0 \to {\textstyle \frac{1}{2}} \to 0,~~~~ 0 \to {\textstyle \frac{1}{2}} \to 1,
      \nonumber \\
   1 \to {\textstyle \frac{1}{2}} \to 0,~~~~\mbox{and}~~~~ 1 \to {\textstyle \frac{1}{2}} \to 1 \;.
 \end{eqnarray}
To obtain correct expressions for the average spacing between 
peaks, one should weight these sequences with the appropriate 
probability of occurrence.

Before proceeding with the calculations, we note several general 
properties.   First, ensemble averaged corrections to the ``odd'' 
and ``even'' spacings are of the same magnitude and opposite sign, 
Fig.~\ref{fig:ch3_su}(b).  Second, the mesoscopic fluctuations 
of both corrections are equal.  Indeed, the shift in position of 
an ``even-odd'' ($n \to n \!+\! 1$) peak maximum, 
Fig.~\ref{fig:ch3_su}(b), is determined by the interplay between 
the $0\!\to\!\frac{1}{2}$ and $1\!\to\!\frac{1}{2}$ spin sequences.  
Likewise, the shift of the ``odd-even'' ($n \!-\! 1\to n$) peak is 
determined by the $\frac{1}{2}\!\to\! 0$ and $\frac{1}{2}\!\to\! 1$ 
spin sequences.  Now if we sweep the back-gate voltage in the 
opposite direction and write the same peak as $n\to n \!-\! 1$, 
then the corresponding spin sequences are exactly the same as they 
were in the first case: $0\!\to\!\frac{1}{2}$ and 
$1\!\to\!\frac{1}{2}$.  From this symmetry argument one can 
conclude that (i) the ensemble averaged shifts of the ``even-odd'' 
and ``odd-even'' peaks are of the same magnitude and in the 
opposite directions~\cite{Kam00} and (ii) the mesoscopic 
fluctuations of both shifts are equal.

Thus, to simplify the calculations we study only the ``even'' 
spacing case.  This corresponds to the two spin sequences given 
in Eq.~(\ref{eq:ch3_seqeven}).  First, we calculate corrections 
to the spacing between peaks for both spin sequences.  A complete 
calculation for the doublet-triplet-doublet spin sequence is in 
the next section.  Second, we elaborate on how to put these 
spacings together in the final expression for an ``even'' spacing.  
Finally, we calculate GOE and GUE ensemble averaged corrections 
to the spacing and the rms fluctuations.

\section{Doublet-Triplet-Doublet Spin Sequence: 
Calculation of the Spacing Between Peaks}
\label{sec:ch3_case}

\begin{figure}
\begin{center}
\includegraphics[width=8.0cm]{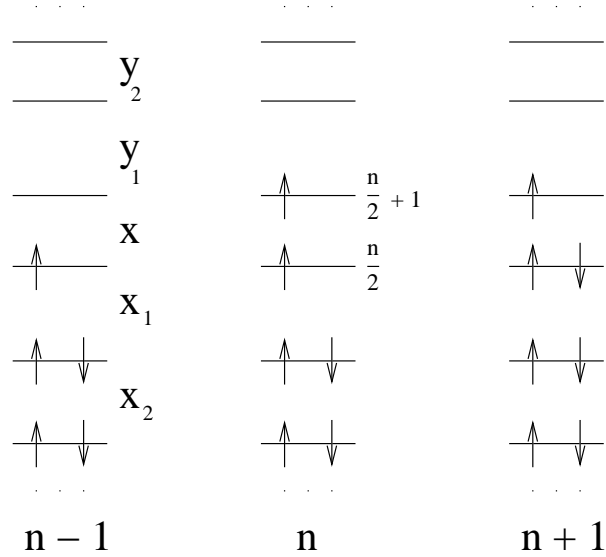}
\caption{Occupation of the QD levels in the ground 
state in three consecutive valleys with total electron number 
$n-1$, $n$, and $n+1$, respectively.  A doublet-triplet-doublet 
spin sequence is shown.  The variables $x$, $(x_{1}, x_{2}, \dots)$, 
and $(y_{1}, y_{2}, \dots)$ denote the energy level spacings in 
the QD normalized by mean level spacing $\Delta$.  For example, 
$x \!=\! \left(\varepsilon_{\frac{n}{2}+1}
-\varepsilon_{\frac{n}{2}}\right)/\Delta$.}
\label{fig:ch3_ll}
\end{center}
\end{figure}

Let us find the correction to the spacing between peaks for 
a doublet-triplet-doublet spin sequence.  The corresponding 
electron occupation of the quantum dot in three consecutive 
valleys with $n \!-\! 1$, $n$, and $n \!+\! 1$ electrons is 
shown in Fig.~\ref{fig:ch3_ll}.

For the isolated QD the position of the $n \!-\! 1\to n$ 
conductance peak maximum is determined by
\begin{eqnarray}
E^{(0)}_{n-1,S=\frac{1}{2}}\big({\mathcal N}^{(0)}_{n-\frac{1}{2}}\big)=
E^{(0)}_{n,S=1}\big({\mathcal N}^{(0)}_{n-\frac{1}{2}}\big)
\label{eq:ch3_energy0}
\end{eqnarray}
where $E_{n-1,S=\frac{1}{2}}^{(0)}$ and $E_{n,S=1}^{(0)}$ are 
the ground state energies of dot Hamiltonian [Eq.~(\ref{eq:ch3_dot})].
The corrections due to dot-lead tunneling are different for 
the doublet and triplet states. The resultant shift in peak 
position is given by~\cite{Kam00}
\begin{equation}
{\mathcal N}^{(2)}_{n-\frac{1}{2}}=\frac{1}{2E_C}
\left[E^{(2)}_{n,S=1}\big({\mathcal N}^{(0)}_{n-\frac{1}{2}}\big)
-E^{(2)}_{n-1,S=\frac{1}{2}}\big({\mathcal N}^{(0)}_{n-\frac{1}{2}}\big)\right] \;.
\label{eq:ch3_secondn2}
\end{equation}
Note that for the second-order correction to the position, 
the ground state energies are taken at the gate voltage 
obtained in the zeroth-order calculation, Eq.~(\ref{eq:ch3_energy0}).

Analogous equations hold for the $n \to n\!+\!1$ conductance peak. 
The spacing between these two conductance peaks is then defined as 
\begin{equation}
{\mathcal U}_{n,S=1}=
{\mathcal N}_{n+\frac{1}{2}}\big(1\to {\textstyle \frac{1}{2}}\big)
-{\mathcal N}_{n-\frac{1}{2}}\big({\textstyle \frac{1}{2}}\to 1\big).
\label{eq:ch3_totalspacing}
\end{equation}

\subsection{Zeroth Order: Isolated Quantum Dot}

For the doublet-triplet $n \!-\! 1\to n$ sequence 
Eq.~(\ref{eq:ch3_energy0}) gives
\begin{equation}
{\mathcal N}^{(0)}_{n-\frac{1}{2}}
\big({\textstyle \frac{1}{2}} \to 1\big) = n - \frac{1}{2}
+ \frac{1}{2E_C}
\left( \varepsilon_{\frac{n}{2}+1}-\frac{5}{4}J_{\rm s}
- \frac{T}{2} \ln\frac{3}{2}\right)
\label{eq:ch3_zerothn0}
\end{equation}
where the last temperature dependent term is the entropic 
correction to the condition of equal energies~\cite{Gla88}. 
For the position of the $n\to n\!+\!1$ peak maximum we obtain
\begin{equation}
{\mathcal N}^{(0)}_{n+\frac{1}{2}}\big(1 \to {\textstyle \frac{1}{2}}\big)=n+\frac{1}{2}
+\frac{1}{2E_C}\left(
\varepsilon_{\frac{n}{2}}+\frac{5}{4}J_{\rm s}
+\frac{T}{2}\ln\frac{3}{2}\right).
\end{equation}
Thus the spacing between peaks in zeroth order is
\begin{eqnarray}
{\mathcal U}^{(0)}_{n,S=1}(x)=1+\frac{5j-2x}{2\xi}
+\frac{T}{2E_C}\ln\frac{3}{2},
\label{eq:ch3_u0x}
\end{eqnarray}
where $j \!=\! J_{\rm s}/\Delta$ and 
$x \!=\! \left(\varepsilon_{\frac{n}{2}+1}
-\varepsilon_{\frac{n}{2}}\right)/\Delta$ 
(see Fig.~\ref{fig:ch3_ll}).  Similarly, for the 
doublet-singlet-doublet spin sequence the spacing is
\begin{eqnarray}
{\mathcal U}^{(0)}_{n,S=0}(x)=1+
\frac{2x-3j}{2\xi}-\frac{T}{2E_C}\ln 2.
\label{eq:ch3_u1x}
\end{eqnarray}
Note that in both cases ${\mathcal U}_{n}^{(0)}$ depends only 
on the spacing $x$.

\subsection{Second Order: Contribution From Virtual Processes}

\begin{figure}
\begin{center}
\includegraphics[width=13.4cm]{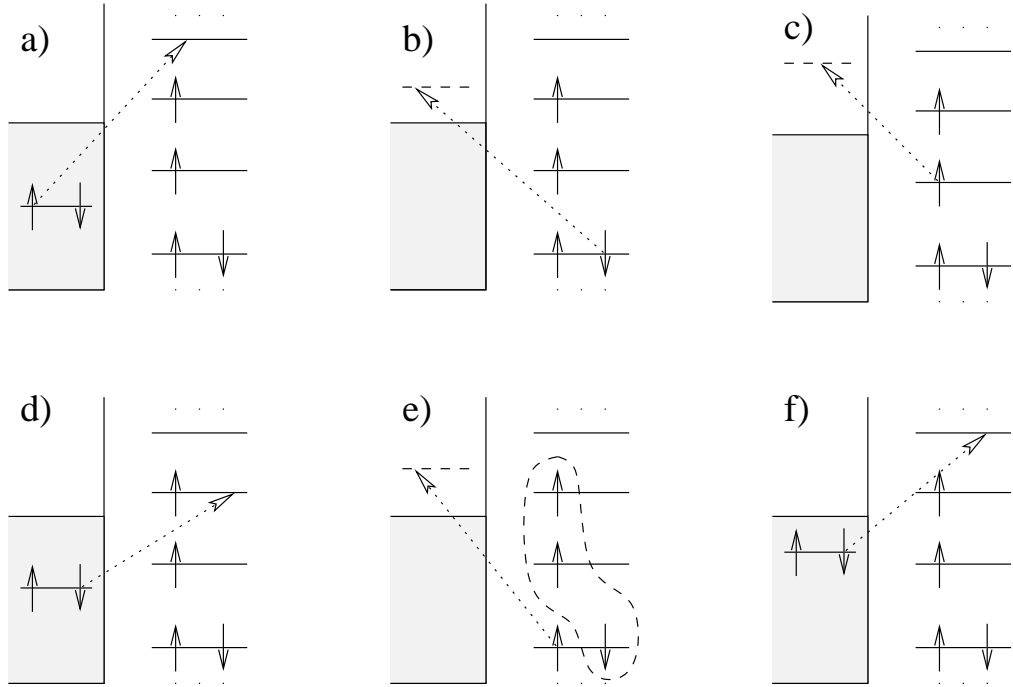}
\caption{Six distinct types of the virtual processes contribute 
to the ground state energy correction for the QD in the triplet 
state.  Only tunneling processes in (or out of) the left lead 
are shown.  In the first four cases spin of the dot in the 
virtual state ${\mathcal S}$ has a definite value.  In the last 
two cases (e) and (f) QD spin in the virtual state has two 
allowed values: ${\mathcal S}=\frac{1}{2}$ and ${\mathcal S}=\frac{3}{2}$ 
with the probabilities $w_{\mathcal S}$ given by Eq.~(\ref{eq:ch3_ws}).  
Electron structure of the virtual state corresponding to two 
allowed values of ${\mathcal S}$ is circled by dashed line in panel (e).}
\label{fig:ch3_s1}
\end{center}
\end{figure}

Let us consider in detail the second-order correction to the 
ground state energy of the triplet for subsequent use in 
(\ref{eq:ch3_secondn2}):
\begin{equation}
E^{(2)}_{n,S=1}\left({\mathcal N}\right)={\sum_m}'
\frac{\left|\left<\psi^{(0)}_{m}\left| H_{\rm tun}
\right|\psi^{(0)}_{n,S=1}\right>\right|^{2}}
{E^{(0)}_{n,S=1}\left({\mathcal N}\right)-E^{(0)}_{m}\left({\mathcal N}\right)},
\label{eq:ch3_e2}
\end{equation}
where the sum is over all possible virtual states.  $E^{(0)}$ 
and $\left|\psi^{(0)}\right>$ are the eigenvalues and eigenvectors 
of $H_{\rm dot}$, Eq.~(\ref{eq:ch3_dot}).

Different terms in Eq.~(\ref{eq:ch3_e2}) have different structure 
depending on the type of  virtual state involved; six possibilities 
are shown in Fig.~\ref{fig:ch3_s1}.  To take into account all 
virtual processes, we sum over all energy levels in the QD and 
integrate over states in each lead. To simplify the calculation 
even further, we assume (just for a moment) that $T \!=\! 0$ so 
that the Fermi distribution in the leads is a step function.  
Later we will see how $T$ reappears as a lower cutoff within 
a logarithm. 

Following the order of terms in Fig.~\ref{fig:ch3_s1}, the 
second-order correction to the triplet ground state energy is
\begin{eqnarray}
E^{(2)}_{n,S=1}({\mathcal N})=
&-&\sum^{\infty}_{k=\frac{n}{2}+2}\sum_{{\bf p}}
\frac{\theta (-\varepsilon_{\bf p})\left|t_{k{\bf p}}\right|^2}
{\left(\varepsilon_k-\varepsilon_{\bf p}\right)
+2E_C(n+\frac{1}{2}-{\mathcal N})-\frac{7}{4}J_{\rm s}}
\nonumber \\
&-&\sum_{k=1}^{\frac{n}{2}-1}\sum_{{\bf p}}
\frac{\theta (\varepsilon_{\bf p})\left|t_{k{\bf p}}\right|^2}
{\left(\varepsilon_{\bf p}-\varepsilon_k\right)
-2E_C(n-\frac{1}{2}-{\mathcal N})-\frac{7}{4}J_{\rm s}}
\nonumber \\
&-&\sum^{\frac{n}{2}+1}_{k=\frac{n}{2}}\sum_{{\bf p}}
\frac{\theta (\varepsilon_{\bf p})\left|t_{k{\bf p}}\right|^2}
{\left(\varepsilon_{\bf p}-\varepsilon_k\right)
-2E_C(n-\frac{1}{2}-{\mathcal N})+\frac{5}{4}J_{\rm s}}
\nonumber \\
&-&\sum_{k=\frac{n}{2}}^{\frac{n}{2}+1}\sum_{{\bf p}}
\frac{\theta (-\varepsilon_{\bf p})\left|t_{k{\bf p}}\right|^2}
{\left(\varepsilon_k-\varepsilon_{\bf p}\right)
+2E_C(n+\frac{1}{2}-{\mathcal N})+\frac{5}{4}J_{\rm s}}
\nonumber \\
&-&\sum_{{\mathcal S}=\frac{1}{2},\frac{3}{2}}w_{\mathcal S}
\left[
\sum^{\frac{n}{2}-1}_{k=1}\sum_{{\bf p}}
\frac{\theta (\varepsilon_{\bf p})\left|t_{k{\bf p}}\right|^2}
{\left(\varepsilon_{\bf p}-\varepsilon_k\right)
-2E_C(n-\frac{1}{2}-{\mathcal N})+f_{\mathcal S}J_{\rm s}}
\right.
\nonumber \\
&+&\left.\sum_{k=\frac{n}{2}+2}^{\infty}\sum_{{\bf p}}
\frac{\theta (-\varepsilon_{\bf p})\left|t_{k{\bf p}}\right|^2}
{\left(\varepsilon_k-\varepsilon_{\bf p}\right)
+2E_C(n+\frac{1}{2}-{\mathcal N})+f_{\mathcal S}J_{\rm s}}
\right]
\nonumber \\
&+&\left\{\mbox{similar terms for the right lead:}~{\bf p}\to{\bf q}\right\},
\label{eq:ch3_e2t}
\end{eqnarray}
where ${\mathcal S}$ is the spin of the QD in the virtual state.  
One can easily find ${\mathcal S}$ for the first four processes, 
Figs.~\ref{fig:ch3_s1}(a)-(d), and so calculate the denominators 
for the first four terms in (\ref{eq:ch3_e2t}): the values are 
$\frac{3}{2}$, $\frac{3}{2}$, $\frac{1}{2}$, and $\frac{1}{2}$, 
respectively.  In the last two cases, Figs.~\ref{fig:ch3_s1}(e)-(f), 
the QD spin in the virtual state can take two values, 
${\mathcal S}=\frac{1}{2}$ or $\frac{3}{2}$; it does so with the 
following probabilities
\begin{eqnarray}
w_{\frac{1}{2}} = {\textstyle \frac{2}{3}}~~~~\mbox{and}~~~~
w_{\frac{3}{2}} = {\textstyle \frac{1}{3}} \;.
\label{eq:ch3_ws}
\end{eqnarray}
The corresponding contributions to the energy correction must be 
weighted accordingly.  In addition, the energy difference in the 
denominators depends on ${\mathcal S}$; to account for this dependence, 
we introduce an additional function
\begin{eqnarray}
f_{\mathcal S} \equiv 2-{\mathcal S}\left({\mathcal S}+1\right),
\end{eqnarray}
appearing in the denominators of the fifth and sixth terms in 
(\ref{eq:ch3_e2t}).

Let us integrate over the continuous energy levels in the lead. 
The sum can be replaced by an integral,
$\sum_{\bf p}\,\cdots~\longrightarrow
\int d\varepsilon~\nu_{L}(\varepsilon )\,\cdots\; $,
where $\nu_{L}$ is the density of states in the left lead.  
Taking the dot-lead contacts to be point-like, the tunneling 
matrix elements $\{t_{k{\bf p}}\}$ depend on the momentum in 
the lead ${\bf p}$ only weakly; hence, 
$t_{k{\bf p}} \approx t_{kL}$.  In addition, as the leads are 
formed from 2DEG, their density of states is roughly independent 
of energy.  We assume that it is constant in the energy band of 
$2E_C$ near the Fermi surface.  Then the result of integrating 
over the energy spectrum in the lead (in schematic form) for 
the first term in Eq.~(\ref{eq:ch3_e2t}) is
\begin{eqnarray}
\sum_{\bf p}\frac{\theta(-\varepsilon_{\bf p})
\left|t_{k{\bf p}}\right|^{2}}
{\epsilon_k -\varepsilon_{\bf p}}
~\longrightarrow~
\nu_L\left|t_{kL}\right|^{2}\ln
\left|\frac{\varepsilon}{\epsilon_k}
\right|_{\varepsilon\to\infty}.
\end{eqnarray}
This expression diverges, but when we calculate an {\it observable}, 
e.g. the shift in the position of the peak maximum 
[Eq.~(\ref{eq:ch3_secondn2})], we encounter the energy difference 
between corrections to the triplet and doublet states.  The result 
for the shift is, therefore, finite:
\begin{eqnarray}
\left.\left(
\ln\left|\frac{\varepsilon}{\epsilon_k}\right|
-\ln\left|\frac{\varepsilon}{\epsilon'_k}\right|
\right)\right|_{\varepsilon\to\infty}
=\ln\left|\frac{\epsilon'_k}{\epsilon_k}\right|.
\label{eq:ch3_finite}
\end{eqnarray}

In a similar fashion one can calculate the second-order correction 
to the ground state energy of the doublet.  The difference of these 
energies at the gate voltage corresponding to the peak maximum 
in zeroth order [Eq.~(\ref{eq:ch3_zerothn0})], needed in  
Eq.~(\ref{eq:ch3_secondn2}), then follows.  There is one resonant 
term, proportional to $\ln\left({2E_C}/{T}\right)$,  in which the 
lower cutoff $T$ appears because of the entropic term in 
Eq.~(\ref{eq:ch3_zerothn0}). 
Alternatively, $T$ would appear as the natural cutoff for the 
resonant term upon reintroduction of the Fermi-Dirac distribution 
for the occupation numbers in the leads.

For a point-like dot-lead contact, the tunneling matrix element 
is proportional to the value of the electron wave function 
in the QD at the point of contact: 
$t_{k\alpha}\propto\psi_k({\bf r}_{\alpha})$, where $\alpha =L$ 
or $R$. Here, we neglect the fluctuations of the electron wave 
function in the large lead.  Thus, the following identity is valid
\begin{eqnarray}
\nu_{\alpha}\left|t_{k\alpha}\right|^2
=\frac{\Delta}{4\pi^{2}}~g_{\alpha}
\frac{\left|\psi_{k}\left({\bf r}_{\alpha}\right)\right|^{2}}
{\left<\left|\psi_{k}\left({\bf r}_{\alpha}\right)\right|^{2}\right>},
\label{eq:ch3_ttopsi}
\end{eqnarray}
where the average in the denominator is taken over the statistical 
ensemble.  Note that by taking the ensemble average of both sides 
of (\ref{eq:ch3_ttopsi}), one arrives at the standard golden rule 
expression for the dimensionless conductance.

In our calculations we take advantage of the fact that
$J_{\rm s} \!<\! \Delta \!\ll\! E_C$ and neglect terms that are 
of order $1/\xi \!=\! \Delta/2E_C$.
Sums like
\begin{eqnarray}
-\frac{1}{2}\sum_{k=\frac{n}{2}+2}^{\infty}
\ln\left(1+\frac{2J_{\rm s}}
{\varepsilon_{k}-\varepsilon_{\frac{n}{2}+1}}\right)
\end{eqnarray}
are split using
$\sum_{k=\frac{n}{2}+2}^{\infty}\cdots
\!=\! \sum_{k=\frac{n}{2}+2}^{\frac{n}{2}+\xi +1}\cdots
\!+\! \sum_{k=\frac{n}{2}+\xi +2}^{\infty}\cdots$,
and so the last term, which is $O(1/\xi )$, is dropped. 
Likewise, expressions like
\begin{eqnarray}
\frac{2}{3}
\sum_{k=\frac{n}{2}+2}^{\infty}
\ln\left(1+\frac{3}{2}~
\frac{J_{\rm s}}{\varepsilon_{k}-\varepsilon_{\frac{n}{2}+1}+2E_C}
\right)
\end{eqnarray}
are of order $O(1/\xi )$ and so neglected.

Thus, for the second order correction to the position of the 
peak maximum, we obtain
\begin{eqnarray}
&& {\mathcal N}^{(2)}_{n-\frac{1}{2}}\left(\frac{1}{2}\to 1\right)
=\frac{1}{4\pi^{2}\xi}\sum_{\alpha =L,R}
g_{\alpha}
\nonumber \\
&\times&
\left[
-
2\sum_{k\ne\frac{n}{2},\frac{n}{2}+1}
\mbox{sign}\left(\frac{n}{2}-k\right)
\frac{\left|\psi_{k}\left({\bf r}_{\alpha}\right)\right|^{2}}
{\left<\left|\psi_{k}\left({\bf r}_{\alpha}\right)\right|^{2}\right>}
\ln\left(\frac{2E_C}{\left|\varepsilon_{\frac{n}{2}+1}-\varepsilon_{k}\right|}+1\right)
\right.
\nonumber \\
&-&
\left.
\frac{\left|\psi_{\frac{n}{2}}\left({\bf r}_{\alpha}\right)\right|^{2}}
{\left<\left|\psi_{\frac{n}{2}}\left({\bf r}_{\alpha}\right)\right|^{2}\right>}
\ln\left(\frac{2J_{\rm s}}{\varepsilon_{\frac{n}{2}+1}-\varepsilon_{\frac{n}{2}}}-1\right)
\right.
+
\left.
\frac{1}{2}
\frac{\left|\psi_{\frac{n}{2}+1}\left({\bf r}_{\alpha}\right)\right|^{2}}
{\left<\left|\psi_{\frac{n}{2}+1}\left({\bf r}_{\alpha}\right)\right|^{2}\right>}
\left(\ln\frac{E_C}{J_{\rm s}}+\ln\frac{2E_C}{T}\right)
\right.
\nonumber \\
&+&
\left.
\frac{4}{3}
\sum_{k=\frac{n}{2}-\xi}^{\frac{n}{2}-1}
\frac{\left|\psi_{k}\left({\bf r}_{\alpha}\right)\right|^{2}}
{\left<\left|\psi_{k}\left({\bf r}_{\alpha}\right)\right|^{2}\right>}
\ln\left|1-\frac{3J_{\rm s}}{\varepsilon_{\frac{n}{2}+1}-\varepsilon_{k}}\right|
\right.
\nonumber \\
&-&
\left.
\frac{1}{2}
\sum_{k=\frac{n}{2}+2}^{\frac{n}{2}+\xi}
\frac{\left|\psi_{k}\left({\bf r}_{\alpha}\right)\right|^{2}}
{\left<\left|\psi_{k}\left({\bf r}_{\alpha}\right)\right|^{2}\right>}
\ln\left(1+\frac{2J_{\rm s}}{\varepsilon_{k}-\varepsilon_{\frac{n}{2}+1}}\right)
+O\left(\frac{1}{\xi}\right)
\right],
\end{eqnarray}
where $2J_{\rm s}>\varepsilon_{\frac{n}{2}+1} \!-\! \varepsilon_{\frac{n}{2}}>0$ because the total spin of the QD with $n$ electrons 
is equal to 1.

In a similar fashion one can find the shift in the position of 
the other peak maximum, 
${\mathcal N}^{(2)}_{\frac{n}{2}+1}\left(1\to\frac{1}{2}\right)$.  
Then, according to (\ref{eq:ch3_totalspacing}), the difference of 
these two shifts yields the second-order correction to the spacing 
for the doublet-triplet-doublet spin sequence:
\begin{eqnarray}
{\mathcal U}_{n,S=1}^{(2)}
&=&\frac{1}{4\pi^{2}\xi}\sum_{\alpha =L,R}
g_{\alpha}
\left\{
2\sum_{k\ne\frac{n}{2},\frac{n}{2}+1}
\mbox{sign}\left(\frac{n}{2}-k\right)
\frac{\left|\psi_{k}\left({\bf r}_{\alpha}\right)\right|^{2}}
{\left<\left|\psi_{k}\left({\bf r}_{\alpha}\right)\right|^{2}\right>}
\right.
\nonumber \\
&\times&
\left[
\ln\left(\frac{2E_C}{\left|\varepsilon_{\frac{n}{2}+1}
-\varepsilon_{k}\right|}+1\right)
-\ln\left(\frac{2E_C}{\left|\varepsilon_{\frac{n}{2}}
-\varepsilon_{k}\right|}+1\right)
\right]
\nonumber \\
&+&
\left(
\frac{\left|\psi_{\frac{n}{2}}\left({\bf r}_{\alpha}\right)\right|^{2}}
{\left<\left|\psi_{\frac{n}{2}}\left({\bf r}_{\alpha}\right)\right|^{2}\right>}
+\frac{\left|\psi_{\frac{n}{2}+1}\left({\bf r}_{\alpha}\right)\right|^{2}}
{\left<\left|\psi_{\frac{n}{2}+1}\left({\bf r}_{\alpha}\right)\right|^{2}\right>}
\right)
\nonumber \\
&\times&
\left[
\ln\left(\frac{2J_{\rm s}}{\varepsilon_{\frac{n}{2}+1}
-\varepsilon_{\frac{n}{2}}}-1\right)
-\frac{1}{2}\ln\frac{E_C}{J_{\rm s}}
-\frac{1}{2}\ln\frac{2E_C}{T}
\right]
\nonumber \\
&+&
\left.
\sum_{k=\frac{n}{2}-\xi}^{\frac{n}{2}-1}
\frac{\left|\psi_{k}\left({\bf r}_{\alpha}\right)\right|^{2}}
{\left<\left|\psi_{k}\left({\bf r}_{\alpha}\right)\right|^{2}\right>}
\left[
\frac{1}{2}
\ln\left(1+\frac{2J_{\rm s}}{\varepsilon_{\frac{n}{2}}-\varepsilon_{k}}\right)
-\frac{4}{3}
\ln\left|1-\frac{3J_{\rm s}}{\varepsilon_{\frac{n}{2}+1}-\varepsilon_{k}}\right|
\right]
\right.
\nonumber \\
&+&
\sum_{k=\frac{n}{2}+2}^{\frac{n}{2}+\xi}
\frac{\left|\psi_{k}\left({\bf r}_{\alpha}\right)\right|^{2}}
{\left<\left|\psi_{k}\left({\bf r}_{\alpha}\right)\right|^{2}\right>}
\left[
\frac{1}{2}
\ln\left(1+\frac{2J_{\rm s}}{\varepsilon_{k}-\varepsilon_{\frac{n}{2}+1}}\right)
-\frac{4}{3}
\ln\left|1-\frac{3J_{\rm s}}{\varepsilon_{k}-\varepsilon_{\frac{n}{2}}}\right|
\right]
\nonumber \\
&+&
\left.
O\left(\frac{1}{\xi}\right)
\right\}.
\label{eq:ch3_u2triplet}
\end{eqnarray}
A potential complication is that the addition of two electrons 
to the quantum dot ($n \!-\! 1\to n\to n \!+\! 1$) may scramble 
the energy levels and wave functions of the 
QD~\cite{Bla97,Val98,Usa02}. Since the number of added electrons 
is small, we assume that the same realization of the Hamiltonian, 
Eq.~(\ref{eq:ch3_hamilt}), is valid in all three valleys~\cite{Pat98,Ste97}.

For the second-order correction to the spacing for the 
doublet-singlet-doublet spin sequence we similarly obtain
\begin{eqnarray}
{\mathcal U}_{n,S=0}^{(2)}
&=&\frac{1}{4\pi^{2}\xi}\sum_{\alpha =L,R}
g_{\alpha}
\left\{
-2\sum_{k\ne\frac{n}{2},\frac{n}{2}+1}
\mbox{sign}\left(\frac{n}{2}-k\right)
\frac{\left|\psi_{k}\left({\bf r}_{\alpha}\right)\right|^{2}}
{\left<\left|\psi_{k}\left({\bf r}_{\alpha}\right)\right|^{2}\right>}
\right.
\nonumber \\
&\times&
\left[
\ln\left(\frac{2E_C}{\left|\varepsilon_{\frac{n}{2}+1}-\varepsilon_{k}\right|}+1\right)
-\ln\left(\frac{2E_C}{\left|\varepsilon_{\frac{n}{2}}-\varepsilon_{k}\right|}+1\right)
\right]
\nonumber \\
&+&
\left(
\frac{\left|\psi_{\frac{n}{2}}\left({\bf r}_{\alpha}\right)\right|^{2}}
{\left<\left|\psi_{\frac{n}{2}}\left({\bf r}_{\alpha}\right)\right|^{2}\right>}
+\frac{\left|\psi_{\frac{n}{2}+1}\left({\bf r}_{\alpha}\right)\right|^{2}}
{\left<\left|\psi_{\frac{n}{2}+1}\left({\bf r}_{\alpha}\right)\right|^{2}\right>}
\right)
\nonumber \\
&\times&
\left[
\frac{3}{2}
\ln\left(1-\frac{2J_{\rm s}}{\varepsilon_{\frac{n}{2}+1}-\varepsilon_{\frac{n}{2}}}\right)
-2\ln\left(
\frac{2E_C}{\varepsilon_{\frac{n}{2}+1}-\varepsilon_{\frac{n}{2}}}+1
\right)
+\ln\frac{2E_C}{T}
\right]
\nonumber \\
&+&
\frac{3}{2}
\sum_{k=\frac{n}{2}-\xi}^{\frac{n}{2}-1}
\frac{\left|\psi_{k}\left({\bf r}_{\alpha}\right)\right|^{2}}
{\left<\left|\psi_{k}\left({\bf r}_{\alpha}\right)\right|^{2}\right>}
\ln\left(1-\frac{2J_{\rm s}}{\varepsilon_{\frac{n}{2}+1}-\varepsilon_{k}}\right)
\nonumber \\
&+&
\left.
\frac{3}{2}
\sum_{k=\frac{n}{2}+2}^{\frac{n}{2}+\xi}
\frac{\left|\psi_{k}\left({\bf r}_{\alpha}\right)\right|^{2}}
{\left<\left|\psi_{k}\left({\bf r}_{\alpha}\right)\right|^{2}\right>}
\ln\left(1-\frac{2J_{\rm s}}{\varepsilon_{k}-\varepsilon_{\frac{n}{2}}}\right)
+O\left(\frac{1}{\xi}\right)
\right\},
\label{eq:ch3_u2singlet}
\end{eqnarray}
where $\varepsilon_{\frac{n}{2}+1} \!-\! \varepsilon_{\frac{n}{2}} > 2J_{\rm s} \ge 0$ because the total spin of the QD with $n$ 
electrons is equal to $0$ in this case.

Unlike the zeroth-order spacings, the second-order corrections 
are functions of many energy level spacings as well as the wave 
functions at the dot-lead contact points:
${\mathcal U}_{n,S}^{(2)}=
{\mathcal U}_{n,S}^{(2)}(x,{\bf X},{\bf Y};
\{Z_{k\alpha}\})$,
where $x$, ${\bf X} = (x_{1}, x_{2}, \dots )$, and 
${\bf Y} = (y_{1}, y_{2}, \dots )$ are the energy level spacings 
in the QD normalized by the mean level spacing $\Delta$ 
(see Fig.~\ref{fig:ch3_ll}) and 
\begin{eqnarray}
Z_{k\alpha} \equiv
\frac{\left|\psi_{k}\left({\bf r}_{\alpha}\right)\right|^{2}}
{\left<\left|\psi_{k}\left({\bf r}_{\alpha}\right)\right|^{2}\right>} \;.
\end{eqnarray}

The expressions for ${\mathcal U}^{(2)}$ suggest that the main 
contribution to their fluctuation comes from the fluctuation 
of the energy level $x$ and the wave functions 
$\left\{\psi_{k}\left({\bf r}_{\alpha} \right) \right\}$.  
The other spacings, ${\bf X}$ and ${\bf Y}$, always appear within 
a logarithm; therefore, their contribution to the fluctuation 
of ${\mathcal U}^{(2)}$ is small.  With good accuracy, one can replace 
these levels by their mean value 
\begin{equation}
{\mathcal U}_{n,S}^{(2)} \approx
{\mathcal U}_{n,S}^{(2)}(x,{\bf 1},{\bf 1};
\{Z_{k\alpha}\}) \equiv {\mathcal U}_{n,S}^{(2)}(x;\{Z_{k\alpha}\}) \;.
\end{equation}
Converting to dimensionless units, we find that
\begin{equation}
{\mathcal U}^{(2)}_{n,S}\left(x;\left\{Z_{k\alpha}\right\}\right)
=\frac{1}{4\pi^{2}\xi}
\sum_{\alpha =L,R}g_{\alpha}\Phi_{\alpha ,S}
\left(x;\left\{Z_{k\alpha}\right\}\right),
\label{eq:ch3_u2gen}
\end{equation}
where
\begin{eqnarray}
\Phi_{\alpha ,S=0}
\left(x;\left\{Z_{k\alpha}\right\}\right)
&=&
\left(Z_{\frac{n}{2},\alpha}+Z_{\frac{n}{2}+1,\alpha}\right)
\left[
-\ln \xi +\ln\delta +\frac{1}{2}\ln x+\frac{3}{2}\ln (x-2j)
\right]
\nonumber \\
&+&
2\sum_{l=1}^{\infty}
\left(Z_{\frac{n}{2}-l,\alpha}+Z_{\frac{n}{2}+1+l,\alpha}\right)
\left[
\ln\left(1+\frac{x}{l}\right)-\ln\left(1+\frac{x}{\xi +l}\right)
\right]
\nonumber \\
&+&
\frac{3}{2}\sum_{l=1}^{\xi}
\left(Z_{\frac{n}{2}-l,\alpha}+Z_{\frac{n}{2}+1+l,\alpha}\right)
\ln\left(1-\frac{2j}{x+l}\right)
+O\left(1\right),
\label{eq:ch3_ups0}
\end{eqnarray}
\begin{eqnarray}
&&\Phi_{\alpha ,S=1}
\left(x;\left\{Z_{k\alpha}\right\}\right)
=\left(Z_{\frac{n}{2},\alpha}+Z_{\frac{n}{2}+1,\alpha}\right)
\left[
-\ln \xi -\frac{1}{2}\ln\delta +\frac{1}{2}\ln 2j+\ln\left(\frac{2j}{x}-1\right)
\right]
\nonumber \\
&-&
2\sum_{l=1}^{\infty}
\left(Z_{\frac{n}{2}-l,\alpha}+Z_{\frac{n}{2}+1+l,\alpha}\right)
\left[
\ln\left(1+\frac{x}{l}\right)-\ln\left(1+\frac{x}{\xi +l}\right)
\right]
\nonumber \\
&+&
\sum_{l=1}^{\xi}
\left(Z_{\frac{n}{2}-l,\alpha}+Z_{\frac{n}{2}+1+l,\alpha}\right)
\left[
\frac{1}{2}\ln\left(1+\frac{2j}{l}\right)
-\frac{4}{3}\ln\left|1-\frac{3j}{x+l}\right|
\right]
+O\left(1\right),
\label{eq:ch3_ups1}
\end{eqnarray}
where $\delta =\Delta /T$.  Here, the upper limit in two of 
the sums is infinity because the Fermi energy is the largest 
energy scale.

In summary, the total spacing is
\begin{eqnarray}
{\mathcal U}_{n,S}={\mathcal U}_{n,S}^{(0)}(x)+
{\mathcal U}_{n,S}^{(2)}(x;\{Z_{k\alpha}\}),
\end{eqnarray}
where the first term is given by 
Eqs.~(\ref{eq:ch3_u0x}) and (\ref{eq:ch3_u1x}) and the second by 
(\ref{eq:ch3_u2gen})-(\ref{eq:ch3_ups1}).  The spin of the QD 
in the even valley, $S$, can take two values, $0$ or $1$, 
depending on the spacing $x$.

\section{Ensemble Averaged Correction to the Peak Spacing}
\label{sec:ch3_averaged}

The average and rms correction to the peak spacing can now 
be found by using the known distribution of the single-particle 
quantities $x$ and $\{Z_{k\alpha}\}$.
In what follows, $\left<{\mathcal U}\right>$ denotes the average 
over the wave functions, $\overline{\mathcal U}$ denotes the full 
average over both wave functions and energy levels, and $P(x)$ 
is the distribution of the spacing $x$.
Since $\left<Z_{k\alpha}\right> \!=\! 1$,
$\left<\Phi_{\alpha,S}\right>$ does not depend on $\alpha$, and 
the average ``even'' spacing is\footnote{One may think that an 
additional term due to shift in the border $x = 2j$ between singlet 
and triplet states should be present in this equation as well.  
However, one can show that the contribution of this term to 
$\overline{{\mathcal U}_{n}^{(2)}}$ (as well as to its rms) is small.}
\begin{eqnarray}
\overline{{\mathcal U}_n^{(2)}} &=& \int_{0}^{\infty}dx\, P(x)\left<{\mathcal U}_{n,S}^{(2)}\right>
\label{eq:ch3_aveugen}
\\
& = &
\frac{g_L+g_R}{4\pi^2\xi}
\left(
\int_{2j}^{\infty}dx~P(x)\left<\Phi_{\alpha,S=0}^{(2)}\right>
+\int_{0}^{2j}dx~P(x)\left<\Phi_{\alpha,S=1}^{(2)}\right>
\right) .
\end{eqnarray}
Using the asymptotic formulas
\begin{eqnarray}
\sum\limits_{l=1}^{\infty} & &  \!\!\Big[
\ln\Big(1+\frac{x}{l}\Big)
-\ln\Big(1+\frac{x}{\xi +l}\Big)
\Big]\approx x\,\ln \xi ,
\nonumber \\
\sum\limits_{l=1}^{\xi} & &
\ln\Big(1-\frac{2j}{x+l}\Big)\approx - 2 j\,\ln \xi 
\end{eqnarray}
for $\xi \!\gg\! 1$ in the expressions for 
$\left<\Phi_{\alpha,S}\right>$, we find
\begin{eqnarray}
\left<\Phi_{\alpha ,S=0}\right> 
& = & 2\, (2x-3j-1)\,\ln \xi +2\,\ln\delta \;,
\nonumber
\\
\left<\Phi_{\alpha ,S=1}\right> 
& = & 2\, (-2x+5j-1)\,\ln \xi -\ln\delta \;,
\label{eq:ch3_aveups}
\end{eqnarray}
valid for $\xi ,\delta\gg 1$.  By carrying out the integration 
over the distribution of the spacing $x$, the final expression is
\begin{eqnarray}
\overline{{\mathcal U}^{(2)}_{n}\left(j\right)}
&=& \frac{g_{L}+g_{R}}{4\pi^{2}\xi}
\big[{\mathcal C}(j)\ln \xi +{\mathcal D}(j)\ln\delta +O(1)\big],
\nonumber \\
{\mathcal C}(j)&=&2\big[-8jP_{0}(2j)+4x_{0}(2j)+5j-3\big],
\label{eq:ch3_even2} \\
{\mathcal D}(j)&=&3P_{0}(2j)-1,
\nonumber
\end{eqnarray}
where $P_{0}(2j) = \int_{2j}^{\infty}dx\, P(x)$ and 
$x_{0}(2j) = \int_{2j}^{\infty}dx\, x\, P(x)$.  Note that 
$P_{0}(2j)$ is the probability of obtaining a singlet ground 
state while $x_0(2j)/P_0(2j)$ is the average value of $x$ given 
that the ground state is a singlet.

For the CI model, $j \!=\!0$ and, hence, 
${\mathcal C}(0)\!=\!{\mathcal D}(0)\!=\!2$.  In this limit, then, 
the ensemble averaged correction to the spacing is
\begin{eqnarray}
\overline{{\mathcal U}^{(2)}_{n}\left(0\right)}
=\frac{g_L+g_R}{2\pi^{2}\xi}\ln\frac{2E_C}{T} \;,
\label{eq:ch3_un20}
\end{eqnarray}
in agreement with previous work~\cite{Kam00}. 
The magnitude here is approximately
$0.05$ $\left(g_L+g_R\right)$ $\ln\left(2E_C/T\right)$ 
in units of the mean level spacing.

It is convenient to relate the average change in spacing at 
non-zero $J_{\rm s}$ to that at $J_{\rm s} \!=\! 0$:
\begin{eqnarray}
\delta u(j)
& \equiv & \frac{\;\overline{{\mathcal U}^{(2)}_{n}\left(j\right)}\;}
{\overline{{\mathcal U}^{(2)}_{n}\left(0\right)}}
= \frac{\lambda\,{\mathcal C}(j) + {\mathcal D}(j)}{2(\lambda+1)},
\label{eq:ch3_ave2plot}
\\
\lambda & \equiv & \frac{\ln \xi}{\ln\delta}
=\frac{\ln(2E_C / \Delta)}{\ln (\Delta /T)} \;.
\label{eq:ch3_lambda}
\end{eqnarray}
The dependence of $\delta u$ on $j$ is fully determined by the 
parameter $\lambda$ and the choice of random matrix ensemble.

\begin{figure}
\begin{center}
\includegraphics[width=10.4cm]{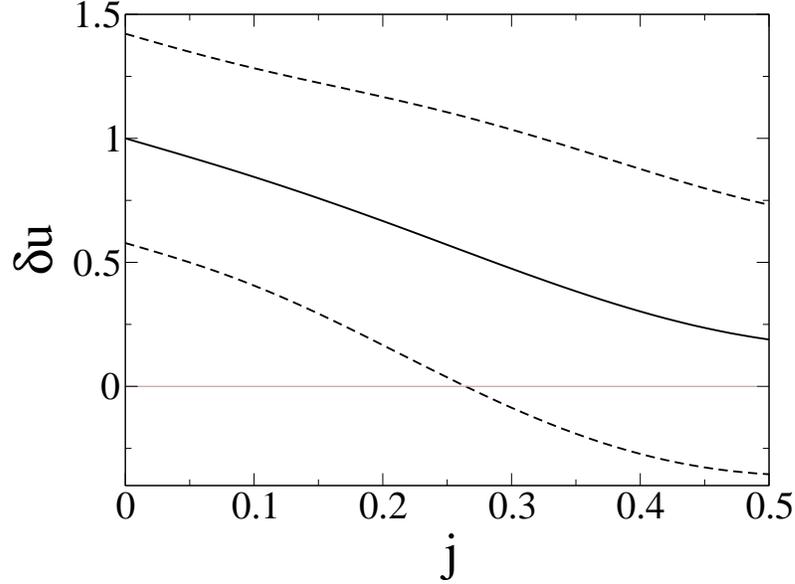}
\caption{Correction to ``even'' peak spacing as a function 
of strength of exchange, $j \!=\! J_{\rm s}/\Delta$, normalized 
by the correction at $j \!=\! 0$. GUE case with $\lambda \!=\! 1$ 
and $g_{L} \!=\! g_{R}$. Solid: Ensemble average. Dashed: Ensemble 
average plus/minus the rms, showing the width of the distribution.}
\label{fig:ch3_ave1}
\end{center}
\end{figure}

\begin{figure}
\begin{center}
\includegraphics[width=10.4cm]{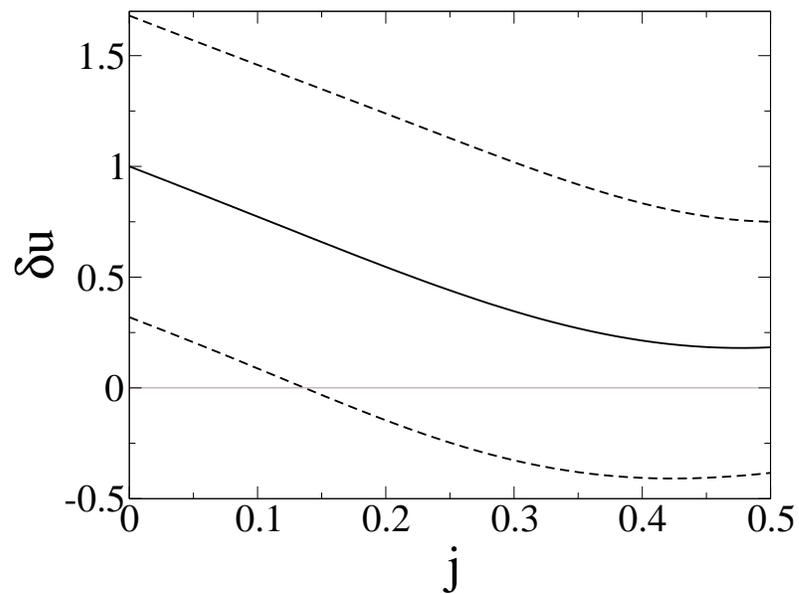}
\caption{The same quantities as in Fig.~\ref{fig:ch3_ave1} 
plotted for $\lambda = 3$.}
\label{fig:ch3_ave3}
\end{center}
\end{figure}

\begin{figure}
\begin{center}
\includegraphics[width=10.4cm]{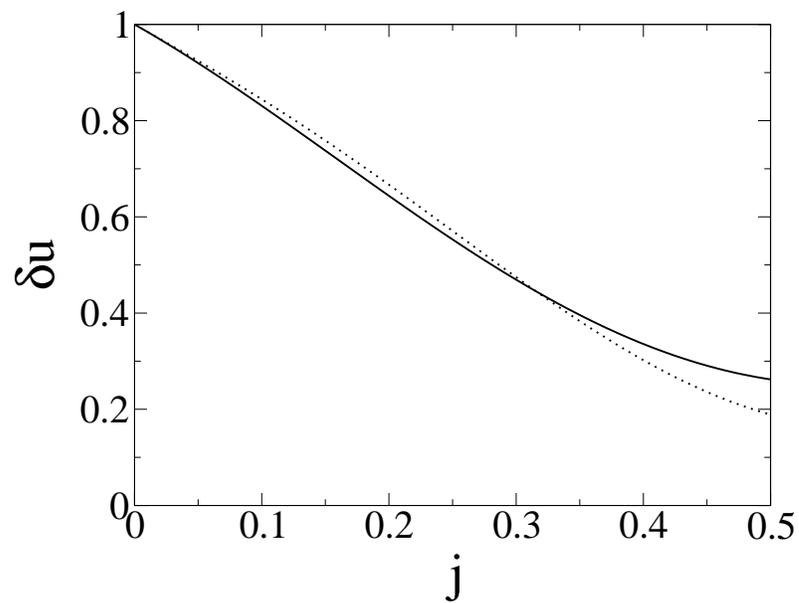}
\caption{Ensemble averaged correction to the ``even'' peak 
spacing as a function of strength of exchange, 
$j \!=\! J_{\rm s}/\Delta$, normalized by the correction at 
$j \!=\! 0$ for $\lambda \!=\! 1$. Solid: GOE. Dotted: GUE.}
\label{fig:ch3_goe}
\end{center}
\end{figure}

Figs.~\ref{fig:ch3_ave1} and \ref{fig:ch3_ave3} show the results 
in the GUE ensemble for $\lambda \!=\! 1$ and $3$, respectively, 
and Fig.~\ref{fig:ch3_goe} shows those for the GOE ensemble at 
$\lambda \!=\! 1$.  In evaluating these expressions, we use the 
Wigner surmise distributions for $P(x)$, which allow an analytic 
evaluation of $P_{0}(2j)$ and $x_{0}(2j)$.  As $j$ increases, 
the average correction to the peak spacing decreases monotonically 
in all three cases.  (Note, however, that our results are not 
completely trustworthy when $0.4 \!<\! j \!<\! 0.5$ because in 
this regime higher spin states should be taken into account.)  
Since $\lambda$ depends on $\xi$ and $\delta$ only logarithmically, 
the qualitative behavior of $\delta u (j)$ is very robust with 
respect to changes in charging energy, mean level spacing, or 
temperature.

\begin{figure}
\begin{center}
\includegraphics[width=10.4cm]{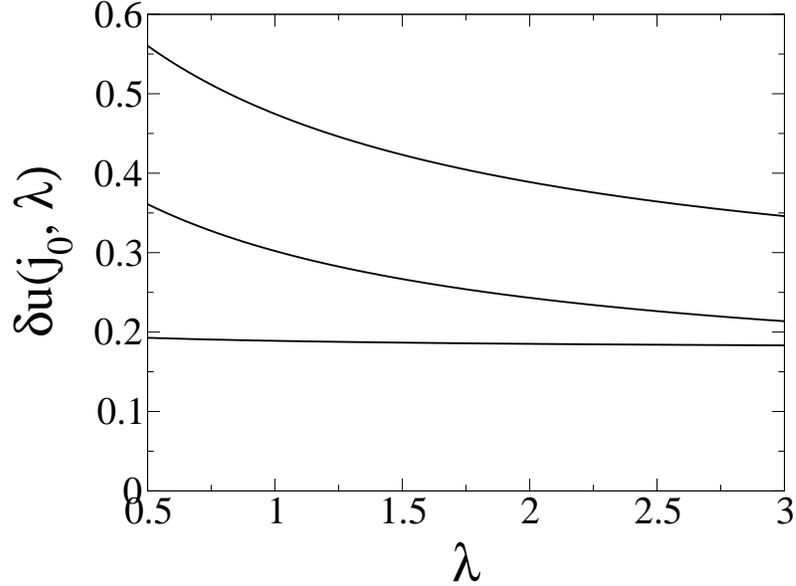}
\caption{GUE ensemble averaged correction to the even peak 
spacing as a function of 
$\lambda \!=\! \ln(2E_C / \Delta) / \ln (\Delta /T)$ at 
$j \!=\! j_{0}$.  The curves correspond to $j_{0} = 0.3$, $0.4$, 
and $0.5$ from top to the bottom.}
\label{fig:ch3_fj}
\end{center}
\end{figure}

Similarly the dependence of $\delta u$ on $\lambda$ at 
$j \!=\! j_{0}$ is
\begin{eqnarray}
\delta u(j_{0},\lambda )
= \frac{\lambda\, {\mathcal C}(j_{0})+{\mathcal D}(j_{0})}{2(\lambda+1)} \;.
\label{eq:ch3_avej0}
\end{eqnarray}
Fig.~\ref{fig:ch3_fj} shows results in the GUE case for several 
values of $j_{0}$.  \textit{Thus for the realistic value 
$j_{0} \!=\! 0.3$, the CEI model gives an average correction 
to the peak spacing which is two to three times smaller than 
the CI model.}

\section{RMS of The Correction to Peak Spacing 
due to Mesoscopic Fluctuations}
\label{sec:ch3_fluctuations}

Mesoscopic fluctuations of the correction to the peak spacing 
are characterized by the variance of ${\mathcal U}^{(2)}$.  It is 
convenient to separate the average over the wave functions 
from that over the spacing $x$, writing
\begin{eqnarray}
\mbox{var}\big({\mathcal U}_{n}^{(2)}\big)
= \sigma^{2}_{Z}\big({\mathcal U}_{n}^{(2)}\big)
+ \sigma^{2}_{x}\big({\mathcal U}_{n}^{(2)}\big),
\label{eq:ch3_sigmamain}
\end{eqnarray}
where
\begin{equation}
\sigma^{2}_{Z}
= \int_{0}^{\infty}dx\, P(x)
\Big<\Big(
{\mathcal U}_{n,S}^{(2)}-\big<{\mathcal U}_{n,S}^{(2)}\big>
\Big)^{2}\Big>
\label{eq:ch3_sigmapsi}
\end{equation}
is the contribution due to wave function fluctuations and 
\begin{equation}
\sigma^{2}_{x}
= \int_{0}^{\infty}\!\!dx\, P(x)
\big<{\mathcal U}_{n,S}^{(2)}\big>^{2}
-\left[
\int_{0}^{\infty}\!\!dx\, P(x)
\big<{\mathcal U}_{n,S}^{(2)}\big>
\right]^{2}
\label{eq:ch3_sigmax}
\end{equation}
is the contribution due to fluctuation of the spacing $x$.

We start by considering the fluctuations of the wave functions.  
As the number of electrons in the dot is large, the distance 
between the left and right dot-lead contacts is large, 
$\left|{\bf r}_{L}-{\bf r}_{R}\right|\gg\lambda_F$ where 
$\lambda_{F}$ is the Fermi wavelength.  Therefore, the wave 
functions at ${\bf r}_{L}$ and ${\bf r}_{R}$ are 
uncorrelated~\cite{Pri95},
\begin{eqnarray}
\left<\left( Z_{kL} - 1 \right)
\left( Z_{k'R} - 1 \right)\right> = 0
\end{eqnarray}
for all $k$ and $k'$.  The fluctuation of ${\mathcal U}^{(2)}$ 
can then be written entirely in terms of the properties of 
a single lead:
\begin{equation}
\Big<\Big(
{\mathcal U}_{n,S}^{(2)}-\big<{\mathcal U}_{n,S}^{(2)}\big>
\Big)^{2}\Big>
=\frac{g_L^2+g_R^2}{\left(4\pi^{2}\xi\right)^2}
\Big<\big(
\Phi_{L,S}-\left<\Phi_{L,S}\right>
\big)^{2}\Big> \;.
\label{eq:ch3_sigups}
\end{equation}
The cross terms here disappear because, according to RMT, 
wave functions of different energy levels are uncorrelated 
even at the same point in space,
\begin{eqnarray}
\big<
\left( Z_{kL} - 1 \right)
\left( Z_{k'L} - 1 \right)
\big>
=\frac{2}{\beta}\,\delta_{kk'}
\label{eq:ch3_zz}
\end{eqnarray}
where $\beta \!=\! 1$ ($\beta \!=\! 2$) for the GOE (GUE) case. 
In fact, only the $k \!=\! \frac{n}{2}$ and 
$k \!=\! \frac{n}{2} \!+\! 1$ terms contribute, as one can see 
by using
\begin{eqnarray}
\sum_{l=1}^{\xi}\,\ln^{2}
\Big( 1 - \frac{2j}{x+l}\Big) = O(1)
\end{eqnarray}
valid for $\xi\gg 1$. Integrating (\ref{eq:ch3_sigups}) over 
the distribution of $x$ according to Eq.~(\ref{eq:ch3_sigmapsi}) 
(keeping in mind $\xi ,\delta \!\gg\! 1$), we obtain
\begin{eqnarray}
\label{eq:ch3_sigmazc}
\sigma_Z^2\Big({\mathcal U}_n^{(2)}\Big)
=\frac{g_L^2+g_R^2}{\beta\left(4\pi^{2}\xi\right)^2}
\big\{
4\ln^{2}\!\xi
+\left[3P_{0}(2j)+1\right]\ln^{2}\!\delta
-4\left[3P_0(2j)-1\right]\ln\xi\ln\delta
\big\}.
\end{eqnarray}

In the contribution to the variance due to fluctuation of 
the level spacing $x$, Eq.~(\ref{eq:ch3_sigmax}), 
$\big<{\mathcal U}_{n,S}^{(2)}\big>$ can be taken from the previous 
section. Since the average eliminates the dependence on the 
lead $\alpha$, we have immediately
\begin{eqnarray}
\big<{\mathcal U}^{(2)}_{n,S}\big>^{2}
= \left(\frac{g_{L}+g_{R}}{4\pi^{2}\xi}\right)^{2}
\left<\Phi_{L,S}\right>^{2}
\label{eq:ch3_uns22}
\end{eqnarray}
where $\left<\Phi_{L,S=0}\right>$ and $\left<\Phi_{L,S=1}\right>$ 
for $\xi ,\delta\gg 1$ are given by Eq.~(\ref{eq:ch3_aveups}).  
Using these expressions in Eq.~(\ref{eq:ch3_sigmax}), we obtain
\begin{equation}
\sigma_x^2
=\left(\frac{g_{L}+g_{R}}{4\pi^{2}\xi}\right)^{2}
\left(
{\mathcal C}_{\xi\xi}\ln^{2}\xi
+{\mathcal C}_{\delta\delta}\ln^{2}\delta
+{\mathcal C}_{\xi\delta}\ln \xi\ln\delta
\right).
\label{eq:ch3_sigmaxnoc}
\end{equation}
Explicit expressions for the coefficients are given below once 
we reach the final result.

The dependence on $g_{L}$ and $g_{R}$  of the two contributions 
to the variance is different. In particular, the contribution 
due to fluctuations of the wave functions 
[Eq.~(\ref{eq:ch3_sigmazc})] is proportional to
\begin{equation}
g_{L}^{2} + g_{R}^{2}
= \frac{\left( g_{L} + g_{R}\right)^{2}}{2}
+ \frac{\left( g_{L} - g_{R}\right)^{2}}{2} \;.
\end{equation}
The first term has the same form as the contribution 
(\ref{eq:ch3_sigmaxnoc}) from fluctuations of $x$. It is 
convenient to write the total variance as a sum of symmetric 
and antisymmetric parts. Our final result for the variance is
\begin{eqnarray}
\mbox{var}\big({\mathcal U}_{n}^{(2)}\big)
= \sigma_{s}^{2}\big({\mathcal U}_{n}^{(2)}\big)
+ \sigma_{a}^{2}\big({\mathcal U}_{n}^{(2)}\big)
\label{eq:ch3_varianceref}
\end{eqnarray}
where
\begin{eqnarray}
\sigma_{s}^{2}\left({\mathcal U}_{n}^{(2)}\right)
& = & \left(\frac{g_L+g_R}{4\pi^{2}\xi}\right)^{2}
\Big(
{\mathcal S}_{\xi\xi}\ln^{2}\xi
+{\mathcal S}_{\delta\delta}\ln^{2}\delta
+{\mathcal S}_{\xi\delta}\ln \xi\ln\delta
\Big)
\label{eq:ch3_sigmasym7}
\\
\sigma_{a}^{2}\left({\mathcal U}_{n}^{(2)}\right)
& =& \left(\frac{g_L-g_R}{4\pi^{2}\xi}\right)^{2}
\Big(
{\mathcal A}_{\xi\xi}\ln^{2}\xi
+{\mathcal A}_{\delta\delta}\ln^{2}\delta
+{\mathcal A}_{\xi\delta}\ln \xi\ln\delta
\Big)
\label{eq:ch3_sigmaasym5}
\end{eqnarray}
with the coefficients $\{ {\mathcal S} \}$ and $\{ {\mathcal A} \}$
given by
\begin{eqnarray}
{\mathcal S}_{\xi\xi}(j)
&=&
\frac{2}{\beta}+16\left({\chi}-1\right)
\nonumber \\
&+&
64\left[2jP_{0}(2j)-x_{0}(2j)\right]
\left\{ 2j\left[1-P_{0}(2j)\right]-\left[1-x_{0}(2j)\right]\right\},
\label{eq:ch3_s1j}
\\
{\mathcal S}_{\delta\delta}(j)
&=&\frac{1}{2\beta}\left[3P_{0}(2j)+1\right]
+9P_{0}(2j)\left[1-P_{0}(2j)\right],
\\
{\mathcal S}_{\xi\delta}(j)
&=&-\frac{2}{\beta}\left[3P_{0}(2j)-1\right]
+24\left\{
x_{0}(2j)\left[1-P_{0}(2j)\right]
\right.
\nonumber \\
&+&
\left.
\left[1-x_{0}(2j)\right]P_{0}(2j)
-4jP_{0}(2j)\left[1-P_{0}(2j)\right]
\right\},
\\
{\mathcal A}_{\xi\xi}(j)
&=&\frac{2}{\beta},
~~{\mathcal A}_{\delta\delta}(j)
=\frac{1}{2\beta}\left[3P_{0}(2j)+1\right],
~~{\mathcal A}_{\xi\delta}(j)
= - \frac{2}{\beta}\left[3P_{0}(2j)-1\right].
\end{eqnarray}
The constant $\chi$ introduced in Eq.~(\ref{eq:ch3_s1j}) is
\begin{eqnarray}
{\chi} = \int_{0}^{\infty}\!\!dx\,x^{2}\,P(x).
\label{eq:ch3_xi}
\end{eqnarray}

For the CI model, $j=0$ and $P_{0}(0)=x_{0}(0)=1$; hence
\begin{eqnarray}
\left.\mbox{var}\left({\mathcal U}_{n}^{(2)}\right)\right|_{j=0}
=\frac{4}{\beta}\frac{g_L^2+g_R^2}{\left(4\pi^{2}\xi\right)^{2}}
\left(\ln \xi -\ln\delta\right)^{2}
+16\left({\chi}-1\right)
\left(\frac{g_L+g_R}{4\pi^{2}\xi}\right)^{2}\ln^{2}\xi \;.
\end{eqnarray}
The first term is due to fluctuation of the wave functions 
at the dot-lead contacts, and the second term comes from the 
fluctuation of the level spacing $x$.  The presence of the 
second term was missed in previous work [see Eq.~(44b) in 
Ref.~\cite{Kam00}]. If $\xi \!=\! \delta$, the first term 
vanishes; nonetheless, due to the second term the variance 
is always finite.  

\begin{figure}
\begin{center}
\includegraphics[width=10.4cm]{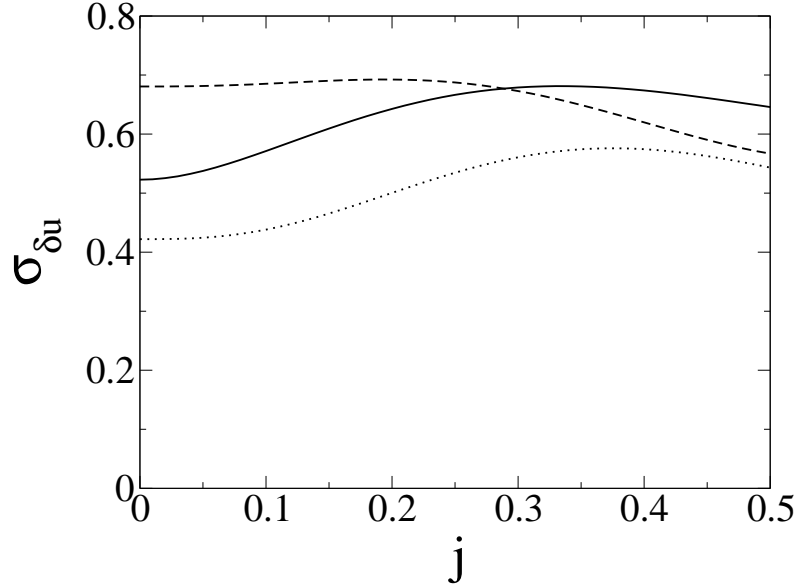}
\caption{The rms of the correction to the peak spacing for 
the symmetric setup $g_{L} = g_{R}$ as a function of 
$j = J_{\rm s}/\Delta$ normalized by the ensemble averaged 
correction at $j = 0$, Eq.~(\ref{eq:ch3_sigmasglgr}). Solid 
(dotted) curve corresponds to the GOE (GUE) at $\lambda = 1$.
Dashed curve corresponds to the GUE at $\lambda=3$.}
\label{fig:ch3_sigma}
\end{center}
\end{figure}

Let us consider a realistic special case of symmetric tunnel 
barriers, $g_{L}=g_{R}$~\cite{Mau99}. Then the asymmetric 
contribution vanishes, and the rms fluctuation of the correction 
to the peak spacing normalized by the average correction at 
$j = 0$ [Eq.~(\ref{eq:ch3_un20})] is
\begin{equation}
\sigma_{\delta u}(j)
= \frac{\sigma_{s}\big({\mathcal U}_{n}^{(2)}\big)}
{\overline{{\mathcal U}_{n}^{(2)}\left(0\right)}}
= \frac{\sqrt{{\mathcal S}_{\xi\xi}(j)\lambda^{2}
+ {\mathcal S}_{\xi\delta}(j)\lambda + {\mathcal S}_{\delta\delta}(j)}}
{2(\lambda+1)}.
\label{eq:ch3_sigmasglgr}
\end{equation}
Fig.~\ref{fig:ch3_sigma} shows this quantity plotted as 
a function of $j$ for both GOE and GUE.  Notice that (i) the rms 
is of the same order as the average, and (ii) its magnitude weakly 
depends on $j$.  To show the magnitude of the fluctuations in the 
correction relative to its average value, we plot two additional 
curves in both Fig.~\ref{fig:ch3_ave1} and Fig.~\ref{fig:ch3_ave3}, 
namely $\delta u \pm \sigma_{\delta u}$.  We find that at the 
realistic value $j \!=\! 0.3$, the correction to the even peak 
spacing is \textit{negative} for a small fraction of the quantum 
dots in the ensemble.

\section{Conclusions}
\label{sec:ch3_conclusions}

In this chapter we studied corrections to the spacings between 
Coulomb blockade conductance peaks due to finite dot-lead tunneling 
couplings.  We considered both GOE and GUE random matrix ensembles 
of 2D quantum dots with the electron-electron interactions being 
described by the CEI model. We assumed 
$T \!\ll\! \Delta \!\ll\! E_{C}$. The $S \!=\! 0$, $\frac{1}{2}$, 
and $1$ spin states of the QD were accounted for, thus limiting 
the applicability of our results to $J_{\rm s} \!<\! 0.5\Delta$.  

The ensemble averaged correction in even valleys is given in 
Eq.~(\ref{eq:ch3_even2}).  The average correction decreases 
monotonically (always staying positive, however) as the exchange 
interaction constant $J_{\rm s}$ increases 
(Figs.~\ref{fig:ch3_ave1}-\ref{fig:ch3_goe}).  The behavior found 
is very robust with respect to the choice of RMT ensemble or change 
in charging energy, mean level spacing, or temperature.  Our results 
obtained in second-order perturbation theory in the tunneling 
Hamiltonian are somewhat similar to the zeroth-order 
results~\cite{Usa01,Usa02} in that the exchange interaction reduces 
even-odd asymmetry of the spacings between peaks.  While the average 
correction to the even spacing is positive, that to the odd peak 
spacing is negative and of equal magnitude.

The fluctuations of the correction to the spacing between Coulomb 
blockade peaks mainly come from the mesoscopic fluctuations of the 
wave functions and energy level spacing $x$ in the QD.  The rms 
fluctuation of this correction is given by 
Eqs.~(\ref{eq:ch3_varianceref})-(\ref{eq:ch3_xi}).  It is of the 
same order as the average value of the correction 
(Figs.~\ref{fig:ch3_ave1} and \ref{fig:ch3_ave3}) and weakly depends 
on $J_{\rm s}$ (Fig.~\ref{fig:ch3_sigma}).  Therefore, for a small 
subset of ensemble realizations, the correction to the peak spacing 
at the realistic value of $j = 0.3$ is of the opposite sign. The 
rms fluctuation of the correction for an odd valley is the same 
as that for an even one.

We are aware of two experiments directly relevant to the results 
here.  First, in the experiment by Chang and co-workers~\cite{Jeo}, 
the corrections to the even and odd peak spacings due to finite 
dot-lead tunnel couplings were measured.  It was found that the 
even (odd) peak spacing increases (decreases) as the tunnel 
couplings are increased.  This is in qualitative agreement with 
the theory, see Eq.~(\ref{eq:ch3_even2}).  The magnitude of the 
effect was measured at different values of the gas parameter 
$r_{s}$ (and, hence $J_{\rm s}$) as well.  Unfortunately, because 
the effect is small and the experimentalists did not focus on this 
issue, one can not from this work draw a quantitative conclusion 
about the behavior of the correction to the peak spacing as a 
function of $J_{\rm s}$.

Second, in the experiment by Maurer and co-workers~\cite{Mau99}, 
the fluctuations in the spacing between Coulomb blockade peaks 
were measured as a function of the dot-lead couplings with 
$g_{L} \!=\! g_{R}$.  Therefore, only the symmetric part 
[Eq.~(\ref{eq:ch3_sigmasym7})] would contribute to the total 
variance.  Ref.~\cite{Mau99} reported results for two dots: a 
small one with area 0.3\,$\mu$m$^{2}$ and a large one with area 
1\,$\mu$m$^{2}$.  From the area (excluding a depletion width of 
about 70 nm), we estimate that the large (small) dot contains about 
500 (100) electrons. Measurements on the large QD found larger 
fluctuations upon increasing the dot-lead tunnel coupling, in 
qualitative agreement with the theory. Though the temperature was 
larger than the mean level spacing in the large QD whereas our 
theory is developed for $T \!\ll\! \Delta$, the theory gives about 
the correct magnitude for the peak spacing fluctuations. It is 
inconclusive whether the data is in better agreement with the CI 
or CEI model as the fluctuations are roughly the same 
(Fig.~\ref{fig:ch3_sigma}) in both. In the small QD, there is an 
anomaly for the strongest coupling in the experiment -- the 
fluctuations suddenly decrease. In addition, the experimental 
fluctuations are one order of magnitude larger than the theoretical 
estimate [Eqs.~(\ref{eq:ch3_sigmasglgr}) and (\ref{eq:ch3_un20})]. 
The reason for this discrepancy is not clear at this time. Possible 
contributing factors include scrambling of the electron spectrum 
as the charge state of the dot changes, or the role of the 
fluctuations when the dot is isolated (i.e., fluctuations in 
${\mathcal U}^{(0)}$). In order to assess quantitatively the role 
of dot-lead coupling in the Coulomb blockade, further experiments 
are needed.

%% file: chapters/chapter4.tex
\setlength{\textheight}{8.0in}
\clearpage
\chapter{Introduction to Quantum Computation with Quantum Dots}
\label{ch:ch4}
\thispagestyle{botcenter}
\setlength{\textheight}{8.4in}

\section{Overview}

This chapter serves as an introduction to 
the field of quantum computation (QC).
The plan of the chapter is as follows.
In Section~\ref{sec:ch4_qcintro} we explain
why QC field deserves our attention.
In Section~\ref{sec:ch4_mathbasis} we outline
the mathematical basis of quantum computations.
In Section~\ref{sec:ch4_divince_criteria} we list
and briefly explain DiVincenzo's criteria 
of physical implementation. 
Finally, in Section~\ref{sec:ch4_spin_qubits}
we briefly mention 
liquid state NMR quantum computing proposal 
but mostly concentrate on the spintronic quantum dot 
proposal put forth by Loss and DiVincenzo~\cite{Los98}.

\section{Introduction to Quantum Computation}
\label{sec:ch4_qcintro}

Quantum computation (QC) holds out tremendous promise for
efficiently solving some of the most difficult problems
in computational science: integer factorization~\cite{Sho94}, 
discrete logarithms, and modeling of quantum mechanical 
systems~\cite{Fey82}. 
These problems are intractable on any present or future 
conventional computer. 

The fact that quantum systems can perform a computation 
was realized in 1982 by Paul Benioff and Richard Feynman.
In 1994 Peter Shor proved that quantum computer
can perform prime factoring in polynomial time as compared
to the classical one which requires exponential time
to solve this problem~\cite{Sho94}.
The prime factoring problem is defined as follows:
given number $N$ which we know is a product of 
two primes $P$ and $Q$, find $P$ and $Q$.
Table~\ref{tab:qc_vs_cc} gives one a flavor on how 
computation times scale with the complexity of the 
factoring problem for classical and quantum computers.
As one can see for the 400-digits prime factoring problem 
even relatively small quantum computer will do much better
than a network of hundreds classical computers. 
This is attained due to quantum parallelism 
discussed in the next section.

\begin{table}
\label{tab:qc_vs_cc}
\begin{center}
\begin{tabular}{|c|c|c|}
\hline
& Quantum Computer & Classical Computer: \\
& of 70 Qubits & Network of \\
& (Shor's Algorithm) & Hundreds Workstations \\
\hline
130-Digits Number & 1 Month & 1 Month \\
Factoring & & \\
\hline
400-Digits Number & 3 Years & $10^{10}$ Years \\
Factoring & & (Age of The Universe) \\
\hline
\end{tabular}
\end{center}
\caption{Quantum computer vs. 
network of hundreds classical computers:
Scaling of the computational time with
the complexity of the factoring problem.}
\end{table}

Shor's spectacular result attracted to the field 
of QC many physicists and computer scientists alike. 
Many proposals for physical realization of QC 
have been put forth since then.
New experimental techniques were developed to control 
phase coherence of the quantum mechanical systems 
on a nano scale.

Since prime factoring is the most secure encryption
scheme in wide use today, the field of QC enjoyed
significant funding from government agencies 
in the USA, EU, and Australia.

\section{Mathematical Basis of Quantum Computations}
\label{sec:ch4_mathbasis}

In this section we review the mathematical basis
of quantum computations.

\subsection{Quantum Bits and Entangled States}

\begin{figure}
\begin{center}
\includegraphics[width=11.0cm]{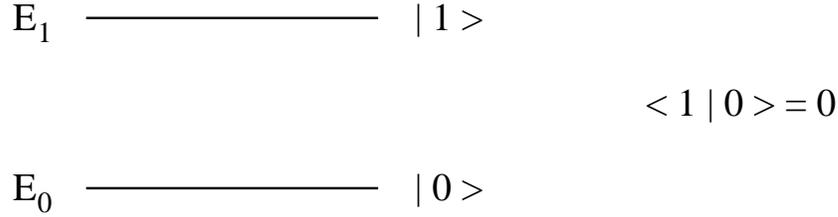}
\caption{Schematic picture of the quantum bit.}
\label{fig:ch4_qubit}
\end{center}
\end{figure}

The bits of information in a quantum computer 
are formed by two orthogonal states 
of a two-level quantum mechanical system 
$\left| 0 \right>$ and $\left| 1 \right>$, 
Fig.~\ref{fig:ch4_qubit}. 
To distinguish between classical and quantum bits
quantum bit was given a special name: {\it qubit}.
%
%
For example, a nucleus of spin-$\frac{1}{2}$ 
(or an electron) can serve as a natural qubit:
$\left| 0 \right> = \left| \downarrow \right>$, 
$\left| 1 \right> = \left| \uparrow \right>$.
In contrast to the classical bit, 
qubit has the following properties~\cite{Mer99}: \\
(i) The states $\left| 0 \right>$ and $\left| 1 \right>$
do not constitute all possible states of a qubit.
In fact, it can be in an arbitrary superposition state:
$\left| \chi \right> 
= \alpha \left| 0 \right> + \beta \left| 1 \right>$,
where $\alpha$ and $\beta$ are complex numbers with
$\left| \alpha \right|^{2} + \left| \beta \right|^{2} 
= 1$ constraint. \\
(ii) {\it Entangled state} of two qubits is possible,
{\it e.g.}:
\begin{equation}
\left| \psi \right> = \frac{1}{\sqrt{2}}
\left(
\left| 0 \right>_{1} \otimes \left| 0 \right>_{2}
+ \left| 1 \right>_{1} \otimes \left| 1 \right>_{2}
\right).
\end{equation}
In this state each individual qubit fails to have 
any state of its own. The information is, therefore,
encoded only in the correlation between the qubits
1 and 2. Two qubits in an entangled state are also 
called {\it EPR pair}~\cite{Ein35}.

\subsection{Steps in Quantum Computation and Quantum Gates}

To perform quantum computation we should~\cite{Mer99} \\
(i) Prepare a collection of $n$ qubits in the appropriate
initial state at $t = 0$: $\left| \psi (0) \right>$. 
The initial state is usually taken to be
\begin{equation}
\left| \psi (0) \right> 
= \left| 0 \right>_{1} \otimes
\left| 0 \right>_{2} \otimes 
\, \dots \,
\otimes \left| 0 \right>_{n}
\label{eq:ch4_00000}
\end{equation}
or a state constructed from this one by application of
the appropriate unitary transformations. \\
(ii) Let that state evolve under the appropriate unitary 
transformation $U$. $U$ can be realized as the ordinary
time evolution of a physical system with appropriately
constructed time-dependent Hamiltonian $H(t)$:
\begin{equation}
\left| \psi (t) \right> = U \left| \psi (0) \right>
= T e^{-i\int_{0}^{t}d\tau H(\tau )} \left| \psi (0) \right>.
\end{equation}
(iii) Extract the result of the computation by appropriate
measurement on the final state. Typical final measurement 
is made on a subset of individual qubits.

Among all possible unitary transformations there are
those acting on a single qubit, a pair of qubits,
a set of three qubits, etc. 
Let us call unitary transformations acting only on
a single qubit or a pair of qubits {\it quantum gates}.
In 1995 Barenco and coworkers proved that any quantum 
computation algorithm can be decomposed in a series of 
{\it quantum gates}: single qubit 
and two-qubit XOR gates~\cite{Bar95}.

\subsection{Quantum Parallelism}

The computational basis of $n$-qubit quantum computer 
is formed by $2^{n}$ states like this one:
\begin{equation}
\left| 0 \right>_{1} \otimes \left| 0 \right>_{2} \otimes 
\left| 0 \right>_{3} \otimes \left| 0 \right>_{4} \otimes 
\, \dots \,
\otimes \left| 0 \right>_{n-3} \otimes \left| 1 \right>_{n-2}
\otimes \left| 1 \right>_{n-1} \otimes \left| 1 \right>_{n}.
\label{eq:ch4_compbasvec}
\end{equation}
Each qubit in this ket-vector can be in one of the two 
possible eigenstates. Let us denote ket-vector 
like the one in Eq.~(\ref{eq:ch4_compbasvec}) 
simply as $\left| x \right>$, where $x$ is the integer 
between $0$ and $2^{n}-1$, whose binary expansion gives
the corresponding ket-vector. For example, in the case
of Eq.~(\ref{eq:ch4_compbasvec}) the binary expansion is
$0000 \dots 0111$. It corresponds to $\left| x \right> 
= \left| 7 \right>$. 

The concept of {\it quantum parallelism} 
can be illustrated as follows~\cite{Mer99}. 
Consider certain two-qubit entangled state
and act on it with a quantum gate $\,{\mathcal U}$.
Since unitary transformations are linear
we obtain:
\begin{equation}
{\mathcal U} \left( 
\frac{1}{\sqrt{3}} \left| 0 \right>_{1} \otimes \left| 0 \right>_{2}
+\sqrt{\frac{2}{3}} \left| 1 \right>_{1} \otimes \left| 1 \right>_{2}
\right)
= \frac{1}{\sqrt{3}}~{\mathcal U} \left| 0 \right>_{1} \otimes \left| 0 \right>_{2}
+\sqrt{\frac{2}{3}}~{\mathcal U} \left| 1 \right>_{1} \otimes \left| 1 \right>_{2}.
\label{eq:ch4_qpar}
\end{equation}
Two important remarks should be made about Eq.~(\ref{eq:ch4_qpar}):
(i) as a result of quantum computation we obtained 
the same superposition in the final state as we had
in the initial state and 
(ii) we performed quantum computation on all 
input states simultaneously! This property of
quantum computation is called {\it quantum parallelism}.
Quantum parallelism gives quantum computer the potential
to do tricks that a classical computer can perform only
with significantly greater computational effort.

Suppose we have a unitary transformation ${\mathcal U}_{f}$,
associated with a function $f$. Function $f$ replaces 
a computational basis state $\left| x \right>$ by 
$\left| f(x) \right>$, where $f(x)$ can take the following 
integer values: $f(x) = 0, 1, 2, \dots , 2^{n}-1$.
Since ${\mathcal U}_{f}$ is a unitary transformation,
it is invertible. Hence, function $f$ is bijective --
its action is nothing but a permutation of integers $\{ x \}$.
This restriction on the nature of $f$ is excessively strong.
Therefore, let us double the number of qubits and define
${\mathcal U}_{f}$ on the $2n$-qubit space as follows:
\begin{equation}
\left| x \right> \otimes \left| 0 \right>
\, \stackrel{{\mathcal U}_{f}}{\longrightarrow} \,
\left| x \right> \otimes \left| f(x) \right>.
\end{equation}
Let us call the first set of $n$-qubits {\it input register}
and the second set of $n$-qubits {\it output register}~\cite{Mer99}.

To illustrate the concept of {\it quantum parallelism} 
even further let us take the {\it input register}
to be in the state
\begin{equation}
\left| s \right> = \frac{1}{2^{n/2}}
\sum_{x=0}^{2^{n}-1} \left| x \right>
= \frac{1}{\sqrt{2}} \left( 
\left| 0 \right>_{1} + \left| 1 \right>_{1}
\right) \otimes
\, \dots \,
\otimes \frac{1}{\sqrt{2}} \left( 
\left| 0 \right>_{n} + \left| 1 \right>_{n}
\right).
\end{equation}
This input state can be obtained from the state in 
Eq.~(\ref{eq:ch4_00000}) by subjecting each qubit to
the one-qubit gate that ``rotates'' each ``spin''
90 degrees about the $y$-axis. Under the action
of ${\mathcal U}_{f}$
\begin{equation}
\left| s \right> \otimes \left| 0 \right>
= \frac{1}{2^{n/2}} \sum_{x=0}^{2^{n}-1} 
\left| x \right> \otimes \left| 0 \right>
\, \stackrel{{\mathcal U}_{f}}{\longrightarrow} \,
\frac{1}{2^{n/2}} \sum_{x=0}^{2^{n}-1} 
\left| x \right> \otimes \left| f(x) \right>.
\label{eq:ch4_massive_paral}
\end{equation}
Notice what just happened: in one run we calculated
$f(x)$ on all $2^{n}$ possible inputs. If we would
have 100 qubits in our input register, then $f(x)$
would have been simultaneously calculated on 
$2^{100}\approx 10^{30}$ different inputs.
This {\it massive quantum parallelism} is 
well outside the reach of any classical computer.

\subsection{Important Remarks}

Useful information has to be extracted from 
the output of the quantum computation, 
see Eq.~(\ref{eq:ch4_massive_paral}).
To illustrate how the measurement process works 
let us perform a measurement on the final state 
in Eq.~(\ref{eq:ch4_massive_paral}).
Namely, let us measure the individual states of
all qubits in the input and output registers.
With the equal probability of $2^{-n}$
one can find any state of the input register.
Let us assume that we find $\left| x_{0} \right>$ 
as a result of the measurement on the input register.
Then the result of the measurement on the output 
register is $\left| f(x_{0}) \right>$.
Thus, we calculated the value of $f(x_{0})$
at the randomly chosen $x_{0}$. 
All other information has been lost
because we collapsed the whole superposition
in Eq.~(\ref{eq:ch4_massive_paral})
onto a single component
$\left| x_{0} \right> \otimes \left| f(x_{0}) \right>$.

If Eq.~(\ref{eq:ch4_massive_paral}) was the only 
{\it quantum algorithm} one could come up with,
one would be no better off than 
just doing single classical computation
on a randomly chosen input.
Therefore, the goal of designing quantum algorithms
is to have easily accessible data of interest
in the final state and to make sure that the data
can not be generated by a classical algorithm
without enormously greater computational effort.
The first algorithm which satisfies these conditions
was created by Peter Shor for the prime factoring problem.

The other difficulty is to maintain the coherence
of an entangled state of many qubits. 
Unfortunately, it is impossible to completely 
isolate our qubits from the environment
during a quantum computation. 
The coupling to the environment leads to 
{\it decoherence} or collapse of the quantum state
which is especially severe for 
the entangled states of many qubits. 
The solution to this challenge is 
to use quantum error-correcting codes~\cite{Ste98}.
We know that these codes can work only for 
the sufficiently small decoherence rates.
Therefore, one of the main problems in physical
realization of the quantum computer is how to
reduce decoherence due to different couplings 
to the environment.

\section{DiVincenzo's Criteria of Physical Implementation}
\label{sec:ch4_divince_criteria}

Five DiVincenzo's criteria~\cite{DiV99} 
serve as a general guidance 
to the physical implementation of 
a working scalable quantum computer.
They naturally follow from
the mathematical basis of QC discussed in 
the previous section 
and can be formulated as follows:
\\
{\bf (i)}~A collection of the quantum mechanical two-level
systems (qubits) is needed. Each qubit should be
separately identifiable and externally addressable.
One should be able to add qubits at will.
\\
{\bf (ii)}~It should be possible to completely (or with
very high accuracy) decouple qubits from one another.
One should be able to set the state of each qubit 
to $\left| 0 \right>$ in the beginning of each computation.
\\
{\bf (iii)}~ Long decoherence times are required. 
For the error correcting-codes to be applicable 
decoherence time should be at least 4 orders
of magnitude larger than the ``clock time''
of the quantum gates.
\\
{\bf (iv)}~Logic operations should be doable.
For example, two-body Hamiltonians involving
nearby qubits should be under independent and
precise external control.
One should be able to 
smoothly (on the time scale of one clock cycle)
turn these Hamiltonians on and off.
The integral of the pulse should be
controlled with the accuracy of at least 
1 part in $10^{4}$.
\\
{\bf (v)}~Projective quantum measurements on the qubits
must be doable. It is useful, though not necessary,
for these measurements 
to be doable fast (that is, within few clock cycles) 
and have high quantum efficiency (otherwise quantum
computation has to be done in an ensemble style).

\section{Quantum Computing Proposals}
\label{sec:ch4_spin_qubits}

\subsection{Liquid State NMR}

Since the original work by Peter Shor
many quantum computing proposals were put forth
in different areas of physics. 
At the time of writing 
liquid state nuclear magnetic resonance (NMR)
proposal is well ahead of its competition.
In the experiment by 
Vandersypen and coworkers~\cite{Van01}
the custom-synthesized molecule 
(used as the quantum computer)
contained five $^{19}F$ and two $^{13}C$ 
spin-$\frac{1}{2}$ nuclei as qubits.
The quantum gates were realized by a sequence of
spin-selective radio-frequency pulses.
The readout was performed via NMR spectroscopy.
For the first time the simplest instance of 
Shor's algorithm: factorization of 15 
($15 = 3 \times 5$) was demonstrated~\cite{Van01}.

Unfortunately, liquid state NMR proposal lacks 
scalability, see the first DiVincenzo's criterion. 
Indeed, to realize more complicated algorithms 
one has to design and synthesize new molecules. 

\subsection{Spintronic Quantum Dot Proposal
by Loss and DiVincenzo}

The experimental progress in solid state 
QC proposals falls significantly behind 
liquid state NMR proposal. 
Nonetheless, once the universal gates 
(one- and two-qubit quantum gates)
become sufficiently error free 
the scaling of the quantum computer 
is going to be more or less straightforward.
Among different solid state QC proposals
one can distinguish three major categories:
(i)~superconducting microcircuits containing
Josephson junctions~\cite{Mak01,Dev04}, 
(ii) nuclear spins of the impurity atoms in
semiconductor QDs~\cite{Kan98}, and 
(iii) electron spin- or charge-qubits in
semiconductor QDs, see Chapters~\ref{ch:ch5} 
and \ref{ch:ch7} for the references. 
Charge qubit in the semiconductor double QD 
are discussed in Chapter~\ref{ch:ch7}. 

\begin{figure}
\begin{center}
\includegraphics[width=12.8cm]{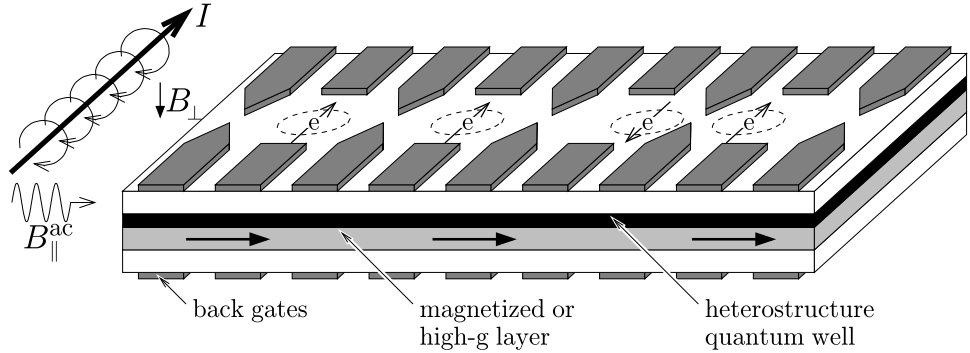}
\caption{Schematic picture of the spintronic 
QD proposal by Loss and DiVincenzo.
2D electron gas is confined on the interface
of AlGaAs/GaAs heterostructure. The electrodes
deposited on top of the heterostructure control
the shape of the QDs and their tunnel couplings
to each other. Back gates control the electron
numbers in each dot. Magnetized layer is needed
to realize one-qubit operations. 
Copyright by G.~Burkard, H.-A.~Engel, and D.~Loss, 
from Ref.~\cite{Bur00}.}
\label{fig:ch4_qdarray}
\end{center}
\end{figure}

Here we describe the original spintronic quantum dot 
proposal by Loss and DiVincenzo~\cite{Los98,DiV99}.
They consider an array of coupled 2D lateral 
quantum dots, see Fig.~\ref{fig:ch4_qdarray}. 
Each QD should contain odd number of electrons
so that spin of an electron on the last occupied
orbital serves as a natural qubit:
$\left| 0 \right> = \left| \downarrow \right>$
and $\left| 1 \right> = \left| \uparrow \right>$.

Let us briefly go over five DiVincenzo's criteria
to see if this system satisfies them and what is
needed to be improved as compared to the modern
experimental technologies. 

{\bf Criterion (i).} 
We have a collection of two-level 
quantum mechanical systems and we can add 
to the system as many new qubits as we want. 

For the qubit to be separately identifiable 
the corresponding quantum dot should contain
odd number of electrons with sufficiently
high accuracy. This electron number can be
adjusted by back-gate electrode voltages 
shown in Fig.~\ref{fig:ch4_qdarray}.
To make sure that
the number of electrons is odd and hence 
spin of the QD is equal to $\frac{1}{2}$ 
one can employ zero-bias anomaly measurement
in transport through the QD which signals
a Kondo effect. 

{\bf Criterion (ii).} 
To set the state of each qubit to $\left| 0 \right>$ 
it is sufficient to place the spins in the magnetic 
field of several Tesla at the liquid-He temperature.
For this $\left| 0 \right> \otimes \left| 0 \right> 
\otimes \dots \otimes \left| 0 \right>$ state 
to be stable upon removal of the magnetic field
the qubits must be decoupled from one another
to a very high accuracy.
This can be achieved by setting the pairs of electrodes 
between neighboring QDs to high negative voltages
so that the orbital wave-functions overlap is negligible. 

{\bf Criterion (iii).} 
The experiments on the precession of the electron spin 
in a variety of semiconductor materials were performed by 
Kikkawa and Awschalom~\cite{Kik98}.
In some structures they found spin decoherence times 
$\tau_{\phi}$ up to hundreds of nanoseconds, 
that is, at least one order of magnitude
larger than charge decoherence times. 
These results are quite encouraging.
However spin decoherence times are expected 
to be very device- and structure-specific.
Therefore, more experiments 
especially on the QD array setups are necessary.

One must emphasize that 
what really matters is the ratio $\tau_{\phi}/\tau_{s}$, 
where $\tau_{s}$ denotes the clock cycle time 
which will be defined below.

{\bf Criterion (iv).} 
To physically realize one-qubit quantum gates
it must be possible to subject a specified qubit
to the localized magnetic field 
of proper direction and strength.
This can be done by using 
a scanned magnetic particle or
by the use of magnetized barrier material
that the electron can be inserted in and out of
by electric gating, see Fig.~\ref{fig:ch4_qdarray}. 
The experimental realization of both of these ideas 
is going to be quite challenging, especially,
taking into account the fact that the required
accuracy is 1 part in $10^{4}$.

Fortunately, according to the paper by
DiVincenzo and coworkers~\cite{DiV00}
one can completely eliminate one-qubit quantum gates
at the expense of increasing the size of 
the logical qubit from one to three QDs.
In this scheme two-body Hamiltonians is
the only building block necessary to realize
quantum computations.

These two-body Hamiltonians, 
or two-qubit quantum gates if we go back 
to the original Loss and DiVincenzo proposal, 
can be realized by increasing the voltages
on the electrodes between neighboring QDs.
When these voltages are increased 
the potential barrier between neighboring QDs
is lowered and the electron wave-functions overlap,
see two right-most QDs in Fig.~\ref{fig:ch4_qdarray}.
This leads (with high accuracy) to the effective 
spin-spin Heisenberg interaction
$J\,{\bf S}_{i}\,{\bf S}_{i+1}$ with
$J \approx 4\,t^{2}/U$, where 
$t$ is the tunneling matrix element 
determined by the overlap of the electron
wave-functions in neighboring QDs and
$U$ is the Coulomb on-site repulsion energy.
The corresponding ``clock time'' can be estimated
as follows: $\tau_{s} \sim 1/J$ ($\hbar = 1$).

For the error-correcting codes to be applicable
Heisenberg spin-spin interaction 
$J\,{\bf S}_{i}\,{\bf S}_{i+1}$
has to be valid with the accuracy of $10^{-4}$.
Besides, to avoid correlated errors 
$J$ should be sufficiently small
in the ``off'' state of the two-body Hamiltonian.
One major source of errors
is the mesoscopic fluctuations. 
They appear in both the tunneling matrix element $t$ 
and the relative position of the energies 
of the last occupied orbitals in neighboring QDs.
In Chapter~\ref{ch:ch5} we find that the constraint 
of having small errors implies keeping accurate control, 
at the few percent level, of several electrode voltages.

The other technical problem is to control
the area under the two-qubit gate pulse 
with the accuracy of $10^{-4}$.
Especially, taking into account the fact 
that the pulse should be adiabatic 
on the time scale of tens of picoseconds 
so that electrons are not excited to the 
higher-lying energy levels. 
This issue is also discussed in details
in Chapter~\ref{ch:ch5}.

{\bf Criterion (v):} 
A significant progress has been made recently 
in the one electron spin readout technique~\cite{Elz04,Han04}. 
Kouwenhoven and coworkers have demonstrated single-shot 
read-out of single- and two-electron spin states in a QD. 
An electron was allowed to escape from the QD or not
depending on its spin state (the tunneling was 
energetically allowed only for the higher energy state). 
The charge of the QD was then measured 
using a quantum point contact located near the QD. 
With this technique they obtained 65\% measurement 
visibility (the corresponding fidelity was 82.5\%).
In the other slightly modified scheme they obtained 
even better values for the visibility and fidelity: 
80\% and 90\%, respectively.
Thus, high quantum efficiency of the measurement
is reached.

The readout time in the experiment was about $0.5\,$ms,
whereas the energy relaxation time $T_{1} \sim 0.85\,$ms.
Although not in principle necessary, it would be useful
for the implementation of the error-correcting codes
to reduce the readout time to about one hundreds
of $T_{1}$ time.

In conclusion, more experiments are on the way.
In particular, Kouwenhoven and coworkers are planning
to realize swapping of the spin states in adjacent 
quantum dots.
Since two-body Hamiltonian is the main building
block of the spin based quantum computer
these experiments will provide a valuable inside
into possible directions of the future research.

%% file: chapters/chapter5.tex
\setlength{\textheight}{8.0in}
\clearpage
\chapter{Spin Qubits in Multi-Electron Quantum Dots}
\label{ch:ch5}
\thispagestyle{botcenter}
\setlength{\textheight}{8.4in}

\section{Overview}

In this chapter
we study the effect of mesoscopic fluctuations on the magnitude of
errors that can occur in exchange operations on quantum dot
spin-qubits. Mid-size double quantum dots, with an odd number of
electrons in the range of a few tens in each dot, are investigated
through the constant interaction model using realistic parameters. It
is found that the constraint of having short pulses and small errors
implies keeping accurate control, at the few percent level, of several
electrode voltages. In practice, the number of independent parameters
per dot that one should tune depends on the configuration and ranges
from one to four.

The work in this chapter was done in collaboration
with Eduardo~R. Mucciolo and Harold~U. Baranger.

\section{Introduction}

Since the discovery that quantum algorithms can solve certain
computational problems much more efficiently than classical
ones \cite{Sho94,Gro96}, attention has been devoted to the physical
implementation of quantum computation (QC). Among the many proposals,
there are those based on the spin of electrons in laterally confined
quantum dots (QD) \cite{Los98}, which may have great potential for
scalability and integration with current technologies. For any
successful proposal, one must be able to perform single- and
double-qubit operations much faster than the decoherence time. 
In fact, all logical operations required for QC can be realized 
if these elementary operations are sufficiently error free~\cite{Bar95}.

Single qubit operations involving a single QD will likely require
precise engineering of the underlying material or control over local
magnetic fields~\cite{DiV99}; both have yet to be achieved in
practice. Two-qubit operations, in contrast, are already within
experimental reach. They can be performed by sending electrical pulses
to modulate the potential barrier between adjacent QDs. That permits
direct control over the effective, Heisenberg-like, exchange
interaction between the qubit spins, which is created by the overlap
between the electronic wave-functions of the QDs~\cite{Los98}. These
operations are important elements in forming a basic two-qubit gate
such as the controlled-not~\cite{DiV95} and in the propagation
of quantum information through QD arrays~\cite{Los98}. In fact, using
three QDs instead of just one to form a logical qubit would allow one
to perform all logical operations entirely based on the exchange
interaction~\cite{DiV00}. Thus, exchange operations will likely
play a major role in the realization of QD qubits. A quantitative
understanding of errors that occur during an exchange operation will
help in designing optimal systems.

The first proposal for a QD spin qubit~\cite{Los98} relied on having 
a single electron in a very small laterally confined QD. One advantage
of such a system is that the Hilbert space is nominally
two-dimensional. Leakage from the computational space involves
energies of order either the charging energy or the single-particle
excitation energy, both of which are quite large in practice ($\sim
1$~meV $\sim 10$~K). Working adiabatically -- such that the inverse of
the switching time is much less than the excitation energy -- assures
minimal leakage. The large excitation energy implies that pulses of
tens of picoseconds would be both well within the adiabatic regime and
below the dephasing time $\tau_\phi$ (which is typically in the
nanosecond range since orbital degrees of freedom are involved). 
However, in practice, it is difficult to fabricate very
small tunable devices~\cite{MarPC}. Moreover, one-electron QDs
may offer little possibility of gate tuning due to their rather
featureless wave functions.

Alternatively, a qubit could be formed by the top most ``valence''
electron in a QD with an odd number of electrons~\cite{Hu01a}. In this
case, electrons filling the lower energy states should comprise an
inert shell, leaving as the only relevant degree of freedom the spin
orientation of the valence electron. Large QDs with 100-1000
electrons, while much simpler to fabricate than single electron QDs,
are unsuitable because the excitation energy is small ($\sim
50\,\mu$eV $\sim 0.6$~K), leading to leakage or excessively slow
exchange operations. On the other hand, mid-size QDs, with 10-40
electrons, are sufficiently small to have substantial excitation
energies, yet both reasonable to fabricate and tunable through plunger
electrodes. For these dots, a careful analysis of errors is necessary.

Perhaps the best example of an exchange operation is the swap of the
spin states of the two qubits. For instance, it causes up-down spins
to evolve to down-up. Maximum entanglement between qubits occurs when
half of a swap pulse takes place -- a square-root-of-swap operation.
Several authors have treated the problem of swap errors in QD
systems \cite{Bur99,Hu00a,Hu00b,Hu01b,Sch01,Bra02}. A
primary concern was the occurrence of double occupancy (when both
valence electrons move into the same QD) during and after the
swap. However, no study so far has considered another intrinsic
characteristic of electronic states in multi-electron QDs, namely,
their marked dependence on external perturbations such as electrode
voltage or magnetic field. This sensitivity gives rise to strong
sample-to-sample fluctuations arising from the phase-coherent orbital
motion~\cite{Kou97}. These features can make the precise control
of energy levels, wave functions, and inter-dot couplings a difficult
task.

In this chapter we study errors and error rates that can take place
during the exchange operation of two spin qubits based in
multi-electron QDs. We consider realistic situations by taking into
account an extra orbital level and fluctuations in level positions and
coupling matrix elements. These lead to deviations from a
pre-established optimal swap operation point, especially when a
single-particle level falls too close to the valence electron
level. Reasons for such fluctuations can be, for instance, (i) the
lack of a sufficient number of tuning parameters (i.e., plunger
electrodes), or (ii) the cross-talk between the tuning electrodes. Our
results set bounds on the amount of acceptable detuning for mid-size
QD qubits.

This chapter is organized as follows. In Section~\ref{sec:ch5_model}, we
introduce and justify the model Hamiltonian. The states involved in
the exchange operation are presented in Section~\ref{sec:ch5_error}, where we
also discuss the pulses and the parameters involved in the exchange
operations. In Section~\ref{sec:ch5_results} we present the results of our
numerical simulations. We also discuss the impact of mesoscopic
effects on errors and put our analysis in the context of actual
experiments. Finally, in Section~\ref{sec:ch5_conclusion} we draw our
conclusions.

\section{Model System}
\label{sec:ch5_model}

We begin by assuming that the double QD system can be described by the
Hamiltonian~\cite{Ale02}
\begin{equation}
\label{eq:ch5_hamilton}
H = H_A + H_B + H_{AB},
\end{equation}
where
\begin{equation}
\label{eq:ch5_CImodel}
H_\alpha = \sum_{j,\sigma} \epsilon_j^\alpha\, n_{\alpha,j\sigma} +
\frac{U_\alpha}{2} \sum_{j,\sigma} n_{\alpha,j\sigma} \Big(
\sum_{k,\sigma^\prime} n_{\alpha,k\sigma^\prime} - 1 \Big),
\end{equation}
$\alpha = A,B$, and
\begin{equation}
\label{eq:ch5_hopping}
H_{AB} = \sum_{j,k,\sigma} \left( t_{jk}\, a_{j\sigma}^\dagger
b_{k\sigma} + \text{h.c.} \right).
\end{equation}
Here, $n_{A,j\sigma} = a^\dagger _{j\sigma} a_{j\sigma}$ and
$n_{B,k\sigma} = b^\dagger _{k\sigma} b_{k\sigma}$ are the number
operators for the single-particle states in the QDs (named $A$ and
$B$), $\epsilon^\alpha_j$ denotes the single-particle energy levels,
$t_{jk}$ are the tunneling amplitudes between the dots, and
$U_{\alpha}$ is the charging energy ($\sigma = \uparrow, \downarrow$
and $j,\,k$ run over the single-particle states). Typically, for mid-
to large-size QDs, the charging energy is larger than the mean level
spacing.

In the literature of Coulomb blockade phenomena in closed QDs, the
Hamiltonian in Eq.~(\ref{eq:ch5_CImodel}) is known as the constant
interaction model. It provides an excellent description of
many-electron QDs, being supported by both microscopic calculations
and experimental data \cite{Kou97,Ale02,Ull01a,Usa01,Usa02,Usa03,Ull01b}.
The reasoning behind its success can be understood from two
observations. First, mid- to large-size QDs, with more than ten
electrons, behave very much like conventional disordered metals in the
diffusive regime. Wavelengths are sufficiently small to resolve
irregularities in the confining and background potentials, leading to
classical chaos and the absence of shell effects in the energy
spectrum. In this case, the single-particle states obey the statistics
of random matrices, showing complex interference patterns and
resembling a random superposition of plane waves. This is in contrast
with the case of small, circularly symmetric, few-electron QDs, where
shell effects are pronounced~\cite{Hu01a,Kou01}.

Second, for realistic electron densities, the QD linear size is larger
than the screening length of the Coulomb interactions. In the presence
of random plane waves, the screened interaction can then be broken up
into a leading electrostatic contribution characterized by the QD
capacitance plus weak inter-particle residual
interactions~\cite{Ale02,Ull01b}. This description becomes more
accurate as the number of electrons gets larger since the residual
interactions become weaker. The electron bunching is reduced as wave
functions become more uniformly extended over the QD. Also, the
increase in the number of oscillations in the wave functions leads to
a self-averaging of the residual interactions. In this limit, one
arrives at the so-called ``universal Hamiltonian" for QDs, containing
only single-particle levels, the charging energy, and a mean-field
exchange term~\cite{Kou97,Ale02}. This Hamiltonian can be
derived explicitly via a random-phase approximation treatment of the
Coulomb interaction and the use of random-matrix wave
functions~\cite{Ale02,Ull01b}.

According to these arguments, interaction effects beyond the charging
energy term are omitted in Eq. (\ref{eq:ch5_CImodel}). In addition, the
intra-dot exchange interaction, which tends to spin polarize the QD,
is also neglected. The reason for that is the following. One can show
that the intra-dot exchange term only affect states where there is
double occupancy of a level. Thus, the exchange interaction constant
always appears side-by-side with the charging energy. But in
multi-electron dots, the exchange energy (which is at most of order of
the mean level separation) is much smaller than the charging
energy. Thus, intra-dot exchange effects are strongly suppressed by
the charging energy. We have verified that their inclusion does not
modify appreciably our final results. We expect the exchange
interaction to become important for two-qubit operations only in the
case of small QDs with only a few electrons, when all energy scales
(including the mean level spacing) are of the same order.

Thus, the simple picture where single-particle states are filled
according to the Pauli principle up to the top most level is an
appropriate description of multi-electron 
dots \cite{Kou97,Ale02,Ull01a,Usa01,Usa02,Usa03,Ull01b}.
In order to define the spin-$\frac{1}{2}$
qubits, both QDs should contain an odd number of electrons (say,
$2N_{A}-$1 and $2N_{B}-$1). The QD spin properties are then dictated
by the lone, valence electron occupying the highest level. The
remaining electrons form an inert core, provided that operations are
kept sufficiently slow so as not to cause particle-hole excitations to
other levels.

Experimentally, the two-qubit exchange operations also require the
capability of isolating the QDs from each other, so that a direct
product state can be prepared, such as
\begin{equation}
|i\rangle = |N_A,\uparrow\rangle_A\otimes |N_B,\downarrow\rangle_B,
\end{equation}
where the kets represent only the spin of the valence electron on each
QD.

\section{Errors in Exchange Operations}
\label{sec:ch5_error}

We focus our study on errors that appear after a full swap operation
(which should result in no entanglement). Although it could in
principle seem more sensible to look at the square-root-of-swap
operation (which creates entanglement and is therefore a building
block of logical gates), error magnitudes for the latter are
straightforwardly related to those of the full swap operation, as we
will show. We leave the discussion of the square root of swap to
Section~\ref{sec:ch5_results}.

The ideal full swap operation exchanges the valence electrons of the
QD system. For instance, it takes the product state $|i\rangle$ into
\begin{equation}
|f\rangle = \hat{U}_\text{sw}\,
|i\rangle = |N_A,\downarrow\rangle_A\otimes
|N_B,\uparrow\rangle_B.
\end{equation}
Physically, the full swap can be implemented by starting with isolated
QDs, turning on the inter-dot coupling for a time $T$ (the pulse
duration), and then turning it off, isolating the QDs again. For
weakly coupled QDs ($|t| \ll U,\delta\epsilon$), one finds $T \approx
(\pi/4)\, U/|t|^{2}$, where $t$ and $U$ here represent typical values
for the coupling matrix element and the charging energy, respectively
(throughout we assume $\hbar = 1$). To quantify the amount of error
that takes place during the operation, we use the probability of not
reaching $|f \rangle$ asymptotically, namely,
\begin{equation}
\label{eq:ch5_error}
\varepsilon = 1 - | \langle f | \psi ( + \infty ) \rangle|^2.
\end{equation}
%

\begin{figure}
\begin{center}
\includegraphics[width=10.0cm]{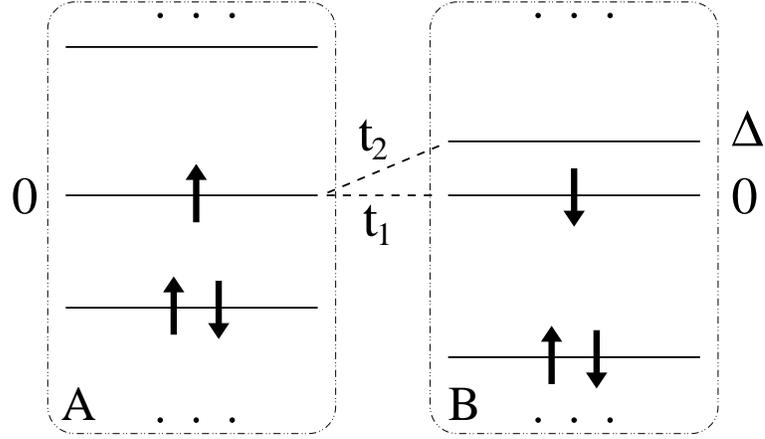}
\caption{Schematic disposition of energy levels of a system of 
two QD spin qubits (only levels close to the top occupied state 
are shown). The dashed lines indicate the most probable transitions 
that can occur during the exchange operation.}
\label{fig:ch5_1}
\end{center}
\end{figure}

We solved numerically the time-dependent Schr\"odinger equation that
derives from Eq. (\ref{eq:ch5_hamilton}) for the particular but
nevertheless realistic case shown in Fig.~\ref{fig:ch5_1}. We assumed that
voltage tuning allows one to place the top most electron of QD $A$
into an isolated single-particle state of energy $\epsilon^{A}_{N_A} =
0$ aligned with the energy of the top most electron in QD $B$,
$\epsilon^{B}_{N_B} = 0$. However, limited tuning ability leaves an
adjacent empty state close in energy in QD $B$: $\epsilon^{B}_{N_B+1}
= \Delta$. Therefore, while we can approximately neglect all levels
but one in QD $A$, for QD $B$ we needed to take two levels into
account, having hopping matrix elements denoted by $t_1=t_{N_{A}N_B}$
and $t_2=t_{N_A,N_B+1}$.\footnote{Though we assume time-reversal 
symmetry (real $t_{1}$ and $t_{2}$), we have checked that 
its violation does not modify our results.} To facilitate the analysis, 
we assumed that the dots have the same capacitance, $C$, so that
$U_A=U_B=U=e^2/C$.

The Hamiltonian of Eq. (\ref{eq:ch5_hamilton}) conserves total
spin. Assuming that filled inner levels in both QDs are inert (forming
the ``vacuum'' state $| 0 \rangle$), we can span the $S_z=0$ Hilbert
subspace with nine two-electron basis states. According to their
transformation properties, they can be divided into ``singlet''
\begin{eqnarray}
|S_l \rangle & = & \frac{1}{\sqrt{2}} \left(
b_{N_B+l-1,\downarrow}^\dagger a_{N_A\uparrow}^\dagger -
b_{N_B+l-1,\uparrow}^\dagger a_{N_A\downarrow}^\dagger \right) |0
\rangle, \nonumber \\ |D_l \rangle & = &
b_{N_B+l-1,\downarrow}^\dagger b_{N_B+l-1,\uparrow}^\dagger | 0
\rangle, \\ |D_3 \rangle & = & \frac{1}{\sqrt{2}} \left(
b_{N_B+1,\downarrow}^\dagger b_{N_B\uparrow}^\dagger -
b_{N_B+1,\uparrow}^\dagger b_{N_B\downarrow}^\dagger \right) |
0\rangle, \nonumber \\ |D_4 \rangle & = & a_{N_A\downarrow}^\dagger
a_{N_A\uparrow}^\dagger | 0 \rangle, \nonumber
\end{eqnarray}
and ``triplet''
\begin{eqnarray}
| T_l \rangle & = & \frac{1}{\sqrt{2}} \left(
b_{N_B+l-1,\downarrow}^\dagger a_{N_A\uparrow}^\dagger +
b_{N_B+l-1,\uparrow}^\dagger a_{N_A\downarrow}^\dagger \right) | 0
\rangle, \nonumber \\ |D_5 \rangle & = & \frac{1}{\sqrt{2}} \left(
b_{N_B+1,\downarrow}^\dagger b_{N_B\uparrow}^\dagger +
b_{N_B+1,\uparrow}^\dagger b_{N_B\downarrow}^\dagger \right) |
0\rangle,
\end{eqnarray}
classes, with $l=1,2$. 

The final states that correspond to an error have either double
occupancy ($|D_k\rangle$, $k=1,\ldots,5$), or an electron in the
$(N_B+1)$-level of QD $B$ ($|S_2\rangle$ and
$|T_2\rangle$).\footnote{While such states can lead to the full 
spin swap (with probability equal to $1/2$), they will induce 
severe detuning errors in subsequent operations.} In addition, 
a return to the initial state
is also considered an error. It is worth noticing the difference
between our treatment of the problem and that of
Ref.~\cite{Sch01}. In our case, errors come mainly from
either ending in the excited single-particle state after the operation
is over (i.e., states $|S_2\rangle$ and $|T_2\rangle$), or from
``no-go" defective operations. In Ref.~\cite{Sch01},
errors come from having double occupancy in the final state. Double
occupancy errors can be exponentially suppressed by adiabatically
switching the pulse on and off on time scales larger than the inverse
charging energy \cite{Los98,Hu00a,Hu00b,Sch01}. Making pulses
adiabatic on the time scale of the inverse mean level spacing for
multi-electron quantum dots is more challenging, especially because
the spacings fluctuate strongly both from quantum dot to quantum dot
and upon variation of any external parameter (mesoscopic
fluctuations). Therefore, multi-electron quantum dots require extra
tunability to get around such problems.

Very small errors, below $10^{-6}$--$10^{-4}$, can, in principal, be
fixed by the use of error correction algorithms~\cite{Ste98}. The
pulses, therefore, should be sufficiently adiabatic for errors to
remain below this threshold. We adopted the following pulse shape:
\begin{equation}
\label{eq:ch5_pulse}
v(t) = \frac{1}{2} \left( \tanh\frac{t+T/2}{2\tau}-
\tanh\frac{t-T/2}{2\tau} \right),
\end{equation}
where $\tau$ is the switching time. The pulse will remain both
well-defined and adiabatic provided that $T \gg \tau \gg \text{max} \{
\Delta^{-1}, U^{-1} \}$. Notice that this pulse is equivalent to that
adopted in Ref.~\cite{Sch01} up to exponential accuracy,
$O(e^{-T/\tau})$, with $T\gg\tau$. There is no particular reason to
believe that either performs better than the other; our choice was
dictated by technical convenience.

\begin{figure}
\begin{center}
\includegraphics[width=11.0cm]{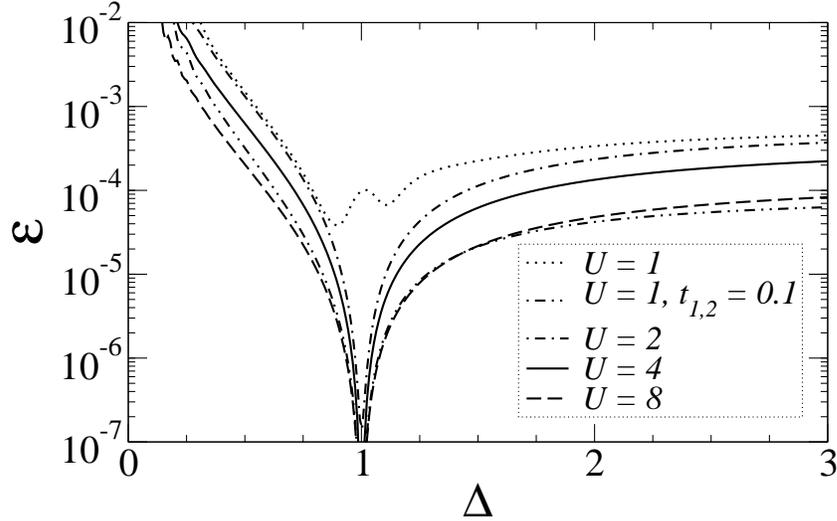}
\caption{Full swap error as a function of upper level detuning in
quantum dot $B$. The pulse width is optimized for $\Delta = 1$, $\tau
= 6$, and $t_{1,2} = 0.2$. Results for different charging energies are
shown. Interference between different quantum mechanical paths in the
device causes a sharp minimum.}
\label{fig:ch5_2}
\end{center}
\end{figure}

\section{Results}
\label{sec:ch5_results}

We used a standard numerical method, the so-called Richardson
extrapolation~\cite{Pre92}, to solve the Schr\"odinger equation for
$|\psi(t)\rangle$. The first step in our analysis was to find the
optimal value of $T$ which minimized the full swap error, as defined
in Eq. (\ref{eq:ch5_error}), for a given set of parameters $U$, $t_1$, and
$\tau$ (we used $\Delta=1$ and took $t_2 = t_1$). The second step was
to study how this minimal error depends on $\tau$. There is actually
an optimal interval for $\tau$, since small switching times spoil
adiabaticity, while large ones compromise the pulse shape (when $T$ is
relatively short). Empirically, we find that errors related to
switching times become negligible once $\tau$ reaches values of about
$\tau_0 = 4\, \text{max} \{ \Delta^{-1}, U^{-1} \}$, provided that
$\tau \ll T$. In what follows, we fix $\tau \ge \tau_0$.

\subsection{Mesoscopic Effects}

Figure \ref{fig:ch5_2} shows the full swap error as a function of $\Delta$
when $T$ is fixed to its optimal value for $\Delta = 1$. Such a
situation would arise experimentally if the pulse is optimized for a
certain configuration, but a fluctuation in level spacing
occurs. Notice the sharp increase in error as $\Delta$
decreases. While increasing $\tau$ reduces this error (by making the
switching more adiabatic), very small level spacings would be
problematic, since $\tau$ can not be larger than $T$ without
sacrificing pulse shape and effectiveness. In order to make space for
an adiabatic switching time for small $\Delta$, one would also have to
increase pulse duration. This is clear in the case of $U = 1$ (see
Fig.~\ref{fig:ch5_2}): Even moderate couplings, $t_{1,2} = 0.2$, lead to
larger errors, which can then be suppressed by decreasing $t_{1,2}$ by
a factor of two; however, that causes a fourfold increase in pulse
width which may be problematic in terms of decoherence.

\begin{figure}
\begin{center}
\includegraphics[width=11.0cm]{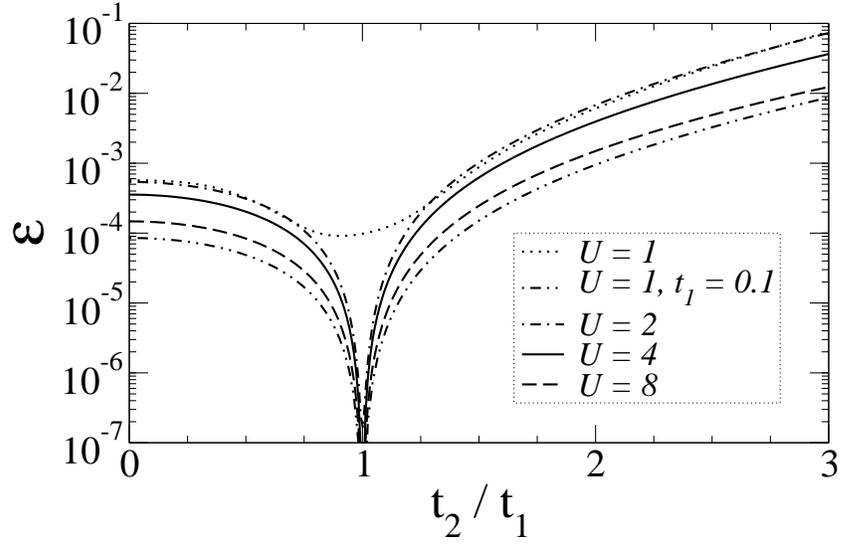}
\caption{Full swap error as a function of detuning in the coupling
constant $t_2$. The parameters used in the pulse width optimization
are the same as in Fig.~\ref{fig:ch5_2}. Results for different charging
energies are shown.}
\label{fig:ch5_3}
\end{center}
\end{figure}

The dependence of errors on fluctuations in the coupling amplitude
$t_2$ is shown in Fig.~\ref{fig:ch5_3}. Again, the pulse duration used is
the optimal value obtained when $t_2 = t_1 = 0.2$. As expected, the
error grows as $t_2$ increases. Errors related to large values of
$t_2$ can also be minimized by increasing the switching time, but the
same issues raised above appear.

\begin{figure}
\begin{center}
\includegraphics[width=10.0cm]{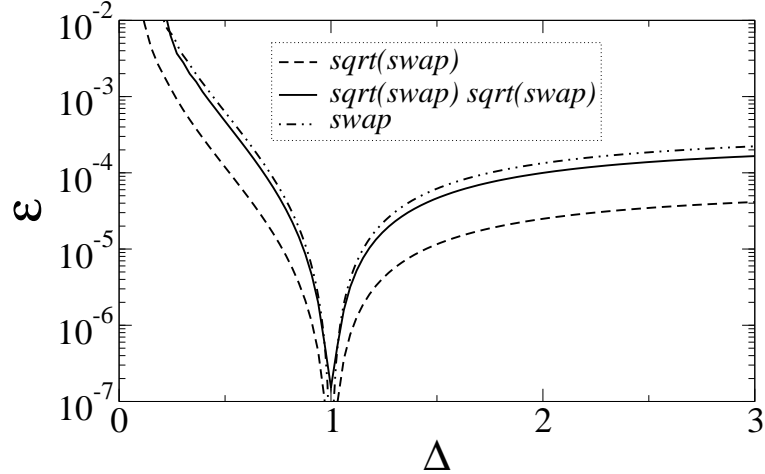}
\caption{Comparison of error resulting from a full swap operation and
two consecutive square root of swap operations. The error is plotted
as a function of upper level detuning in quantum dot $B$. The pulse
width is optimized for $\Delta =1$, $t_{1,2}=0.2$, $U=4$, and $\tau
=6$. The error for a single square-root-of-swap operation is also
shown.}
\label{fig:ch5_6}
\end{center}
\end{figure}

Figure~\ref{fig:ch5_6} presents the error for two situations involving
pulses with duration of about $T/2$, corresponding to the
square-root-of-swap operation. The cases shown are: (i) one, and (ii)
two consecutive square-root-of-swap pulses. For comparison, the curve
corresponding to a full swap pulse is also shown. The error for the
square-root-of-swap operation is given by Eq. (\ref{eq:ch5_error}) with
$|f\rangle$ replaced by
\begin{equation}
\label{eq:ch5_fsqrt}
\left| f^\prime \right> = \frac{1-i}{2}\left| S_{1} \right>
+ \frac{1+i}{2}\left| T_{1} \right> .
\end{equation}
One can observe from Fig.~\ref{fig:ch5_6} that error rates are nearly the
same after a full swap operation and after two consecutive square root
of swap operations. This insensitivity of the error to the pulse
duration led us to concentrate our effort on the full swap operations.

\begin{figure}
\begin{center}
\includegraphics[width=10.0cm]{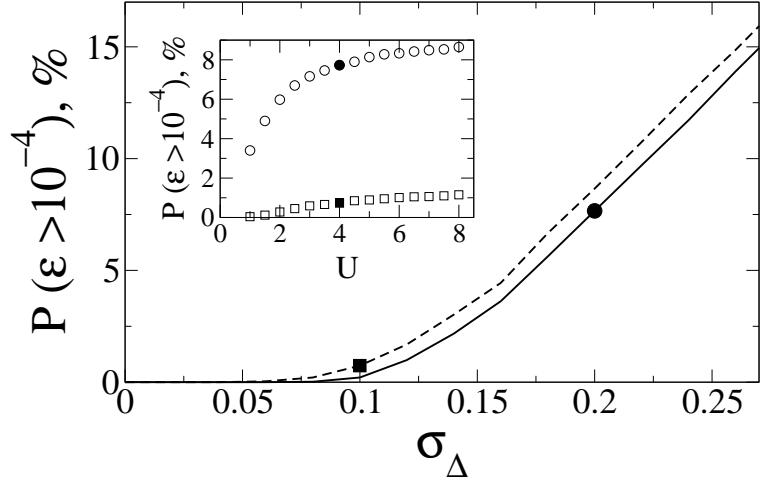}
\caption{Probability of having excessively large full swap errors 
(percentage) as a function of level spacing detuning. The solid (dashed) 
line corresponds to $\sigma_{\Gamma_2}/\bar{\Gamma}_2 = 0$ ($0.1$) 
at $U=4$ and $t_1=0.2$. The inset shows how the probability varies for 
a fixed width pulse ($T \approx 90$), but different charging energies $U$, 
when there is a large level-position detuning:
$\sigma_\Delta = 0.2$, $\sigma_{\Gamma_2} = 0$ (circles) and moderate
level-position and coupling detunings: $\sigma_\Delta = 0.1$,
$\sigma_{\Gamma_2}/\bar{\Gamma}_2 =0.1$ (squares).}
\label{fig:ch5_5}
\end{center}
\end{figure}

In order to establish an upper bound for QD tuning accuracy, we have
performed simulations where both $\Delta$ and $t_2$ were allowed to
vary. The spacing between the levels in QD $B$ was taken from a
Gaussian distribution centered at $\bar{\Delta}=1$, with standard
deviation $\sigma_\Delta$ ($\epsilon_{N_B}^{B}=0$ was kept fixed). For
the coupling amplitude, we generated Gaussian distributed level widths
$\Gamma_2 = 2\pi\, t_2^2/\bar{\Delta}$, with average $\bar{\Gamma}_2 =
2\pi\, t_1^2/\bar{\Delta}$ and standard deviation $\sigma_{\Gamma_2}$.
The pulses had their widths optimized for the typical case where
$\Delta = \bar{\Delta} = 1$, $U = 4$, $t_{1,2} = 0.2$, and $\tau =
6$. For fixed values of $\bar{\Delta}$, $\sigma_\Delta$,
$\bar{\Gamma}_2$, and $\sigma_{\Gamma_2}$, we generated 10,000
realizations of $\Delta$ and $\Gamma_2$ and each time calculated the
error after the application of the full swap pulse. In Fig.~\ref{fig:ch5_5} 
we show how the probability of having an error larger
than the $10^{-4}$ threshold depends on the energy level accuracy,
$\sigma_\Delta$. Two cases are considered, namely, plain and limited
control of the inter-dot coupling constant ($\sigma_{\Gamma_2} = 0$
and $0.1\, \bar{\Gamma}_2$, respectively). The data indicates that
frequent, non-correctable errors will happen if an accuracy in
$\Delta$ of better than 10 percent is not achieved.

\subsection{Relevance for Real Quantum Dots}

To make a quantitative estimate of the impact of these results, let us
consider the double QD setup of Jeong and coworkers~\cite{Jeo01}. In
their device, each QD holds about 40 electrons and has a lithographic
diameter of 180~nm (we estimate the effective diameter to be around
120~nm, based on the device electron density). The charging energy and
mean level spacing of each QD are approximately 1.8~meV and 0.4~meV,
respectively (thus $U/\bar{\Delta} \approx 4.5$). If we allow for a
maximal inter-dot coupling of $t_{1,2} \approx 0.2\, \bar{\Delta}$
(which yields a level broadening of about $0.25\, \bar{\Delta}$), we
find minimal full swap pulse widths of about 100~ps. These values
match those used in Fig.~\ref{fig:ch5_5}. For this case, switching times
of 10~ps would be long enough to operate in the adiabatic regime and
also provide an efficient and well-defined pulse shape. Thus, the
combined times should allow for 8-10 consecutive full swap gates
before running into dephasing effects related to orbital degrees of
freedom~\cite{Fuj03}. While these numbers are yet too small for
large-scale quantum computation, they could be sufficient for the
demonstration of QD spin qubits. Based on Fig.~\ref{fig:ch5_5}, we find
that accuracies in $\Gamma_2$ of about 10\% would make operations only
limited by dephasing, and not by fluctuation-induced errors. However,
as shown in the inset, even for small QDs (typically having small
$U/\bar{\Delta}$ ratios), the occurrence of large errors is quite
frequent when level detuning is large.

An important issue for multi-electron QDs is their strong mesoscopic,
sample-to-sample fluctuations in energy level position and
wave-function amplitudes. Our results so far indicate how big an
effect a given change in energy or wave-function will produce; now we
go further and discuss how mesoscopic fluctuations more generally
affect a collection of qubits.

In experiments, several electrodes are placed around the QD
surroundings and their voltages are used to adjust the lateral
confining potential, the inter-dot coupling, and the coupling between
QDs and leads. These voltages are external parameters that can be used
to mitigate the effects of mesoscopic fluctuations by tuning energy
levels and wave-functions to desired values. Having that in mind, our
results indicate two different scenarios for QD qubit implementations.

First, if one is willing to characterize each QD pair separately and
have them operate one by one, mesoscopic fluctuations will be
irrelevant. It will be possible, with a single parameter per QD, say,
to isolate and align energy levels reasonably well. Errors can be
further minimized by decreasing the inter-dot coupling (thus
increasing $T$). But since QDs are not microscopically identical, each
pair of QDs will require a different pulse shape and
duration. Multi-electron QDs are tunable enough, easy to couple, and
much easier to fabricate than one-electron dots; therefore,
multi-electron QDs are most appropriate for this case.

Second, if the goal is to achieve genuine scalability, one has to
operate qubits in a similar and uniform way, utilizing a single pulse
source. In this case, $T$ and $\tau$ should be the same for all QD
pairs. Based on our results above, one should strive to maximally
separate the top most occupied state from all other states, occupied
or empty, so as to reduce the possibility of leakage during operations
with a fixed duration. At the same time, it is important to reduce
inter-pair cross-talk induced by capacitive coupling between
electrodes, as well as all inter-dot couplings except between the top
most states of each QD. One should bear in mind that not all
electrodes act independently -- in most cases a search in a
multidimensional parameter space has to be carried out. Thus, four
tuning parameters per QD may be necessary to achieve the following
goals: (i) find isolated, single-occupied energy level (two
parameters); (ii) align this level with the corresponding level in an
adjacent QD (one parameter); (iii) control the inter-dot coupling (one
parameter). For parameters involved in (i) and (ii), an accuracy of a
few percent will likely be required. Finally, control over the
inter-dot coupling parameter, (iii), must allow for the application of
smooth pulse shapes in the picosecond range. Our simulations also show
that the pulse width must be controlled within at least $0.5\%$
accuracy. Although these requirements seem quite stringent, recent
experiments indicate that they could be met~\cite{Che04}.

\section{Conclusions}
\label{sec:ch5_conclusion}

In summary, our analysis indicate that mid-size QDs, with ten to a few
tens of electrons, while not allowing for extremely fast gates, are
still good candidates for spin-qubits. They offer the advantage of
being simpler to fabricate and manipulate, but at the same time
require accurate, simultaneous control of several parameters. Errors
related to detuning and sample-to-sample fluctuations can be large,
but can be kept a secondary concern with respect to dephasing effects
provided that a sufficient number of independent electrodes or tuning
parameters exists.

%% file: chapters/chapter6.tex
\setlength{\textheight}{8.0in}
\clearpage
\chapter{Time Evolution of the Reduced Density Matrix}
\label{ch:ch6}
\thispagestyle{botcenter}
\setlength{\textheight}{8.4in}

\section{Introduction}

One of the most fundamental problems in theoretical
physics is that of a quantum mechanical system coupled
to the bosonic (or fermionic) bath~\cite{Leg87,Wei99}. 
One such example is presented in the next chapter 
where we consider double QD charge qubit
coupled to the acoustic phonon bath.
Formal solution to the problem can be found
[see Eqs.~(\ref{eq:ch6_sigma_closed}) and 
(\ref{eq:ch6_S_real}) below],
nonetheless to do any practical calculation
one has to 
(a)~make simplifying assumptions like 
Born and Markov approximations in the derivation 
of the Redfield equation~\cite{Red57,Red65} or 
(b)~implement sophisticated numerical integration 
procedure like QUAPI (quasiadiabatic propagator 
path integral)~\cite{Mak94,Mak95a,Mak95b,Mak95c,Tho02}.

Time evolution of the isolated quantum mechanical system
is governed by the Schr\"{o}dinger equation.
Therefore, (i)~its energy is conserved
and (ii)~its phase coherence is preserved over time.
Coupling to the bath leads to the energy relaxation
-- decoherence and loss of phase coherence -- dephasing.
These processes are important for quantum computations
because they may potentially destroy 
the phase coherence of both 
unentangled single qubit states and 
especially entangled states of many qubits.

In our treatment of the problem 
we choose to go along the lines of Born and Markov 
approximations which eventually lead us to 
the Redfield equation~\cite{Red57,Red65} because 
(i)~we are interested in time evolution of the 
reduced density matrix at sufficiently long times and
(ii)~this formalism allows us 
to obtain analytical expressions for both 
qubit decoherence and dephasing rates.
These analytical expressions contain valuable information 
about the dependence of the rates on different parameters
of the double QD setup.

In this chapter 
we give formal derivation of the Redfield equation 
making simplifying assumptions along the way.
We employ the projection operator technique~\cite{Arg64}
as it provides the most transparent way of the derivation.

The plan of this chapter is as follows.
In Section~\ref{sec:ch6_problem} we formulate the problem.
In Section~\ref{sec:ch6_gen_solution} we find 
formal solution of the problem.
In Section~\ref{sec:ch6_born_approx} we assume that 
the quantum mechanical system and the bath 
are unentangled at $t = 0$ and 
also make Born approximation.
Finally, in Section~\ref{sec:ch6_redfield}
we make Markov approximation and 
finalize the derivation of the Redfield equation.

\section{Formulation of the Problem}
\label{sec:ch6_problem}

The Hamiltonian of the total system is
\begin{equation}
\label{eq:ch6_h}
H(t) = H_{S}(t) + H_{B} + V,
\end{equation}
where $H_{S}(t)$ is the Hamiltonian of the system,
$H_{B}$ is the Hamiltonian of the bath, and 
$V$ is the Hamiltonian of their interaction.
We assume that it can be factorized as follows:
\begin{equation}
\label{eq:ch6_v}
V = \sum_{i} K_{i}\Phi_{i},
\end{equation}
where $\{ K_{i} \}$ are the operators 
in the system's of interest Hilbert space and 
$\{ \Phi_{i} \}$ are the operators in the
bath Hilbert space.
Time evolution of the total system obeys 
Schr\"odinger equation for the density matrix:
\begin{equation}
\label{eq:ch6_sch_eq}
i \frac{d}{dt} \rho (t) = [ H(t), \rho (t)].
\end{equation}
Our goal is to find similar equation for the
time evolution of the reduced (system's of interest) 
density matrix $\sigma (t)$.

\section{General Solution}
\label{sec:ch6_gen_solution}

Let us write $\rho (t)$ as follows:
\begin{equation}
\label{eq:ch6_rho_sigma_eta}
\rho (t) = f (H_{B}) \sigma (t) + \eta (t),
\end{equation}
where
\begin{equation}
\label{eq:ch6_bath}
f (H_{B}) = \frac{e^{-\beta H_{B}}}
{tr_{b} e^{-\beta H_{B}}}.
\end{equation}
One can always do this because no assumption
is made about $\eta (t)$. Now let us substitute
Eq.~(\ref{eq:ch6_rho_sigma_eta}) into 
Eq.~(\ref{eq:ch6_sch_eq}). We obtain
\begin{eqnarray}
\label{eq:ch6_total}
i\left[ f(H_{B}) {\dot \sigma} (t) + {\dot \eta}\right]
= f(H_{B}) [H_{S}(t),\sigma (t)]
+ [H_{S}(t),\eta (t)] \\ \nonumber
+ [H_{B},\eta (t)]
+ [V,f(H_{B})\sigma (t)] + [V,\eta (t)].
\end{eqnarray}
To find the equation on $\sigma (t)$ let us trace 
left and right hand sides of Eq.~(\ref{eq:ch6_total})
over bath degrees of freedom. The following identities
are useful in simplifying the traces:
\begin{equation}
\label{eq:ch6_identities}
tr_{b}\eta (t) = 0,~~
tr_{b}f(H_{B}) = 1,~~
tr_{b}[Vf(H_{B})] = \sum_{i} K_{i} tr_{b}[\Phi_{i}f(H_{B})] = 0.
\end{equation}
The last identity should be valid for the most 
practical applications (it is certainly valid
for the phonons coupled to the electric charge,
for example). Then the equation for the reduced
density matrix becomes
\begin{equation}
\label{eq:ch6_sigma_dot}
i {\dot \sigma} (t) = [H_{S}(t),\sigma (t)]
+ tr_{b} [V,\eta (t)],
\end{equation}
where
\begin{equation}
tr_{b} [V,\eta (t)] 
= \sum_{i}[K_{i},tr_{b}\{\Phi_{i}\eta (t)\}].
\end{equation}
However there is one problem: Eq.~(\ref{eq:ch6_sigma_dot})
contains $\eta (t)$ as well. Therefore, one has to find
the second (coupled to the first one) equation on $\eta (t)$.
This can be done by acting with the projection operator 
$P = 1 - f(H_{B})tr_{b}$ on both sides of Eq.~(\ref{eq:ch6_total}).
A little bit more involved calculation leads to the following
result:
\begin{equation}
\label{eq:ch6_eta_dot}
i {\dot \eta} (t) = [H_{S}(t),\eta (t)]
+ [H_{B},\eta (t)] + [V,\eta (t)]
+ [V,f(H_{B})\sigma (t)] - f(H_{B})tr_{b}[V,\eta (t)].
\end{equation}
One can write Eqs.~(\ref{eq:ch6_sigma_dot}) and 
(\ref{eq:ch6_eta_dot}) in a more compact form as follows:
\begin{eqnarray}
\label{eq:ch6_sigma_dot_short}
&& {\dot \sigma} (t) = -i\, {\mathcal H}_{S}(t)\sigma (t)
-i\, tr_{b}{\mathcal V}\eta (t) \\
\label{eq:ch6_eta_dot_short}
&& {\dot \eta} (t) = -i\, [{\mathcal H}_{S}(t) 
+ {\mathcal H}_{B} + P{\mathcal V}]\eta (t)
-i\, {\mathcal V}f(H_{B})\sigma (t),
\end{eqnarray}
where ${\mathcal H}_{S}$, ${\mathcal H}_{B}$, and
${\mathcal V}$ are the Liouvillian operators:
${\mathcal H}_{S}(t)\sigma (t)=[H_{S}(t),\sigma (t)]$,
${\mathcal V}\eta (t)=[V,\eta (t)]$, etc.

Our further plan is as follows. 
Firstly, let us solve Eq.~(\ref{eq:ch6_eta_dot_short}),
that is, find the solution for $\eta (t)$ -- it is
going to have some functional dependence on $\sigma (t)$.
Secondly, let us substitute the solution in 
Eq.~(\ref{eq:ch6_sigma_dot_short}) and simplify
the expression obtained. 

Eqs.~(\ref{eq:ch6_sigma_dot_short}) and (\ref{eq:ch6_eta_dot_short})
have the following form:
\begin{eqnarray}
\label{eq:ch6_eta_de}
&& {\dot \eta} (t) = {\mathcal A}(t)\eta (t) + B(t) \\
\label{eq:ch6_eta_de_ic}
&& \eta (0) = \eta_{0}.
\end{eqnarray}
This differential equation has a standard solution 
which can be written as follows
\begin{equation}
\label{eq:ch6_eta_sol}
\eta (t) = {\mathcal S}(t,0)\eta_{0}
+ \int_{0}^{t}dt'\, {\mathcal S}(t,t')B(t'),~~
\mbox{where}~~
{\mathcal S}(t,t') = {\hat T}e^{\int_{t'}^{t}dx\, {\mathcal A}(x)}
\end{equation}
and ${\hat T}$ is a time-ordering operator. 
How do we understand ${\mathcal S}(t,0)\eta_{0}$ term
in Eq.~(\ref{eq:ch6_eta_sol}), for example?
It should be understood as follows:
\begin{equation}
\label{eq:ch6_S_matr_act}
{\mathcal S}(t,0)\eta_{0} 
= {\hat T}e^{\int_{0}^{t}dx\, {\mathcal A}(x)}\eta_{0} 
= {\hat T}e^{\int_{0}^{t}dx\, A(x)}\, \eta_{0}\,
{\hat T}_{-}e^{-\int_{0}^{t}dx\, A(x)},
\end{equation}
where ${\hat T}_{-}$ is an anti-time-ordering operator.
Eq.~(\ref{eq:ch6_eta_sol}) solves Eq.~(\ref{eq:ch6_eta_dot_short}).
Let us substitute the solution 
in Eq.~(\ref{eq:ch6_sigma_dot_short}).
Thus, we obtain closed form differential equation for $\sigma (t)$:
\begin{equation}
\label{eq:ch6_sigma_closed}
{\dot \sigma}(t) = -i\, {\mathcal H}_{S}(t)\sigma (t)
-i\, tr_{b}{\mathcal VS}(t,0)\eta_{0}
-tr_{b}{\mathcal V}
\int_{0}^{t}dt'{\mathcal S}(t,t')
{\mathcal V}f(H_{B})\sigma (t'),
\end{equation}
where
\begin{equation}
\label{eq:ch6_S_real}
{\mathcal S}(t,t') = {\hat T}
e^{-i \int_{t'}^{t}dx[{\mathcal H}_{S}(x)
+ {\mathcal H}_{B} + P{\mathcal V}]}.
\end{equation}
An important note is due at this time. 
At $t = 0$: $\rho (0) = f(H_{B})\sigma_{0} + \eta_{0}$.
Therefore, if system and bath are not entangled at $t = 0$
then $\eta_0 = 0$ and 
second term in Eq.~(\ref{eq:ch6_sigma_closed}) simply vanishes.

\section{$\eta_{0} = 0$ and Born Approximations}
\label{sec:ch6_born_approx}

Eq.~(\ref{eq:ch6_sigma_closed}) may look simple 
but it is difficult to solve
mainly due to complex structure of the ${\mathcal S}$ matrix.
Let us make a couple of simplifying assumptions.
Firstly, let us assume that $\eta_{0} = 0$ then
second term in Eq.~(\ref{eq:ch6_sigma_closed}) vanishes.
Secondly, let us assume that the system 
is weakly coupled to the bath ($V$ is small)
and keep only the leading term 
in powers of $V$ expansion of the third term
(Born approximation):
\begin{equation}
\label{eq:ch6_S_eq_S0}
{\mathcal S}(t,t') 
= {\mathcal S}_{0}(t,t') + O ({\mathcal V}),~~
\mbox{where}~~
{\mathcal S}_{0}(t,t')
= {\hat T}e^{-i \int_{t'}^{t}dx[{\mathcal H}_{S}(x)
+ {\mathcal H}_{B}]}.
\end{equation}
Thus, in $\eta_{0} = 0$ and Born approximations,
Eq.~(\ref{eq:ch6_sigma_closed}) becomes
\begin{equation}
\label{eq:ch6_sigma_born}
{\dot \sigma}(t) = -i\, {\mathcal H}_{S}(t)\sigma (t)
-tr_{b}{\mathcal V}
\int_{0}^{t}dt'{\mathcal S}_{0}(t,t')
{\mathcal V}f(H_{B})\sigma (t'),
\end{equation}
where the first term on the right hand side is the conventional 
Liouvillian term corresponding to the time reversible dynamics 
and the second term is a dissipative term (or time irreversible 
part).

The dissipative term can be further simplified. 
First, let us simplify ${\mathcal S}_{0}$:
\begin{equation}
\label{eq:ch6_S0_simpl}
{\mathcal S}_{0}(t,t')
= {\mathcal S}_{S}(t,t')e^{-i(t-t'){\mathcal H}_{B}},~~
\mbox{where}~~
{\mathcal S}_{S}(t,t')
= {\hat T}e^{-i \int_{t'}^{t}dx{\mathcal H}_{S}(x)}.
\end{equation}
Then, the dissipative term becomes
\begin{equation}
\label{eq:ch6_dissipative_term}
- \int_{0}^{t} dt'\, tr_{b}
\left[ V, {\mathcal S}_{S}(t,t')
\left[ e^{-i(t-t')H_{B}} V e^{i(t-t')H_{B}},
f(H_{B}) \sigma (t')\right] \right].
\end{equation}
Further calculations should be performed 
in the following order: 
(i)~substitute $V = \sum_{i} K_{i} \Phi_{i}$; 
(ii)~take advantage of $tr_{b}$ -- define 
bath correlation functions:
\begin{eqnarray}
\label{eq:ch6_bath_corr_func_1}
&& B_{ij}(t) = tr_{b}
\left[ \Phi_{i}(t) \Phi_{j} f(H_{b}) \right], \\
\label{eq:ch6_bath_corr_func_2}
&& B_{ij}(-t) = tr_{b}
\left[ \Phi_{i} \Phi_{j}(t) f(H_{b}) \right], \\
\label{eq:ch6_bath_corr_func_3}
&& \left[ B_{ji}(-t) \right]^{\dagger}
= B_{ij}(t);
\end{eqnarray}
(iii) to transform $\sigma (t-y)$ to $\sigma (t)$ 
use:\footnote{Eq.~(\ref{eq:ch6_ty_to_t}) implies 
the validity of the Markovian approximation as well.}
\begin{equation}
\label{eq:ch6_ty_to_t}
{\mathcal S}_{S}(t,t-y)\, \sigma (t-y)
= \sigma (t) + O({\mathcal V}^{2}).
\end{equation}

The final form of the equation 
for the time evolution of the reduced density matrix 
in $\eta_{0} = 0$ and Born approximations is
\begin{equation}
\label{eq:ch6_sigma_final_form}
{\dot \sigma}(t)
= -i \left[ H_{S}(t), \sigma (t)\right]
+ \sum_{j} \left[ \Lambda_{j}(t)\, \sigma (t), K_{j}\right]
+ \mbox{H.c.},
\end{equation}
where
\begin{eqnarray}
\Lambda_{j}(t)
&=& \sum_{i} \int_{0}^{t}dy\, B_{ji}(y)\,
{\mathcal S}_{S}(t,t-y)\, K_{i} \\
\label{eq:ch6_lambda_mart}
&=& \sum_{i} \int_{0}^{t}dy\, B_{ji}(y)\,
{\hat T}e^{-i\int_{t-y}^{t}dx\, H_{S}(x)}\, K_{i}\,
{\hat T}_{-}e^{i\int_{t-y}^{t}dx\, H_{S}(x)}.
\end{eqnarray}
The structure of $\Lambda$-matrices is quite complex. 
Moreover, at each time slice in the integration
of Eq.~(\ref{eq:ch6_sigma_final_form}) one has 
to numerically calculate $\Lambda$-matrices
all over again.

\section{Markov Approximation on Top: Redfield Equation}
\label{sec:ch6_redfield}

One conventional way to simplify Eq.~(\ref{eq:ch6_lambda_mart})
and make Eq.~(\ref{eq:ch6_sigma_final_form}) local in time
is to assume that {\it bath has no memory}. That is,
bath correlation time $\tau_{c}$ is the smallest
time scale in the problem. This is called Markov approximation. 
In this approximation one can only find 
evolution of the system at sufficiently
long times: $t > \tau_{c}$.
Since $B_{ji}(y)\to 0$ for $y > \tau_{c}$,
only small values of $y$ contribute 
in Eq.~(\ref{eq:ch6_lambda_mart}). Hence,
\begin{equation}
\int_{t-y}^{t}dx\, H_{S}(x)
\approx y\, H_{S}(t)
\end{equation}
and upper limit of the integration over $y$ 
in Eq.~(\ref{eq:ch6_lambda_mart}) can be 
extended to $\infty$. Thus, $\Lambda$-matrices 
in Born-Markov approximation can be written as follows:
\begin{equation}
\label{eq:ch6_lambda_markov}
\Lambda_{j}(t) = 
\sum_{i}\int_{0}^{\infty}dy\, B_{ji}(y)\, 
e^{-iyH_{S}(t)}\, K_{i}\, e^{iyH_{S}(t)}.
\end{equation}
Eq.~(\ref{eq:ch6_sigma_final_form}) with $\Lambda$-matrices
given by Eq.~(\ref{eq:ch6_lambda_markov}) is local in time
and called Redfield equation~\cite{Red57,Red65,Arg64,Pol94}.

%% file: chapters/chapter7.tex
\setlength{\textheight}{8.0in}
\clearpage
\chapter{Phonon Decoherence of a Double Quantum Dot Charge Qubit}
\label{ch:ch7}
\thispagestyle{botcenter}
\setlength{\textheight}{8.4in}


\def\Xint#1{\mathchoice
   {\XXint\displaystyle\textstyle{#1}}%
   {\XXint\textstyle\scriptstyle{#1}}%
   {\XXint\scriptstyle\scriptscriptstyle{#1}}%
   {\XXint\scriptscriptstyle\scriptscriptstyle{#1}}%
   \!\int}
\def\XXint#1#2#3{{\setbox0=\hbox{$#1{#2#3}{\int}$}
     \vcenter{\hbox{$#2#3$}}\kern-.5\wd0}}
\def\ddashint{\Xint=}
\def\dashint{\Xint-}
\hyphenation{pho-non}

\section{Overview}

In this chapter 
we study decoherence of a quantum dot charge qubit due to coupling to
piezoelectric acoustic phonons in the Born-Markov approximation.
After including appropriate form factors, we find that phonon
decoherence rates are one to two orders of magnitude weaker than was
previously predicted.
We calculate the dependence of the $Q$-factor on lattice temperature,
quantum dot size, and interdot coupling. Our results suggest that
mechanisms other than phonon decoherence play a more significant role
in current experimental setups.

The work in this chapter was done in collaboration
with Eduardo~R. Mucciolo and Harold~U. Baranger.

\section{Introduction}

Since the discovery that quantum algorithms can solve certain
computational problems much more efficiently than classical
ones~\cite{Nie00}, attention has been devoted to the physical
implementation of quantum computation. Among the many proposals, 
there are those based on the electron spin~\cite{Los98,DiV00} 
or charge \cite{Bli00,Tan00,Fed04,Fed00,Bra02,Wu04} in 
laterally confined quantum dots, which may have great potential for 
scalability and integration within current technologies.

Single qubit operations involving the spin of an electron in a quantum
dot will likely require precise engineering of the underlying material
or control over local magnetic fields~\cite{DiV99}; both have
yet to be achieved in practice. In contrast, single qubit operations
involving charge in a double quantum dot (DQD)~\cite{Wie03} are
already within experimental reach~\cite{Fuj03,Fuj04b,Pet04}. They can be
performed either by sending electrical pulses to modulate the
potential barrier between the dots (tunnel pulsing)~\cite{Fed04,Fed00,Wu04}
or by changing the relative position of the energy levels (bias
pulsing)~\cite{Fuj03,Fuj04b}. In both cases one acts on the overlap
between the electronic wave functions of the dots. This permits direct
control over the two low-energy charge states of the system -- the
basis states $|1\rangle$ and $|2\rangle$ of a qubit: Calling $N_{1}$
($N_{2}$) the number of excess electrons in the left (right) dot, we
have that $|1\rangle = (1,0)$ and $|2\rangle = (0,1)$.

\begin{figure}
\begin{center}
\includegraphics[width=6.0cm]{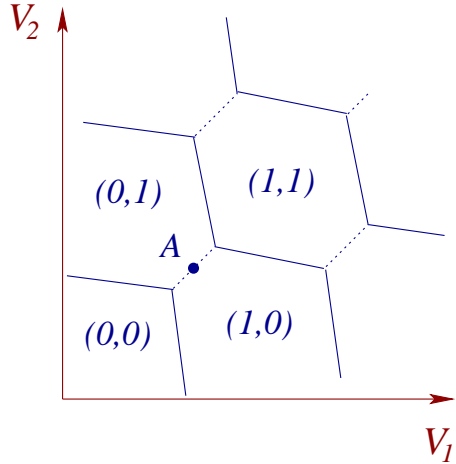}
\caption{Schematic Coulomb blockade stability diagram for a double
quantum dot system at zero bias~\cite{Wie03}. $(N_{1},N_{2})$
denotes the number of excess electrons in the dots for given values of
the gate voltages $V_1$ and $V_2$. The solid lines indicate
transitions in the total charge, while the dotted lines indicate
transitions where charge only moves between dots. The point $A$ marks
the qubit working point.}
\label{fig:ch7_diamonds}
\end{center}
\end{figure}

The proposed DQD charge qubit relies on having two lateral quantum
dots tuned to the $(1,0)\!\leftrightarrow\! (0,1)$ transition line of
the Coulomb blockade stability diagram (see Fig.~\ref{fig:ch7_diamonds}).
Along this line, an electron can move between the dots with no
charging energy cost. An advantage of this system is that the Hilbert
space is two-dimensional, even at moderate temperatures, since
single-particle excitations do not alter the charge configuration.
Leakage from the computational space involves energies of order the
charging energy which is quite large in practice ($\sim\! 1$~meV
$\sim\! 10$~K). In the case of tunnel pulsing, working adiabatically
-- such that the inverse of the switching time is much less than the
charging energy -- assures minimal leakage. The large charging energy
implies that pulses as short as tens to hundreds of picoseconds would
be well within the adiabatic regime. However, the drawback of using
charge to build qubits is the high decoherence rates when compared to
spin. Since for any successful qubit one must be able to perform
single- and double-qubit operations much faster than the decoherence
time, a quantitative understanding of decoherence mechanisms in a DQD
is essential.

In this chapter, we carry out an analysis of phonon decoherence in a DQD
charge qubit. During qubit operations, the electron charge movement
induces phonon creation and annihilation, thus leading to energy
relaxation and decoherence. In order to quantify these effects, we
follow the time dependence of the system's reduced density matrix,
after tracing out the phonon bath, using the Redfield formalism in the
Born and Markov approximations~\cite{Arg64,Pol94}.

Our results show that decoherence rates for this situation are one to
two orders of magnitude weaker than previously estimated. The
discrepancy arises mainly due to the use of different spectral
functions. Our model incorporates realistic geometric features which
were lacking in previous calculations. When compared to recent
experimental results, our calculations indicate that phonons are
likely not the main source of decoherence in current DQD setups.

The chapter is organized as follows. In Section~\ref{sec:ch7_model}, we
introduce the model used to describe the DQD, discuss the coupling to
phonons, and establish the Markov formulation used to solve for the
reduced density matrix. In Section~\ref{sec:ch7_tunnel} we study decoherence
in a single-qubit operation, while in Section~\ref{sec:ch7_bias} we simulate
the bias pulsing experiment of Refs.~\cite{Fuj03,Fuj04b}. Finally,
in Section~\ref{sec:ch7_conclusions} we present our conclusions.

\section{Model System}
\label{sec:ch7_model}

\begin{figure}
\begin{center}
\includegraphics[width=10.0cm]{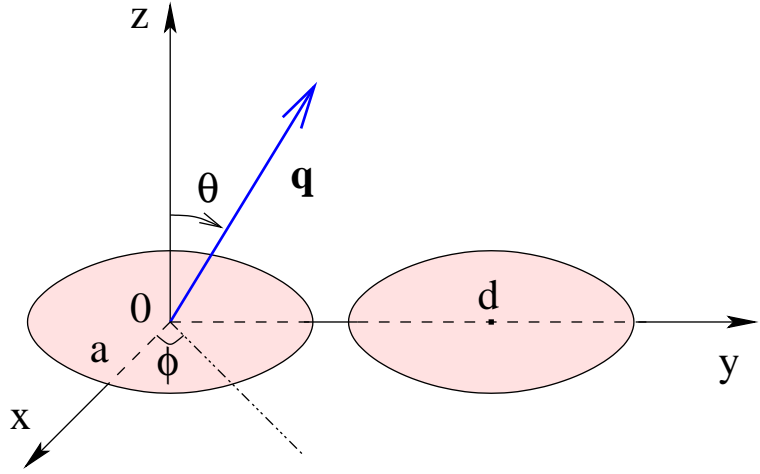}
\caption{Geometry of the double quantum dot charge qubit.}
\label{fig:ch7_geometry}
\end{center}
\end{figure}

We begin by assuming that the DQD is isolated from the leads. 
The DQD and the phonon bath combined can then be described 
by the total Hamiltonian~\cite{Bra02}
\begin{equation}
\label{eq:ch7_hamilton}
H = H_{S} + H_{B} + H_{SB},
\end{equation}
where $H_S$ and $H_B$ are individual DQD and phonon Hamiltonians,
respectively, and $H_{SB}$ is the electron-phonon interaction. We
assume that gate voltages are tuned to bring the system near the
degeneracy point $A$ (Fig.~\ref{fig:ch7_diamonds}) where a single 
electron may move between the two dots with little charging energy 
cost. To simplify the presentation, only one quantum level on each 
dot is included; $E_{1(2)}$ denotes the energy of an excess electron 
on the left (right) QD (possibly including some charging energy). 
Likewise, spin effects are neglected.\footnote{In 
the situation we envision, the total number of electrons 
would be odd: The ground state of each dot in the absence of
the excess electron would have $S \!=\! 0$, so that the dot with the
excess electron would have spin half. Other situations are, of course,
possible, such as the total number of electrons being even with $S=0$
for both dots when the qubit is in state $\left| 1\right>$ and $S
\!=\! 1/2$ for both in state $\left| 2\right>$; in the latter case,
there would be a singlet and triplet state of the DQD with a small
exchange splitting. Such complications do not effect the underlying
physics that we discuss, and so we neglect them.} 
Thus, in the basis $\{\left| 1\right>$, $\left| 2 \right> \}$, 
the DQD Hamiltonian reads
\begin{equation}
\label{eq:ch7_hs}
H_{S} = \frac{\varepsilon (t)}{2}\, \sigma_{z} + v(t)\, \sigma_{x},
\end{equation}
where $\sigma_{z,x}$ are Pauli matrices, $\varepsilon (t) \!=\!  E_1
\!-\! E_2$ is the energy level difference, and $v(t)$ is the tunneling
amplitude connecting the dots. Notice that both $\varepsilon$ and $v$
may be time dependent. The phonon bath Hamiltonian has the usual form
($\hbar = 1$)
\begin{equation}
\label{eq:ch7_hb}
H_{B} = \sum_{\bf q} \omega_{\bf q}\, b^{\dagger}_{\bf q}b_{\bf q},
\end{equation}
where the dispersion relation $\omega_{\bf q}$ is specified below. The
electron-phonon interaction has the linear coupling
form~\cite{Bra02,Bra99},
\begin{equation}
\label{eq:ch7_hsb}
H_{SB} = \sum_{\bf q} \sum_{i=1}^2 \alpha_{\bf q} ^{(i)}\, N_i\,
\left( b_{\bf q}^\dagger + b_{-{\bf q}} \right),
\end{equation}
where $N_i$ is the number of excess electrons in the $i$-th dot and
$\alpha_{\bf q}^{(i)} = \lambda_{\bf q}\, e^{-i {\bf q} \cdot {\bf
R}_i}\, P_i ({\bf q})$, with ${\bf R}_1 \!=\! 0$ and ${\bf R}_2 \!=\!
{\bf d}$ the dot position vectors, see Fig.~\ref{fig:ch7_geometry}. The
dependence of the coupling constant $\lambda_{\bf q}$ on the material
parameters and on the wave vector ${\bf q}$ will be specified
below. The dot form factor is
%
%
%
%
\begin{equation}
P_i({\bf q}) = \int d^3r\, n_i({\bf r})\, e^{-i {\bf q} \cdot {\bf r}},
\end{equation}
where $n_i({\bf r})$ is the excess charge density in the $i$-th dot. 
With no significant loss of generality, we will assume that the form 
factor is identical for both dots and, therefore, drop the $i$ index 
hereafter. In the basis $\{ \left| 1 \right>$, $\left| 2 \right> \}$, 
after dropping irrelevant constant terms, the electron-phonon 
interaction simplifies to
\begin{equation}
H_{SB} = K\, \Phi,
\label{eq:ch7_hsbnew}
\end{equation}
where
\begin{equation}
\label{eq:ch7_KandPhi}
K = \frac{1}{2}\, \sigma_{z} \qquad \mbox{and} \qquad
\Phi = \sum_{\bf q} g_{\bf q}\, \left( b^{\dagger}_{\bf q} + 
b_{- \bf q} \right),
\end{equation}
with $g_{\bf q} \!=\! \lambda_{\bf q}\, P({\bf q})\, \left( 1- e^{-i
{\bf q} \cdot {\bf d}} \right)$. The phonons propagate in three
dimensions, while the electrons are confined to the plane of the
underlying two-dimensional electron gas (2DEG). Notice that the
electron-pho\-non coupling is not isotropic for the DQD
(Fig.~\ref{fig:ch7_geometry}): Phonons propagating along $\phi \!=\! 0$
and any $\theta$ do not cause any relaxation, while coupling is
maximal along $\phi \!=\! \theta \!=\! \pi/2$ direction. We neglect 
any mismatch in phonon velocities at the GaAs/AlGaAs interface, 
where the 2DEG is located.

We now proceed with the Born-Markov-Redfield treatment 
of this system, see Chapter~\ref{ch:ch6}. While 
the Born approximation is clearly justified for weak electron-phonon
interaction, the Markov approximation requires, in addition, that the
bath correlation time is the smallest time scale in the problem. These
conditions are reasonably satisfied for lateral GaAs quantum dots, as
we will argue below.

Let us assume that the system and the phonon bath are disentangled
at $t=0$. Using Eqs.~(\ref{eq:ch7_hs}), (\ref{eq:ch7_hb}), 
and (\ref{eq:ch7_hsbnew}), 
we can write the Redfield equation for the reduced density matrix 
$\rho(t)$ of the DQD (see Chapter~\ref{ch:ch6}),
\begin{equation}
\label{eq:ch7_sigmadot}
{\dot \rho (t)} = -i\left[ H_{S}(t), \rho (t)\right] +
\left\{ \left[\Lambda (t)\rho (t), K\right] + \mbox{H.c.} \right\} \;.
\end{equation}
The first term on the right-hand side yields the Liouvillian evolution
and the other terms yield the relaxation caused by the phonon
bath. The auxiliary matrix $\Lambda$ is defined as
\begin{equation}
\label{eq:ch7_defoflambda}
\Lambda (t) = \int_{0}^{\infty} d\tau\, B(\tau )\, e^{-i\tau
H_{S}(t)}\, K\, e^{i\tau H_{S}(t)}
\end{equation}
where $B(\tau ) \!=\! {\rm Tr}_{b} \{\Phi (\tau )\Phi (0) f(H_{B})\}$
is the bath correlation function, $\Phi (\tau ) \!=\! e^{iH_{B}\tau}\,
\Phi\, e^{-iH_{B}\tau}$, and $f(H_{B}) \!=\! e^{-\beta H_{B}}/{\rm
Tr}_{b}\{ e^{-\beta H_{B}} \}$, with $\beta \!=\! 1/T$ the inverse
lattice temperature ($k_B = 1$).

Using Eq.~(\ref{eq:ch7_hb}) in the definition of the bath correlation
function, we find that the latter can be expressed in the form
\begin{equation}
\label{eq:ch7_bath}
B(\tau ) = \int_{0}^{\infty} \!\! d\omega\, \nu (\omega )\, \{
e^{i\tau\omega} n_{B}(\omega ) + e^{-i\tau\omega} [1+n_{B}(\omega
)]\},
\end{equation}
where $n_{B}(\omega )$ is the Bose-Einstein distribution function and
\begin{equation}
\nu (\omega ) = \sum_{\bf q} |g_{\bf q}|^{2}\, \delta (\omega
-\omega_{\bf q})
\end{equation}
is the spectral density of the phonon bath. 

We now specialize to linear, isotropic acoustic phonons: $\omega_{\bf
q} \!=\! s |{\bf q}|$, where $s$ is the phonon velocity. Moreover, 
we only consider coupling to longitudinal piezoelectric phonons,
neglecting the deformation potential contribution. For bulk GaAs, 
this is justifiable at temperatures below approximately
10~$K$ (see Appendix~\ref{ap:phonons}). Thus,
%
\begin{equation}
\label{eq:ch7_piezo}
|\lambda_{\bf q}|^2 = \frac{g_{\rm ph}\, \pi^2 s^2}{\Omega |{\bf q}|}, 
\end{equation}
where $g_{\rm ph}$ is the piezoelectric constant in dimensionless form
($g_{\rm ph} \!\approx\!  0.05$ for GaAs~\cite{Bra99,Bru93}) and
$\Omega$ is the unit cell volume.

The excess charge distribution in the dots is assumed Gaussian:
\begin{equation}
\label{eq:ch7_chargedensity}
n({\bf r}) = \delta (z)\, \frac{1}{2\pi a^{2}}\exp \left(
-\frac{x^2+y^2}{2a^2}\right).
\end{equation}
This is certainly a good approximation for small dots with few
electrons, but becomes less accurate for large dots. The resulting
form factor reads
\begin{equation}
\label{eq:ch7_formfactor}
P({\bf q}) = e^{-(q_x^2 + q_y^2)a^2/2}.
\end{equation}
Note that this expression differs from that 
in Refs.~\cite{Fed04,Fed00,Wu04}
where a three-dimensional Gaussian charge
density was assumed.

Using Eqs.~(\ref{eq:ch7_piezo}) and (\ref{eq:ch7_formfactor}), 
as well as the DQD geometry of Fig.~\ref{fig:ch7_geometry}, 
we get
\begin{eqnarray}
\label{eq:ch7_specdensity}
\nu (\omega ) = g_{\rm ph} \omega \int_{0}^{\pi /2} d\theta
\sin\theta\exp \left(
-\frac{\omega^{2}a^{2}}{s^{2}}\sin^{2}\theta\right) 
\left[ 1 - J_{0}\left(\frac{\omega
d}{s}\sin\theta\right)\right].
\end{eqnarray}
It is instructive to inspect the asymptotic limits of this
equation. At low frequencies, $\nu (\omega \rightarrow 0) \!\approx\!
g_{\rm ph}\, d^2\, \omega^{3}/6s^2$; thus, the phonon bath is
superohmic. At high frequencies,
\begin{eqnarray}
\label{eq:ch7_omegainfty}
\nu (\omega \rightarrow \infty) \approx \frac{g_{\rm ph}\, s^2}
{a^2\omega} f\left(\frac{d}{a}\right),
\end{eqnarray}
where
\begin{eqnarray}
f \left( \frac{d}{a} \right) = \int_{0}^{\infty} dx~x~e^{-x^2}\left[ 1
- J_{0}\left(\frac{d}{a}x\right)\right].
\end{eqnarray}
Notice that the spectral function does not have the exponential decay
familiar from the spin-boson model, but rather falls off much more
slowly: $\nu (\omega \!\rightarrow\! \infty ) \!\propto\!
\omega^{-1}$. This should be contrasted with the phenomenological
expressions used in Ref.~\cite{Bra02}. 

The characteristic frequency of the maximum in 
$\nu (\omega )$ is $\tau_c^{-1} = s/a$.
For typical experimental setups, $a \!\approx\!  50$ nm 
while $s \!\approx\!  5 \times 10^{3}$ m/s for GaAs, yielding 
$\tau_c \!\approx\!  10$ ps ($\tau_c^{-1} \!\approx\! 65\,\mu$eV). 
Thus, the Markovian approximation can be justified for time scales 
$t > \tau_c$ and if all pulse operations are kept adiabatic 
on the scale of $\tau_c$.

\section{Decay of Charge Oscillations}
\label{sec:ch7_tunnel}

One can operate this charge qubit in two different ways: (i) by
pulsing the tunneling amplitude $v(t)$ keeping $\varepsilon$ constant,
or (ii) by changing the energy level difference $\varepsilon (t)$
keeping $v$ constant (bias pulsing). Tunnel pulsing seems advantageous
as it implies fewer decoherence channels and less leakage. However, a
recent experiment used a bias pulsing scheme~\cite{Fuj03,Fuj04b}.

Our system's Hilbert space is two-dimensional by construction [see
Eq.~(\ref{eq:ch7_hs})], hence there is no leakage to states outside the
computational basis. We can, therefore, use square pulses instead of
smooth, adiabatic ones. This not only allows us to analytically solve
for the time evolution of the reduced density matrix,
Eq.~(\ref{eq:ch7_sigmadot}), but also renders our results applicable to
both tunnel and bias pulsing. Indeed, in both regimes one has
$\varepsilon (t) \!=\! 0$ and $v(t) \!=\! v_{m}$ for $t>0$, taking
that the pulse starts at $t=0$. Let us assume that the excess electron
is initially in the left dot: $\rho_{11}(0) \!=\! 1$ and $\rho_{12}(0)
\!=\! 0$. In this case, since the coefficients on the right-hand side
of (\ref{eq:ch7_sigmadot}) are all constants at $t>0$, we can solve the
Redfield equation exactly (see Appendix~\ref{ap:ch7} for details). 
As $\rho (t)$ has only three real independent components, the solution is
\begin{eqnarray}
\label{eq:ch7_sol1}
\rho_{11}(t) & = & \frac{1}{2}+\frac{1}{2} e^{-\frac{\gamma_{1}}{2}t}
(\cos\omega t +\frac{\gamma_{1}}{2\omega}\sin\omega t),
\\[0.05in]
\label{eq:ch7_sol2}
\mbox{Re}\, \rho_{12}(t) & = & -\frac{1}{2}(1-e^{-\gamma_{1}t})
\tanh\frac{v_{m}}{T},
\\[0.05in]
\label{eq:ch7_sol3}
\mbox{Im}\, \rho_{12}(t) & = & \frac{2v_{m} +
\gamma_{2}}{2\omega}\, e^{-\frac{\gamma_{1}}{2}t} \sin\omega t,
\end{eqnarray}
where
\begin{eqnarray}
\label{eq:ch7_omega}
\omega & = & \left[ 4v_{m}\left( v_{m}+\frac{\gamma_2}{2} \right)
-\frac{\gamma_1^2}{4} \right]^{1/2},
\\
\label{eq:ch7_gamma1}
\gamma_{1} & = & \frac{\pi}{2} \,\nu (2v_{m}) \coth \frac{v_{m}}{T},
\\
\label{eq:ch7_gamma2}
\gamma_{2} & = & -\,\dashint_{0}^{\infty}\frac{dy}{y^{2}-1}
\nu (2v_{m}y) \coth \frac{v_{m}y}{T}.
\end{eqnarray}
Note that $\gamma_{1,2} \!\ll\! v_{m}$.
We extract the customary energy and phase relaxation times, $T_1$ 
and $T_2$, by rotating to the energy eigenbasis
$\{ \left| - \right>$, $\left| + \right> \}$:
\begin{eqnarray}
\rho_{--}(t) & = & \frac{1}{2} - \mbox{Re}\, \rho_{12}(t),
\\
\rho_{-+}(t) & = & -\frac{1}{2} + \rho_{11}(t) + i\, \mbox{Im}\, \rho_{12}(t).
\end{eqnarray}
Then, the damping of the
oscillations in the diagonal matrix elements is the signature of
energy relaxation, while the phonon-induced decoherence is seen in the
exponential decay of the off-diagonal elements. For the DQD, we find
$T_1 \!=\! \gamma_1^{-1}$ and $T_2 \!=\! 2\gamma_1^{-1}$ for the
decoherence time.

The quality factor of the charge oscillations in Eq.~(\ref{eq:ch7_sol1})
is $Q = \omega/ \pi\gamma_{1}$. Using Eqs.~(\ref{eq:ch7_omega}),
(\ref{eq:ch7_gamma1}), and (\ref{eq:ch7_specdensity}), we find that
%
\begin{equation}
\label{eq:ch7_qfactor}
Q \approx \frac{4\tanh (v_{m}/T)} {\pi^{2}g_{\rm ph}} \,
\left\{ \int_{0}^{1} \frac{dx} {\sqrt{1-x}}\, 
e^{- (v_{m}/\omega_a)^2\, x}
\left[ 1 - J_{0} \left( \frac{d}{a} 
\frac{v_{m}}{\omega_a} \sqrt{x} \right) \right] \right\}^{-1},
\end{equation}
%
where $\omega_a = s/2a$. The $Q$-factor depends on the tunneling
amplitude $v_{m}$, lattice temperature $T$, dot radius $a$, and
interdot distance $d$.

\begin{figure}
\begin{center}
\includegraphics[width=11.0cm]{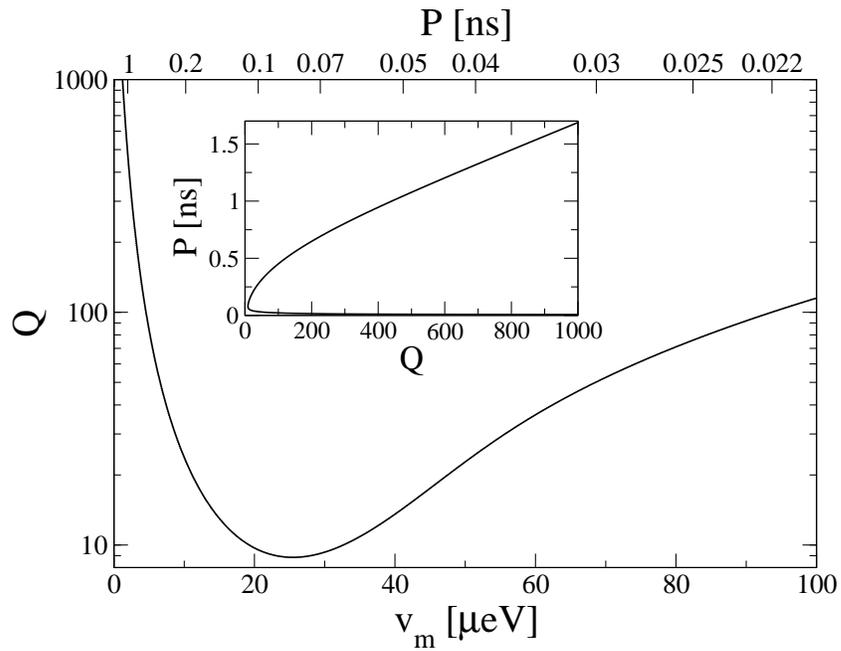}
\caption{The charge oscillation $Q$-factor as a function of the
tunneling amplitude $v_{m}$ (lower scale) and of the oscillation
period $P$ (upper scale) for a GaAs double quantum dot system. The
lattice temperature is $15$ mK and the dot radius and interdot
distance are 60 nm and 180 nm, respectively. The inset shows the
relation between $P$ and $Q$ at small tunneling amplitudes (large
periods).}
\label{fig:ch7_qfactor-Tc}
\end{center}
\end{figure}

Several experimental realizations of DQD systems recently appeared in
the literature \cite{Fuj03,Fuj04b,Pet04,Jeo01,Che04}. In principle,
all these setups could be driven by tunnel pulsing to manipulate
charge and perform single-qubit operations. To understand how the
$Q$-factor depends on the tunneling amplitude $v_{m}$ in realistic
conditions, let us consider the DQD setup of Jeong and
coworkers~\cite{Jeo01}. In their device, each dot holds about 40
electrons and has a lithographic diameter of 180~nm. The effective
radius $a$ is estimated to be around 60~nm based on the device
electron density. Therefore, $d/a \!\approx\!  3$. The lattice base
temperature is 15 mK. Introducing these parameters into
Eq.~(\ref{eq:ch7_qfactor}), one can plot $Q$-factor as a function of
$v_{m}$ or, equivalently, as a function of the period of the charge
oscillations $P = 2\pi /\omega \!\approx\! \pi /v_{m}$. This is shown
in Fig.~\ref{fig:ch7_qfactor-Tc}.

To stay in the tunnel regime $v_{m}$ should be smaller than the mean
level spacing of each QD, approximately 400 $\mu$eV in the
experiment~\cite{Jeo01}. Therefore, in Fig.~\ref{fig:ch7_qfactor-Tc} we
only show the curve for $v_{m}$ up to 100~$\mu$eV. One has to recall
that at these values the Markov approximation used in the Redfield
formulation is not accurate (see end of Section~\ref{sec:ch7_model}), 
and so our results are only an estimate for $Q$. 
For strong tunneling amplitudes, when $25\, \mu{\rm
eV}<v_{m}<100\, \mu{\rm eV}$, the largest value we find for $Q$ is
close to 100. For weak tunneling with $v_{m}<25\, \mu{\rm eV}$, the
situation is more favorable and larger quality factors (thus
relatively less decoherence) can be achieved. Nevertheless, the
one-qubit operation time, which is proportional to the period, grows
linearly with $Q$ in the region of $v_{m} \to 0$, as shown in the
inset of Fig.~\ref{fig:ch7_qfactor-Tc}. Therefore, at a certain point
other decoherence mechanisms are going to impose an upper bound on
$Q$.

\begin{figure}
\begin{center}
\includegraphics[width=11.0cm]{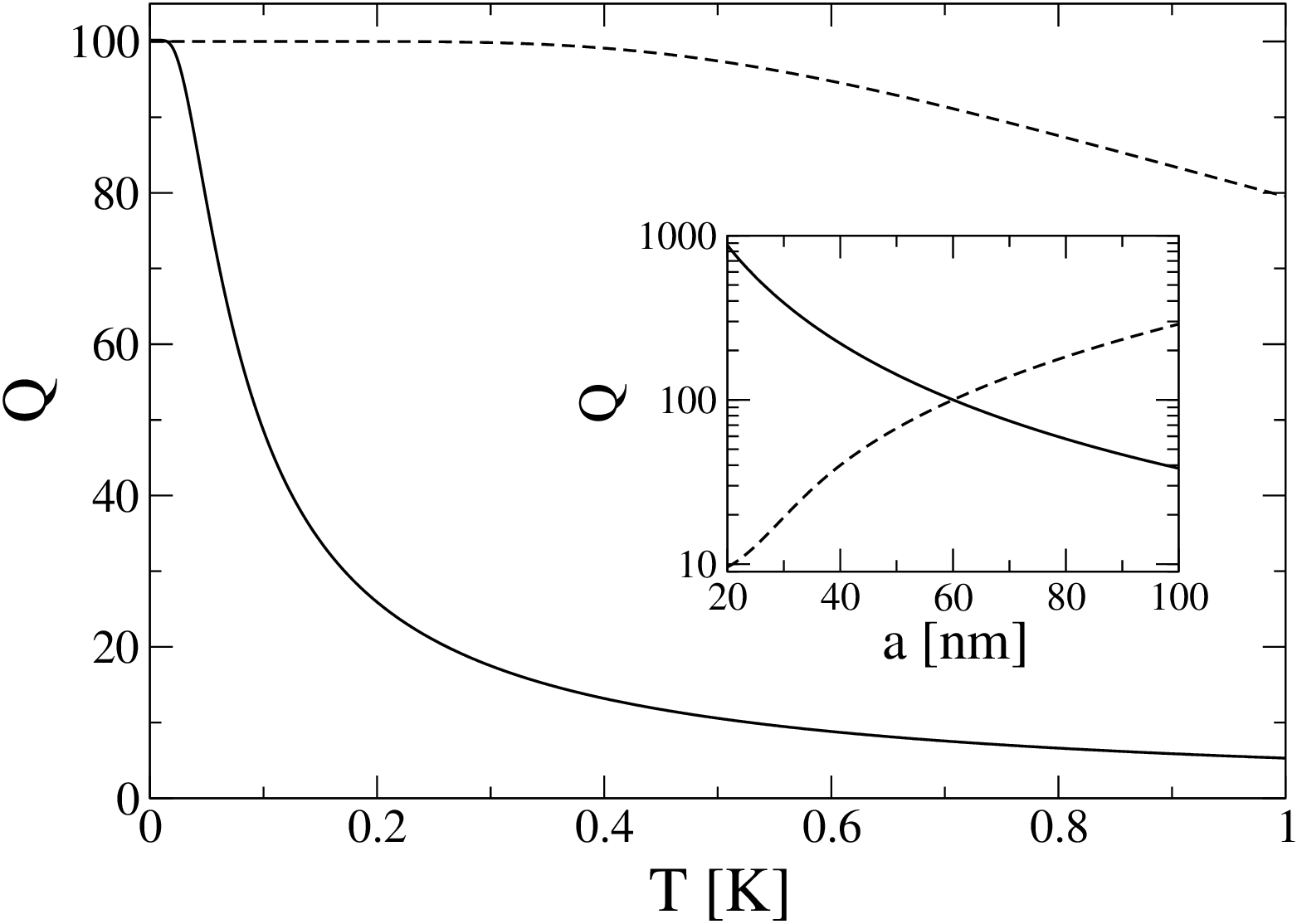}
\caption{The charge oscillation $Q$-factor as a function of the
lattice temperature. Inset: as a function of the dot radius for a
fixed ratio $d/a=3$. The solid (dashed) line corresponds to the weak
$v_{m} \!\simeq\! 53$~mK (strong $v_{m} \!\simeq\! 1.1$~K) tunneling
regime. Other parameter values are equal to those in
Fig.~\ref{fig:ch7_qfactor-Tc}.}
\label{fig:ch7_qfactor-T}
\end{center}
\end{figure}

The minimum of $Q$ in Fig.~\ref{fig:ch7_qfactor-Tc} occurs when $v_{m}$
coincides with the frequency at which the phonon spectral density is
maximum. It corresponds to the energy splitting between bonding and
anti-bonding states of the DQD, $2v_{m}$, being approximately equal to
the frequency of the strongest phonon mode $s/a$: $v_{m} \simeq
\omega_{a}$.

From Fig.~\ref{fig:ch7_qfactor-Tc}, it is evident that one can reach
certain values for the $Q$-factor (say, $Q = 100$) at both weak
($v_{m}\simeq 4.6~\mu$eV $\simeq 53$~mK) and strong ($v_{m}\simeq
93~\mu$eV $\simeq 1.1$~K) tunneling. However, these two regimes are
not equally convenient. From Eq.~(\ref{eq:ch7_qfactor}), it is clear that
the temperature dependence of the $Q$-factor is fully determined by
the bonding-antibonding splitting energy $2v_{m}$: $Q(T) \!=\!
Q(0)\tanh (v_{m}/T)$. We notice that $Q(T) \!\approx\!  Q(0)$ if $T\ll
v_{m}$; therefore, the $Q$-factor is less susceptible to temperature
variations for strong tunneling (Fig.~\ref{fig:ch7_qfactor-T}). Another
parameter that influences the $Q$-factor is the dot radius, which
controls the frequency of the strongest phonon mode, $s/a$. In the
strong tunneling regime (dashed curve in the inset to
Fig.~\ref{fig:ch7_qfactor-T}), one has to increase the QD size to improve
the $Q$-factor. This would reduce the energy level spacing, hence only
moderate improvement in $Q$-factor is possible. In contrast, in the
weak tunneling regime (solid curve in the inset to
Fig.~\ref{fig:ch7_qfactor-T}) one has to reduce the QD size. This can lead
to a significant (up to one order of magnitude) $Q$-factor
improvement.

\section{Bias Pulsing}
\label{sec:ch7_bias}

In a recent experiment~\cite{Fuj03,Fuj04b}, Hayashi and coworkers studied
charge oscillations in a bias-pulsed DQD. In this regime the energy
difference between the left and right-dot single-particle energy
levels is a function of time: $\varepsilon (t) = \varepsilon_0\,
u(t)$. A typical profile used for pulsing is
\begin{eqnarray}
\label{eq:ch7_biaspulse}
u(t) = 1 - \frac{1}{2} \left( \tanh\frac{t+W/2}{2\tau}
-\tanh\frac{t-W/2}{2\tau} \right),
\end{eqnarray}
where $W$ represents the pulse width and $\tau$ controls the rise and
drop times. During bias pulsing, the tunneling amplitude is kept
constant. In Refs.~\cite{Fuj03,Fuj04b}, the difference in energy
levels was induced by applying a bias voltage between left and right
leads (and not by gating the dots separately). For their setup, the
maximum level splitting amplitude was $\varepsilon_0 \!\approx\!  30\,
\mu{\rm eV}$ and $\tau \!\approx\!  15\,{\rm ps}$, corresponding to an
effective ramping time of about 100~ps.\footnote{The value of 
$\tau \!\approx\!  15\,{\rm ps}$ is obtained by fitting 
Eq.~(\ref{eq:ch7_biaspulse}) to the experimental pulse with 
an effective ramping time of $100\,{\rm ps}$, Ref.~\cite{Fuj04a}.} 
The tunneling amplitude was kept constant and estimated as $v \!\approx\!
5\,\mu{\rm eV}$, which amounts to charge oscillations with period $P
\!\approx\!  1\,{\rm ns}$.  The lattice temperature was $20\,{\rm
mK}$. Each quantum dot contained about 25 electrons and the effective
dot radius is estimated to be around $50\,{\rm nm}$ based on the
device electron density. From the electron micrograph of the device
one finds $d \!\approx\!  225\,{\rm nm}$, hence $d/a \!\approx\!
4.5$. When substituting these values into Eq.~(\ref{eq:ch7_qfactor}), one
finds $Q \!\approx\!  54$.

However, from the experimental data one observes $Q \!\approx\! 3$.
Low $Q$-factors were also obtained by Petta and coworkers in an
experiment where coherent charge oscillations in a DQD were detected
upon exciting the system with microwave radiation~\cite{Pet04}. Other
mechanisms of decoherence do exist in these systems, such as
background charge fluctuations~\cite{Ita03} and electromagnetic
noise emerging from the gate voltages. Our results combined with the
recent experiments indicate that these other mechanisms are more
relevant than phonons.

We now turn to yet another possible source of decoherence: Leakage to
the leads when the pulse is on.\footnote{For another source of 
decoherence due to coupling to the leads, see Ref.~\cite{Har04}.} 
To illustrate this alternative source of damping of charge oscillations, 
we simulate the bias-pulsing experiment of Refs.~\cite{Fuj03,Fuj04b} 
by implementing a rate equation formalism similar to that used in
Ref.~\cite{Fuj04a}. The formalism is based on a transport
theory put forward for the strongly biased
limit~\cite{Naz93,Gur96}. First, we find the stationary current
$I_0$ through the DQD structure when the pulse is off (that is, the
bias is applied)~\cite{Gur96}:
\begin{eqnarray}
\label{eq:ch7_i0}
I_{0} = e~\frac{\Gamma_L \Gamma_R}{\Gamma_L + \Gamma_R}
~\frac{v^{2}}{v^2 +\frac{\Gamma_L\Gamma_R}{4} +
\frac{\varepsilon_0^2\Gamma_L\Gamma_R}{(\Gamma_L +\Gamma_R)^2}},
\end{eqnarray}
where $e$ is the elementary charge. $\Gamma_{L(R)}$ is the partial
width of the energy level in the left (right) dot due to coupling to
the left (right) lead (when the bias is applied); in the
experiment~\cite{Fuj03,Fuj04b}, $\Gamma_{L,R} \!\approx\!  30\,\mu{\rm
eV}$. On the other hand, when the pulse is on, the stationary current
is zero. We now apply the pulse $\varepsilon (t)$ and measure the
current $I(t)$. In the experiments, the level widths $\Gamma_{L,R}$
decrease upon biasing the system. To include that effect here, we also
pulse them: $\Gamma_L(t) \!=\!  \gamma_L + (\Gamma_L - \gamma_L) u(t)$
and analogously for $\Gamma_R$, where $\gamma_{L(R)}$ is the residual
leakage to the left (right) lead when the pulse is on. We use
$\gamma_{L,R} \!=\!  0.3\,\mu{\rm eV}$, even though the real leakage
in the experiment was likely much smaller. To obtain the response
current one subtracts the stationary component: $I_{\rm resp}(t) \!=\!
I(t) - I_0 u(t)$.

\begin{figure}
\begin{center}
\includegraphics[width=11.0cm]{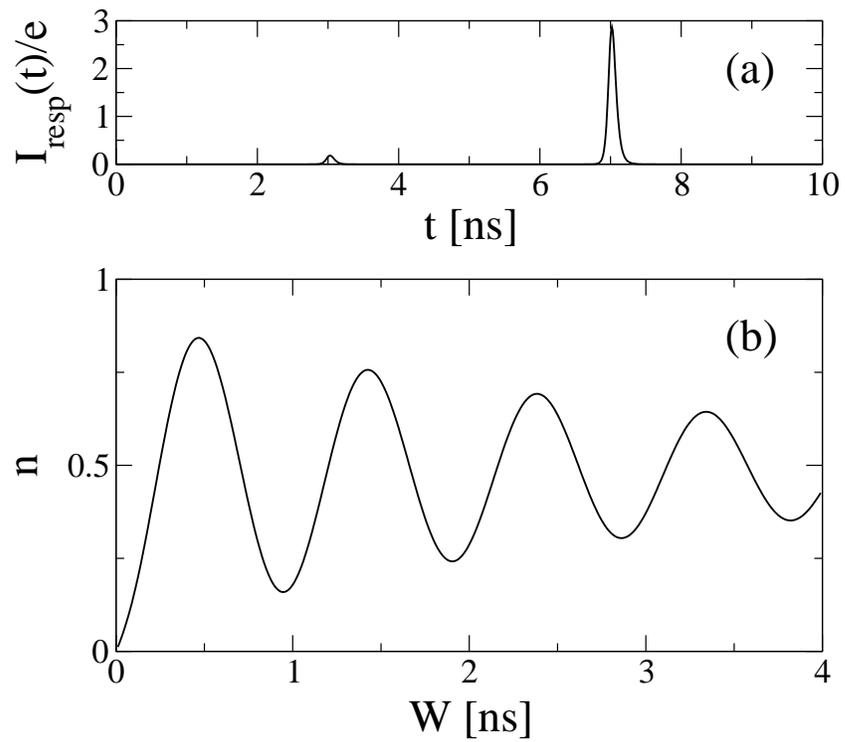}
\caption{(a) The response current $I_{\rm resp}(t)/e$ in ns$^{-1}$ as
a function of time for a pulse with $W \!=\! 4$ ns and $\tau \!=\!
30$ ps. (b) Number of electrons transfered between left and right
leads, as defined in Eq.~(\ref{eq:ch7_ncycle}), as a function of the pulse
width $W$.}
\label{fig:ch7_respcurrent}
\end{center}
\end{figure}

Figure~\ref{fig:ch7_respcurrent}(a) shows the response current for a pulse
of width $W \!=\! 4$~ns and $\tau \!=\! 30$~ps. The latter is
approximately twice as large as in the experiment and is chosen to
enhance the effect. In Refs.~\cite{Fuj03,Fuj04b}, pulses were
applied at a frequency $f \!=\! 100$~MHz. The average number of
electrons transfered from the left to the right lead per cycle minus
that in the stationary regime is~\cite{Fuj04a}
\begin{eqnarray}
\label{eq:ch7_ncycle}
n = \int_0^{1/f}\!dt\; I_{\rm resp}(t)/e.
\end{eqnarray}
(In the simulations there is no need to apply a sequence of pulses.)
Notice that $n$ oscillates as a function of the pulse width $W$ [see
Fig.~\ref{fig:ch7_respcurrent}(b)] as observed in the experiment. 
Two main conclusions can be drawn from our simulation. First, 
the larger $\tau$, the smaller the visibility of the charge
oscillations~\cite{Fuj04a}. Second, the larger the leakage rates
$\gamma_{L,R}$ when the pulse is on, the stronger the damping of the
oscillations. While the damping due to leakage is presumably too weak
an effect to discern in the data presented in
Refs.~\cite{Fuj03,Fuj04b}, the loss of visibility due to finite $\tau$
is likely one of the causes of the small amplitude seen
experimentally.

\section{Conclusions}
\label{sec:ch7_conclusions}

The main conclusion of the chapter is that, under realistic conditions,
phonon decoherence is one to two orders of magnitude weaker than
expected \cite{Bra02,Fed04,Fed00,Wu04}. The analytical expression for
the $Q$-factor given in Eq.~(\ref{eq:ch7_qfactor}) was found using an
expression for the phonon spectral density,
Eq.~(\ref{eq:ch7_specdensity}), which takes into account important
information concerning the geometry of the double quantum dot
system. In a previous work~\cite{Bra02} an approximate,
phenomenological expression, $\nu (\omega )\propto\omega\exp
(-\omega /\omega_{c})$, was utilized in the treatment of charge 
qubits. There is a striking difference between these two expressions
in both the high- and low-frequency limits. Moreover, an arbitrary 
coupling constant was adopted in Ref.~\cite{Bra02} 
to model the electron-phonon interaction while our treatment uses 
a value known to describe the most relevant phonon coupling in GaAs. 
On the other hand, other previous work~\cite{Fed04,Fed00,Wu04}
assumed a spherically symmetric excess charge distribution 
in the dot while we have assumed a two-dimensional pancake form.
These differences account for most of the discrepancy 
between the present and previous results.

Based on these findings we conclude that phonon decoherence is too
weak to explain the damping of the charge oscillations seen in recent
experiments~\cite{Fuj03,Fuj04b,Pet04}. Charge leakage to the leads
during bias pulsing is an additional source of damping, as shown in
Fig.~\ref{fig:ch7_respcurrent}(b); however, for realistic
parameters~\cite{Fuj03,Fuj04b,Fuj04a}, it turns out to be a weak
effect as well. Hence, other decoherence mechanisms, such as
background charge fluctuations or noise in the gate voltages, play the
dominant role~\cite{Ita03}.

There are two distinct ways to operate a double quantum dot charge
qubit: (i) by tunnel pulsing or (ii) by bias pulsing. Tunnel pulsing
seems advantageous due to the smaller number of possible decoherence
channels. In addition, the bias pulsing scheme, in contrast to tunnel
pulsing, introduces significant loss of visibility in the charge
oscillations.

In this chapter we did not attempt to study leakage or loss of fidelity
due to non-adiabatic pulsing, which are both important issues for 
{\it spin}-based quantum dot qubits (see Chapter~\ref{ch:ch5}). 
Moreover, we have not attempted to go beyond the Markov approximation 
when deriving an equation of motion for the reduced density matrix. 
Both of these restrictions in our treatment impose some limitations 
on the accuracy of our results, especially for large tunneling amplitudes.

Finally, it is worth mentioning that some extra insight would be
gained by measuring the $Q$-factor as a function of the tunneling
amplitude $v_{m}$ experimentally. Such a measurement would allow one
to map the spectral density of the boson modes responsible for the
decoherence. This would provide very valuable information about the
leading decoherence mechanisms in double quantum dot systems.

%% file: chapters/conclusion.tex
\setlength{\textheight}{8.0in}
\clearpage
\chapter{Conclusion}
\label{ch:conclusion}
\thispagestyle{botcenter}
\setlength{\textheight}{8.4in}

\section{General Overview}

The continuous minituarization of the integrated 
circuits is going to affect the underlying physics
of the future computers. 
This new physics first came into play as
the effect of Coulomb blockade 
in the electron transport
through the small conducting island. 
Then, as the size of the island $L$ 
continued to shrink further, 
quantum phase coherence length $L_{\phi}$
became larger than $L$ 
leading to the mesoscopic fluctuations
-- fluctuations of the island's quantum mechanical
properties upon small external perturbations.
Quantum coherence of the mesoscopic systems
is essential for building reliable quantum
computer.
Unfortunately, one can not completely 
isolate the system from the environment and 
its coupling to the environment inevitably
leads to the loss of coherence or decoherence.
All these effects are to be thoroughly investigated 
as the potential of the future applications
is enormous.

\section{Main Conclusions of My Work}

In {\bf Chapter~\ref{ch:ch2}} ~I studied Coulomb blockade 
oscillations of the linear conductance through 
a quantum dot weakly coupled to the leads 
via multichannel tunnel junctions. 
To obtain analytic results I have assumed that 
the energy levels in the QD are equally spaced.
The electron-electron interactions in the QD 
have been described by the constant interaction model,
though thermal excitations with all possible spins 
were taken into account.

Firstly, I found the expression for the linear conductance 
in the spinless case [Eq.~(\ref{eq:ch2_g0})].
It is valid at arbitrary values of $E_C$, $\delta E$ and $T$.
Then, in Section~\ref{sec:ch2_edge_states}, I applied 
the spinless case theory result to the problem of 
the transport via quantum dot in the quantum Hall regime.
The quantum dot energy levels in this case are equidistant
with the spacing given by Eq.~(\ref{eq:ch2_qhespacingrealdot}).

Linear conductance in the case of spin-$\frac{1}{2}$ 
electrons at arbitrary values of $E_C$, $\delta E$, and $T$ 
is given by Eq.~(\ref{eq:ch2_spinhalfcond}).
It is plotted in Fig.~\ref{fig:ch2_spinhalfplot}
for the charging energy of the quantum dot
equal to mean level spacing.
Eq.~(\ref{eq:ch2_spinhalfcond}) can be used as 
an experimental data fit to extract the 
charging energy and mean level spacing 
at given temperature.

Though I do not expect my quantitative results 
to precisely describe 
a quantum dot with random energy levels, 
they certainly give correct order of magnitude 
for the conductance oscillations and their generic features.

In {\bf Chapter~\ref{ch:ch3}} ~I studied corrections 
to the spacings between Coulomb blockade peaks
due to finite dot-lead tunnel couplings.
I considered 
both GUE and GOE random matrix ensembles 
of 2D quantum dots.
The electron-electron interactions in the QD 
were described by the constant exchange interaction model. 
$S = 0$, $\frac{1}{2}$, and $1$ spin states of the QD
were accounted for, thus, 
limiting the applicability of my results
to not so large exchange interaction constants: 
$J_{\rm s}<0.5\,\Delta$.
I also assumed that $T \ll \Delta \ll E_{C}$.

I calculated the ensemble averaged correction
to the Coulomb blockade peak spacing in the even valley,
Eq.~(\ref{eq:ch3_even2}).
At $J_{\rm s} = 0$ my result coincide with 
that in Ref.~\cite{Kam00}.
I found that the averaged correction 
decreases monotonically (nonetheless, staying positive) 
as the exchange interaction constant $J_{\rm s}$
is increased, 
see Figs.~\ref{fig:ch3_ave1}, \ref{fig:ch3_ave3}, and \ref{fig:ch3_goe}.
These dependences are very robust with respect to
the choice of RMT ensemble or 
change in the following parameters: 
charging energy, mean level spacing, or temperature. 
The averaged correction to the odd spacing
is of the same magnitude and opposite sign.

I calculated the rms of the correction fluctuations, 
see Eqs.~(\ref{eq:ch3_varianceref})-(\ref{eq:ch3_xi}).
It is of the same order as 
the average value of the correction,
see Figs.~\ref{fig:ch3_ave1} and \ref{fig:ch3_ave3},
and weakly depend on $J_{\rm s}$, see Fig.~\ref{fig:ch3_sigma}.
For a small subset of 
the quantum dot ensemble realizations,
at the realistic value of $J_{\rm s}=0.3\,\Delta$,
the correction is of the opposite sign,
see Figs.~\ref{fig:ch3_ave1} and \ref{fig:ch3_ave3}.
The rms of the correction fluctuations 
in the odd valley is the same as that in the even one.

In the experiment by Jeong and co-workers~\cite{Jeo}
the corrections to the even and odd peak spacings
due to finite dot-lead tunnel couplings were measured.
It was found that the even (odd) peak spacing
increases (decreases) as the tunnel couplings 
are increased. 
This is in the qualitative agreement with 
the theory, see Eq.~(\ref{eq:ch3_even2}).
The magnitude of the effect was measured
at different values of the gas parameter $r_{s}$
(and, hence $J_{\rm s}$) as well. 
Unfortunately, from these measurements
one can not draw any conclusion
about the behavior of the correction 
to the peak spacing as a function of $J_{\rm s}$.
Hence, more experiments are needed
to confirm my theory. 
To measure the ensemble averaged correction
to the peak spacing and its fluctuations 
as a function of $J_{\rm s}$ 
one should fabricate 
a set of 2D lateral quantum dots 
with different values of $r_{s}$.
Then carry out the measurements on each device
under similar conditions.

In {\bf Chapter~\ref{ch:ch5}} ~I studied the effect of 
mesoscopic fluctuations on the magnitude of errors 
that can occur in exchange operations on quantum dot
spin-qubits. 
I considered mid-size double quantum dots, 
with an odd number of electrons in the range of 
a few tens in each dot.
My results indicated two different scenarios 
for quantum dot qubit implementations.

First, if one is willing to characterize each 
quantum dot pair separately 
and have them operate one by one, 
mesoscopic fluctuations will be irrelevant. 
In this scenario, each pair of QDs will require 
a different pulse shape and duration
since QDs are not microscopically identical. 
Multi-electron QDs are tunable enough, easy to couple, 
and much easier to fabricate than one-electron dots; 
therefore, multi-electron QDs are most appropriate 
for this case.

Second, if the goal is to achieve genuine scalability, 
one has to operate qubits in a similar and uniform way, 
utilizing a single pulse source. 
In this case, pulse duration time $T$ and 
switching time $\tau$ [Eq.~(\ref{eq:ch5_pulse})] 
should be the same for all QD pairs. 
Thus, four tuning parameters per QD may be necessary 
to achieve the following goals: 
(i) find isolated, single-occupied energy level (two parameters); 
(ii) align this level with the corresponding level 
in an adjacent QD (one parameter); 
(iii) control the inter-dot coupling (one parameter). 
For parameters involved in (i) and (ii), an accuracy 
of a few percent will likely be required. 
Finally, control over the inter-dot coupling parameter, 
(iii), must allow for the application of smooth pulse shapes 
in the picosecond range. 
Although these requirements seem quite stringent, recent
experiments indicate that they could be met~\cite{Che04}.

In summary, my analysis indicate that mid-size QDs, with ten 
to a few tens of electrons, while not allowing for extremely 
fast gates, are still good candidates for spin-qubits. 
They offer the advantage of being simpler to fabricate and 
manipulate, but at the same time require accurate, simultaneous 
control of several parameters. Errors related to detuning and 
sample-to-sample fluctuations can be large, but can be kept 
a secondary concern with respect to dephasing effects
provided that a sufficient number of independent electrodes 
or tuning parameters exists.

In {\bf Chapter~\ref{ch:ch7}} ~I studied decoherence 
of a quantum dot charge qubit due to coupling to
piezoelectric acoustic phonons in the Born-Markov approximation.
I calculated the dependence of the $Q$-factor on 
lattice temperature, quantum dot size, and interdot coupling. 
The analytical expression for
the $Q$-factor given in Eq.~(\ref{eq:ch7_qfactor}) was found using an
expression for the phonon spectral density,
Eq.~(\ref{eq:ch7_specdensity}), which takes into account important
information concerning the geometry of the double quantum dot system. 
The main conclusion is that, under realistic conditions,
phonon decoherence is one to two orders of magnitude weaker than
expected \cite{Bra02,Fed04,Fed00,Wu04}. 
Based on these findings I conclude that phonon decoherence 
is too weak to explain the damping of the charge oscillations 
seen in recent experiments~\cite{Fuj03,Fuj04b,Pet04}. 
Hence, other decoherence mechanisms, such as
background charge fluctuations or noise in the gate voltages, 
play the dominant role~\cite{Ita03}.

There are two distinct ways to operate a double quantum dot charge
qubit: (i) by tunnel pulsing or (ii) by bias pulsing. Tunnel pulsing
seems advantageous due to the smaller number of possible decoherence
channels. In addition, the bias pulsing scheme, in contrast to tunnel
pulsing, introduces significant loss of visibility in the charge
oscillations.

Finally, it is worth mentioning that some extra insight would be
gained by measuring the $Q$-factor as a function of the tunneling
amplitude $v_{m}$ experimentally. Such a measurement would allow one
to map the spectral density of the boson modes responsible for the
decoherence. This would provide very valuable information about the
leading decoherence mechanisms in double quantum dot systems.

\section{Possible Directions of Future Research}

In my opinion, 
the theory of the Coulomb blockade phenomena
in the quantum dots is in very good shape. 
I believe there is no experimental result 
in the last 5 years
which comes as a surprise.
Moreover, in most instances there is 
a quantitative agreement between experimental
results and the theory.
Therefore, one can say that 
the Coulomb blockade phenomenon 
provide an important tool for analyzing
quantum mechanical properties
of the quantum dot.

The correction to the spacing between
Coulomb blockade peaks
due to dot-lead tunnel couplings
turned out to be very small -- 
both the value averaged over mesoscopic fluctuations
and its rms. 
I expect that it is going to be quite challenging 
to discern this effect in the experiment.

Mesoscopic fluctuations are proved 
to be an important limiting factor 
for implementation of the quantum dot spin-qubits. 
However, it should be possible to demonstrate
basic spin swapping between two quantum dots 
in the laboratory 
provided that enough independent electrodes 
(tuning parameters) exist.
Besides, the commercial pulse generating technique
should be significantly improved
so that shorter pulses with the well-controlled
area could be generated.
The experiments on the exchange operations 
in the quantum dot spin-qubits are 
currently on the way.
The industrial applications require 
the accuracy of $10^{-4}$ and,
therefore, are very much questionable at this point.

I find that phonon decoherence is too weak 
to explain the damping of the charge oscillations 
seen in recent experiments.
Hence, more theoretical work needs to be done 
to quantify other decoherence mechanisms. 
In my opinion, the following mechanisms 
can play the dominant role:
(i)\,background charge fluctuations and/or
(ii)\,noise in the gate voltages.
On the experimental side, 
one should measure the decoherence rate 
as a function of the tunneling amplitude. 
Such a measurement would allow one to map 
the spectral density of the boson modes 
responsible for the decoherence.
In general, more work needs to be done
to reduce charge decoherence in 
double quantum dot setups.

In conclusion, I think the field of quantum computing 
will continue to stimulate both the theoretical research 
and advancements in experimental techniques for years to come.

%% file: chapters/appendix_ch2.tex
\setlength{\textheight}{8.0in}
\clearpage
\chapter{Occupation Numbers in the Canonical Ensemble}
\label{ap:ch2}
\thispagestyle{botcenter}
\setlength{\textheight}{8.4in}


In the canonical ensemble, where the total number of 
particles, $N_e$, is fixed, it is difficult to calculate 
fermionic occupation numbers, Eq.~(\ref{eq:ch2_tough}), directly. 
This is true even in the case of equidistant energy levels
separation.

Fortunately, the bosonization technique allows one to express 
fermionic field annihilation and creation operators,
$\psi$ and $\psi^{\dagger}$,
in terms of the bosonic annihilation and creation,
$a_q$'s and $a_q^{\dagger}$'s, and ladder, $U$ and $U^{\dagger}$,
operators~\cite{Hal81}:
\begin{eqnarray}
\psi^{\dagger}(x) = \frac{1}{\sqrt{L}} e^{-i k_F x}
e^{-i \chi^{\dagger}(x)} U^{\dagger} e^{-i \chi (x)},
\end{eqnarray}
where
\begin{eqnarray}
\chi^{\dagger} (x) = \frac{\pi x}{L} N + i \sum_{q>0}
\sqrt{\frac{2 \pi}{q L}}~e^{-i q x} a_q^{\dagger};
\end{eqnarray}
$L$ is the length of the artificial system; $k_F$ is the Fermi wave vector; 
$N=N_{e}-N_{i}$ is the number of excess electrons operator; 
$q$ is the wave vector; 
and $x$ is the coordinate.
Since these bosons naturally exist in the grand canonical 
ensemble, one can fix total number of excess electrons, $N$,
and calculate occupation  numbers in the bosonic basis,
Eq.~(\ref{eq:ch2_dent}).

Occupation numbers, Eq.~(\ref{eq:ch2_dent}), have the following 
properties:
\begin{eqnarray}
n_j = 1 - n_{1-j},
~~~~n_j e^{j \delta / 2} = n_{-j} e^{-j \delta / 2}.
\end{eqnarray}
They are similar to $n_F(E)=1-n_F(-E)$ and 
$n_F(E)e^{\beta E/2}=n_F(-E)e^{-\beta E/2}$ 
ones of the Fermi-Dirac distribution and reflect the 
electron-hole symmetry. Combining these two properties 
one can get the recursion relation: 
\begin{eqnarray}
n_{j+1} = 1 - e^{j \delta} n_j.
\end{eqnarray}
In the limit of low temperature, $\delta E \gg T$, for $j>0$
we obtain
\begin{eqnarray}
n_j = e^{- j \delta} - e^{-(2j+1)\delta} 
+O\left[ e^{-(3j+3)\delta}\right];
\end{eqnarray}
in the high temperature limit, $T\gg\delta E$:
\begin{eqnarray}
n_j=\frac{1}{e^{\left( j-\frac{1}{2}\right)\delta}+1} 
-\frac{\delta}{8}
\frac{\sinh\left[\left( j-\frac{1}{2}\right)\frac{\delta}{2}\right]}
{\cosh^3 \left[\left( j-\frac{1}{2}\right)\frac{\delta}{2}\right]}
+O\left(\delta^2\right);
\end{eqnarray}
thus, we find first correction to the Fermi-Dirac distribution.
The form of the correction remains valid even for the slightly
non-equidistant energy levels in the dot. 
If this is the case we need to define ratio $\delta$
in the last expression
as $\delta =\overline{\delta E}/T$, where 
$\overline{\delta E}$ is the mean level spacing.

%% file: chapters/appendix_phonons.tex

\setlength{\textheight}{8.0in}
\clearpage
\chapter{On the Electron-Phonon Coupling in GaAs quantum dots}
\label{ap:phonons}
\thispagestyle{botcenter}
\setlength{\textheight}{8.4in}

The material in this appendix is largely based on Ref.~\cite{Muc04}.

\section{Introduction}

There are two types of phonon modes in GaAs/AlGaAs heterostructures
that we need to worry about: deformation potential and piezoelectric.
The former is always present in all semiconductor materials, while the
latter is due to the lack of inversion symmetry in the zinc-blend
lattice structure of GaAs. Optical phonons do not usually couple
strongly to electrons at low temperatures (with some exceptions, like
in the Peierls stability problem). Acoustic phonons, particularly
longitudinal ones, do couple more effectively to electrons at low
temperatures (transversal phonons do not usually couple to electrons
in the absence of Umklapp processes).

\section{Deformation Potential Contribution 
to the \\ Electron-Phonon Coupling}

The deformation potential only makes sense in the long wavelength,
acoustic branch limit (see Section 4.12 in Ref.~\cite{Mad95} 
or Section 6.14 in Ref.~\cite{Zim79}),
where the phonons are associated to compression waves of the ion
displacement field. Let ${\bf u}({\bf r},t)$ be the displacement field
in the continuum limit 
(so, the crystal is considered as a continuum elastic medium). 
Then, the relative volume change due to the compression wave
is given by $\delta V/V = \Delta ({\bf r},t) = \nabla \cdot {\bf
u}$. This volume compression causes a local change in the lattice
constant, which, in turn, leads to a local variation of the conduction
band lower edge, $E_c$:
\begin{equation}
\delta E_c = \frac{\partial E_c}{\partial V}\, \delta V = V\,
\frac{\partial E_c}{\partial V}\,\Delta = \Lambda\, \nabla \cdot 
{\bf u},
\end{equation}
where $\Lambda$ is called the deformation potential constant 
(the upper edge of the valence band would shift in the opposite
direction). Thus, for a plane wave 
${\bf u} = {\bf u}_0\, e^{i({\bf q}\cdot{\bf r} - \omega t)}$, 
we find that
\begin{equation}
\delta E_c = i\, \Lambda\, \left( {\bf u} \cdot {\bf q} \right).
\end{equation}
So, it is clear now that only longitudinal deformation potential
phonons are important. We can put that into a more convenient 
form for the electron-phonon Hamiltonian:
\begin{equation}
H_{\rm el-ph}^{\rm def} = \sum_i \Lambda\, \nabla \cdot {\bf u} 
({\bf r}_i).
\end{equation}
We can rewrite the displacement field in terms of normal modes
(phonons) and introduce bosonic creation and annihilation operators:
\begin{equation}
\label{eq:normalmodes}
{\bf u}({\bf r}) = \sum_{\bf q} Q_{\bf q}\, \hat{\bf e}_{\bf q}\, 
e^{i {\bf q} \cdot {\bf r}},
\end{equation}
with
\begin{equation}
Q_{\bf q} = \sqrt{\frac{\hbar} {2V \rho_{\rm ion} \omega_{\bf q}}}
\left( b_{-{\bf q}}^\dagger + b_{\bf q} \right),
\end{equation}
where $\rho_{\rm ion}$ is the ion mass density, $V$ is a normalization
volume, and ${\bf e}_{\bf q}$ is the polarization unit vector. Also, we
can go from first to the second quantized representation for the electrons
by recalling that, for the Bloch momentum (neglecting Umklapp processes)
\begin{equation}
\langle {\bf k} | e^{i {\bf q} \cdot {\bf r}} | {\bf k}^\prime \rangle
= \delta_{{\bf k},{\bf k}^\prime + {\bf q}}.
\end{equation}
As a result,
\begin{equation}
H_{\rm el-ph}^{\rm def} = \sum_{{\bf k},{\bf q}} M^{\rm def}_{{\bf
k},{\bf q}} \, \left( b_{-{\bf q}}^\dagger + b_{\bf q} \right)\,
c^\dagger_{{\bf k}+{\bf q}} c_{\bf k},
\end{equation}
where the electron-phonon coupling matrix element is given by
\begin{equation}
M^{\rm def}_{{\bf k},{\bf q}} = i\, \Lambda\, \sqrt{\frac{\hbar} 
{2V\rho_{\rm ion} \omega_{\bf q}}}\, \left( {\bf q} \cdot 
\hat{\bf e}_{\bf q} \right).
\end{equation}
Obviously, this is an oversimplification of the problem since the
deformation potential constant is actually a tensor in any non-cubic
crystal. But since we are only interested in the longitudinal modes
and have assumed the medium to be approximately isotropic, the
simplification should be good enough. Notice that $|M^{\rm def}|^2
\propto q$ when $\omega_{\bf q} = s\, q$. Thus the coupling between
electrons and deformation potential phonons is suppressed at long
wavelengths. This is valid, for example, at low temperatures.

\section{Piezoelectric Contribution 
to the \\ Electron-Phonon Coupling}

In III-V compounds, there is an electric polarization field 
associated with the strain (displacement) field:
\begin{equation}
P_i = \sum_{j,k=1}^3 \beta_{ijk} w_{jk},
\end{equation}
where $\beta_{ijk}$ is called the piezoelectric tensor and 
$w_{jk}$ is the strain tensor,
\begin{equation}
w_{ij} = \frac{1}{2} \left( \frac{\partial u_i}{\partial x_j} +
\frac{\partial u_j}{\partial x_i} \right).
\end{equation}
In order to find the associate electric potential energy, 
we need to solve Poisson's equation:
\begin{equation}
-\sum_{i,j=1}^3 \epsilon_{ij} \frac{\partial^2 \varphi} {\partial x_i
 \partial x_j} = - 4\pi\, \sum_{i=1}^3 P_i = - 4\pi \sum_{i,j,k=1}^3
 \beta_{ijk} \frac{\partial ^2 u_k} {\partial x_i \partial x_j}.
\end{equation}
Recalling Eq.~(\ref{eq:normalmodes}) and assuming a normal mode
decomposition of the electric potential, namely,
\begin{equation}
\varphi = \sum_{\bf q} \varphi_{\bf q}\, e^{i{\bf q} \cdot {\bf r}},
\end{equation}
we get 
\begin{equation}
\varphi_{\bf q} = 4\pi\, 
\frac{\sum_{i,j,k=1}^{3}\beta_{ijk}\, q_{i} q_{j}\, 
u_{\bf q}^k}{\sum_{i,j=1}^{3}\epsilon_{ij}\, q_{i}q_{j}}.
\end{equation}
Following the same steps as in the previous section, 
we can then write
\begin{equation}
H_{\rm el-ph}^{\rm piezo} = \sum_{{\bf k},{\bf q}} M_{{\bf k},{\bf
q}}^{\rm piezo}\, \left( b_{-{\bf q}}^\dagger + b_{\bf q} \right)\,
c_{{\bf k}+{\bf q}}^\dagger c_{\bf k},
\end{equation}
where the piezoelectric matrix element is given by
\begin{equation}
M_{{\bf k},{\bf q}}^{\rm piezo} = \frac{4\pi e}{\epsilon} \sqrt{
\frac{\hbar}{2V \rho_{\rm ion} \omega_{\bf q}}} \sum_{i,j,k=1}^3
\beta_{ijk}\, \frac{q_i q_j}{q^2}\, \left( {\bf e_{\bf q}} \right)_k,
\end{equation}
where we have assumed an isotropic dielectric function ($\epsilon_{ij}
= \delta_{ij}\, \epsilon$). Now, notice that $|M^{\rm piezo}|^2
\propto q^{-1}$ for $\omega_{\bf q} = s\, q$. Therefore, the coupling
between electrons and piezoelectric phonons gains importance at low
temperatures. Also notice that $M^{\rm piezo}$ is real
while $M^{\rm def}$ is imaginary. 
Therefore, these two electron-phonon amplitudes 
are out of phase by $\pi/2$ and do not interfere.

\section{Combining Two Mechanisms}

While the piezoelectric coupling constant is intrinsically tensorial,
we can try to use an isotropic approximation in order to estimate its
magnitude. We introduce the substitution
\begin{equation}
\frac{4\pi e}{\epsilon} \left\langle \,\sum_{i,j,k=1}^3 \beta_{ijk}\,
\frac{q_i q_j}{q^2} \right\rangle_{\bf q} = \sqrt{\Theta},
\end{equation}
where $\langle\dots\rangle_{\bf q}$ denotes the average 
over all orientations of ${\bf q}$ at fixed $q = |{\bf q}|$. 
Thus, for the longitudinal acoustic phonons, 
one can write the total electron-phonon coupling as
\begin{equation}
\left| g_{{\bf k},{\bf q}} \right|^2 = \frac{\hbar}{2V \rho_{\rm ion}
\, s\, q}\, \left( \Lambda^2\, q^2 + \Theta \right).
\end{equation}
This expression should be used with caution since for the averaged
piezoelectric contribution it actually mixes longitudinal and
transversal phonon modes. 
Bruus and coworkers~\cite{Bru93} go into some details 
when explaining how the parameter $\Theta$ 
(called the piezoelectric coupling) should actually be calculated 
in terms of phonon velocities and the dominant component 
of the piezoelectric tensor.

It is convenient to rewrite the electron-phonon coupling 
Hamiltonian in terms of the electron density operator 
in the momentum representation:
\begin{equation}
H_{\rm el-ph} = \sum_{\bf q} g_{\bf q}\, \left( b_{-{\bf q}}^\dagger +
b_{\bf q} \right)\, \rho({\bf q}),
\end{equation}
where
\begin{equation}
\rho({\bf q}) = \sum_{\bf q} c_{{\bf k}+{\bf q}}^\dagger c_{\bf k}.
\end{equation}

According to experimental data compiled by Bruus 
and coworkers~\cite{Bru93} for GaAs
$$
\Lambda \approx 2.2 \times 10^{-18}\, {\rm J}~~~~\mbox{and}~~~~
\Theta \approx 5.4 \times 10^{-20}\, {\rm J}^2 {\rm m}^{-2}.
$$
Thus, one can see that the crossover phonon wavelength, where the
piezoelectric contribution begins to dominate over the deformation
potential, is equal to
\begin{eqnarray}
\lambda_c = \frac{2\pi}{q_c} = 2\pi \frac{\Lambda}{\sqrt{\Theta}}
\approx 56\, {\rm nm}.
\end{eqnarray}
If we recall that $s \approx 5 \times 10^3$ m/s for the longitudinal sound
waves in GaAs, we find that $\lambda_c$ corresponds to a temperature
of $T_c \approx 27$ K. 

Thus, in the milli-Kelvin range, 
one can safely neglect the deformation potential contribution.

%% file: chapters/appendix_ch7.tex
\setlength{\textheight}{8.0in}
\clearpage
\chapter{Derivation of Eqs.~(\ref{eq:ch7_sol1})-(\ref{eq:ch7_sol3})}
\label{ap:ch7}
\thispagestyle{botcenter}
\setlength{\textheight}{8.4in}


For $t>0$, Eq.~(\ref{eq:ch7_hs}) is time-independent: $H_{S} = v_{m}\sigma_{x}$. 
Since the $K$ matrix is also time-independent [Eq.~(\ref{eq:ch7_KandPhi})], 
the matrix $\Lambda$ defined by Eq.~(\ref{eq:ch7_defoflambda}) is time-independent 
as well. After some straightforward operator algebra, we find that
\begin{eqnarray}
\label{eq:ch7_lambdatimeindep}
\Lambda = \frac{1}{2} \int_{0}^{\infty} d\tau \, B(\tau )\, e^{-i\tau
v_{m}\sigma_{x}}\, \sigma_{z}\, e^{i\tau v_{m}\sigma_{x}} \\
\label{eq:ch7_lambdatimeindep2}
= \frac{1}{2} \int_{0}^{\infty} d\tau \, B(\tau )
\left[ \sigma_{z} \cos (2v_{m}\tau ) - \sigma_{y} \sin (2v_{m}\tau ) \right].
\end{eqnarray}
One can rewrite Eq.~(\ref{eq:ch7_lambdatimeindep2}) as follows
\begin{equation}
\label{eq:ch7_lambdagamma}
\Lambda = \frac{1}{2} (\gamma_{1} + i\gamma_{3}) \sigma_{z} -
\frac{1}{2} (\gamma_{2} + i\gamma_{4}) \sigma_{y},
\end{equation}
where $\{\gamma_{i}\}$'s are real coefficients:
%
\begin{eqnarray}
\label{eq:ch7_gammas13}
\gamma_{1} + i\gamma_{3}
= \int_{0}^{\infty} d\tau
\, B(\tau )
\cos (2v_{m}\tau ),
\\
\label{eq:ch7_gammas24}
\gamma_{2} + i\gamma_{4}
= \int_{0}^{\infty} d\tau
\, B(\tau )
\sin (2v_{m}\tau ).
\end{eqnarray}
%
%

The density matrix $\rho (t)$ is a $2\times 2$ Hermitian matrix with
unit trace. Hence, it has three real independent components and can be
written as follows:
\begin{equation}
\label{eq:ch7_rhoinsigmas}
\rho = \frac{1}{2} + \sigma_{x}\, \mbox{Re}\, \rho_{12} - \sigma_{y}\,
\mbox{Im}\, \rho_{12} + \sigma_{z}\, (\rho_{11} - \frac{1}{2}).
\end{equation}
Let us substitute Eqs.~(\ref{eq:ch7_rhoinsigmas}) and
(\ref{eq:ch7_lambdagamma}) into the Redfield equation
[Eq.~(\ref{eq:ch7_sigmadot})] and use that $H_{S} = v_{m}\sigma_{x}$ and
$K = \frac{1}{2}\sigma_{z}$. A simple algebraic manipulation leads to
three differential equations,
\begin{eqnarray}
\label{eq:ch7_de1}
\mbox{Re}\, {\dot \rho}_{12} &=& - \gamma_{1}\, \mbox{Re}\, \rho_{12}
+ \frac{\gamma_{4}}{2},
\\
\label{eq:ch7_de2}
{\dot \rho}_{11} &=& -2v_{m}\, \mbox{Im}\, \rho_{12},
\\
\label{eq:ch7_de3}
\mbox{Im}\, {\dot \rho}_{12} &=& 
(2v_{m} + \gamma_{2}) (\rho_{11} - \frac{1}{2})
- \gamma_{1}\, \mbox{Im}\, \rho_{12}.
\end{eqnarray}
The initial conditions are $\rho_{11}(0) \!=\! 1$ and $\rho_{12}(0) \!=\! 0$. 
Eq.~(\ref{eq:ch7_de1}) decouples from Eqs.~(\ref{eq:ch7_de2}) and (\ref{eq:ch7_de3}). 
Its solution is given by Eq.~(\ref{eq:ch7_sol2}), where we used the following 
identity: $\gamma_{4}/\gamma_{1}=-\tanh (v_{m}/T)$. Eqs.~(\ref{eq:ch7_de2}) and
(\ref{eq:ch7_de3}) form a closed system. Their solution is given by
Eqs.~(\ref{eq:ch7_sol1}) and (\ref{eq:ch7_sol3}).

The coefficients $\gamma_{1}$ and $\gamma_{2}$ [Eqs.~(\ref{eq:ch7_gamma1})
and (\ref{eq:ch7_gamma2}), respectively] are calculated using
Eqs.~(\ref{eq:ch7_gammas13}), (\ref{eq:ch7_gammas24}), and (\ref{eq:ch7_bath}).

%% file: chapters/bio.tex
\setlength{\textheight}{8.2in}
\clearpage
\begin{biography}
\ssp
  \addcontentsline{toc}{chapter}{Biography}
  \begin{center}
    {\large{\bf Serguei Vorojtsov}}
  \end{center}
  \subsection*{Personal}
    \begin{itemize}
      \item Born on August 16, 1974 in Russia
    \end{itemize}
  \subsection*{Education}
    \begin{itemize}
      \item Ph.D. in Physics, Duke University, Durham, North Carolina, USA, 2005
      \item M.A. in Physics, Duke University, Durham, North Carolina, USA, 2003
      \item M.S. in Physics, Moscow Institute of Physics and Technology, Russia, 1997
    \end{itemize}
  \subsection*{Positions}
    \begin{itemize}
      \item Research Assistant, Duke University, 2000-2004
      \item Teaching Assistant, Duke University, 1998-2000, 2005
    \end{itemize}
  \subsection*{Publications}

\begin{description}

\item ``Coulomb Blockade Conductance Peak Spacings: 
Interplay of Spin and Dot-Lead Coupling,'' 
Serguei Vorojtsov and Harold~U. Baranger, 
{\it Phys. Rev. B}, vol.~72, p.~165349, 2005

\item ``Phonon Decoherence in Quantum Dot Qubits,'' 
Eduardo~R. Mucciolo, Serguei Vorojtsov, and Harold~U. Baranger,
{\it  Proc. SPIE}, vol.~5815, pp.~53-61, 2005

\item ``Phonon Decoherence of a Double Quantum Dot Charge Qubit,'' 
Serguei Vorojtsov, Eduardo~R. Mucciolo, and Harold~U. Baranger,
{\it Phys. Rev. B}, vol.~71, p.~205322, 2005

\item ``Spin Qubits in Multi-Electron Quantum Dots,''
Serguei Vorojtsov, Eduardo~R. Mucciolo, and Harold~U. Baranger,
{\it Phys. Rev. B}, vol.~69, p.~115329, 2004

\item ``Coulomb Blockade Oscillations of Conductance 
at Finite Level Spacing in a Quantum Dot,'' 
Serguei Vorojtsov, 
{\it Int. J. Mod. Phys. B}, vol.~18, pp.~3915-3940, 2004

\item ``Description of an Arbitrary Configuration Evolution 
in the Conway's Game Life in Terms of the Elementary 
Configurations Evolution in the First Generation,'' 
Serguei Vorojtsov, 
{\it PhysTech Journal}, vol.~3, pp.~108-113, 1997

\item ``Magnetoexciton in Coupled Quantum Wells,'' 
Serguei Vorojtsov, 
{\it Master of Science Diploma Thesis}, MIPT, 1997

\end{description}

\end{biography}